\documentstyle[12pt,epsf]{article}

\catcode`\@=11

\def\section{\@startsection{section}{1}{\z@}
  {3.5ex plus 1.0ex minus 0.2ex}{2.3ex plus .2ex}{\normalsize}}
\def\subsection{\@startsection{subsection}{2}{\z@}
  {3.25ex plus 1.0ex minus 0.2ex}{1.5ex plus 0.2ex}{\normalsize\bf}}
\def\subsubsection{\@startsection{subsubsection}{3}{\z@}
  {3.25ex plus 1.0ex minus 0.2ex}{1.5ex plus 0.2ex}{\normalsize\bf}}

\@addtoreset{equation}{section}

\catcode`\@=12

\newcommand{\skipline}{\vspace{\baselineskip}}

\def\fun#1#2{\lower3.6pt\vbox{\baselineskip0pt\lineskip.9pt
  \ialign{$\mathsurround=0pt#1\hfil##\hfil$\crcr#2\crcr\sim\crcr}}}

\begin{document}

\begin{titlepage}

\begin{center}
Hadron Masses from the Valence Approximation\\
to Lattice QCD
\end{center}

\skipline
\skipline

\begin{center}

F. Butler, H. Chen, 
J. Sexton\footnote{permanent address: Department of Mathematics,
Trinity College, Dublin 2,
Republic of Ireland}, 
A. Vaccarino,\\
and D. Weingarten \\

IBM Research \\
P.O. Box 218, Yorktown Heights, NY 10598
\end{center}

\skipline
\skipline

\begin{center}
ABSTRACT
\end{center}

\begin{quotation}

We evaluate pseudoscalar, vector, spin 1/2 and spin 3/2 baryon masses
predicted by lattice QCD with Wilson quarks in the valence (quenched)
approximation for a range of different values of lattice spacing,
lattice volume and quark mass.  Extrapolating these results to
physical quark mass, then to zero lattice spacing and infinite volume we
obtain values for eight mass ratios.  We also determine the zero lattice
spacing, infinite volume limit of an alternate set of five quantities
found without extrapolation in quark mass. Both sets of
predictions differ from the corresponding observed values by amounts
consistent with the predicted quantities' statistical uncertainties.

\end{quotation}

\skipline
\skipline

\end{titlepage}

\section{INTRODUCTION} \label{sec:intro}

In a recent letter~\cite{Butler93} we summarized a lattice QCD
calculation of the masses of eight low-lying hadrons, extrapolated to
physical quark mass, zero lattice spacing, and infinite volume, using
Wilson quarks in the valence (quenched) approximation.  This
approximation may be viewed as replacing the momentum and frequency
dependent color dielectric constant arising from quark-antiquark vacuum
polarization with its zero momentum, zero frequency limit and might be
expected to be fairly reliable for low-lying baryon and meson
masses~\cite{Weingar81}.  In the present article we describe the hadron
mass calculation of Ref.~\cite{Butler93} in greater detail.

Each of the eight predicted mass ratios which we obtain is within 6\%
of experiment. The difference between the predicted ratios and
experiment is less than 1.6 times the corresponding statistical
uncertainty.  It appears to us reasonable to take these results as
quantitative confirmation of the mass predictions both of QCD and of the
valence approximation. We believe it is unlikely that the valence
approximation would agree with experiment for eight different mass
ratios yet differ significantly from QCD's predictions including the
full effect of quark-antiquark vacuum polarization.

For the four lattices used in our extrapolations to physical mass
ratios, we calculate hadron masses for a range of quark masses above
$0.3 m_s$, where $m_s$ is the strange quark mass. We do not calculate
hadron masses directly at smaller quark mass because below $0.3 m_s$ our
algorithms become unacceptably slow.  For quark masses ranging from
about $0.3 m_s$ to $1.1 m_s$ we find pseudoscalar meson masses squared,
vector meson masses, and baryon masses to be very close to linear
functions of the quark mass.  The masses of nonstrange hadrons are then
found by extrapolating these linear fits down to $m_n$, the average of
the up and down quark masses.  Recent calculations based on chiral
perturbation theory \cite{Bernard, Sharpe} show 
that at sufficiently small quark mass a peculiarity of the valence
approximation might lead to significant deviations from our linear
extrapolations.  In Ref.~\cite{Weingar94} evidence is discussed which
suggests that this potential difficulty occurs mainly at very small
quark mass and probably does not have an effect on our extrapolations
from $0.3 m_s$ down to $m_n$.  In the present article we will also show
that hadron masses in the real world are quite close to linear functions
of valence quark masses over the range of our extrapolations.  The
linearity of observed hadron masses as functions of valence quark masses
is closely related to the success of the Gell-Mann-Okubo mass formula.

An alternate interpretation of our mass calculations can be made,
however, which does not depend on extrapolation in quark mass.  The
linearity of real world hadron masses as a function of quark mass
implies that all masses in each hadron multiplet are determined by the
first two coefficients of a Taylor series expansion around any quark
mass between $m_n$ and $m_s$.  Our data shows that for the valence
approximation this linearity occurs at least between $0.3 m_s$ and $1.1
m_s$. Our results can then be cast as predictions of the first two
Taylor coefficients for each hadron multiplet at some conveniently
chosen quark mass between $0.3 m_s$ and $1.1 m_s$.  Four of our five
predicted values for these coefficients at the point $(m_n + m_s)/2$
differ from experiment by less than 1.6 times the corresponding
statistically uncertainty. The fifth prediction differs by 2.0 times
its statistical uncertainty.  The predicted constant terms are within
6.5\% of experiment with statistical uncertainties of up to 3.3\%.  The
predicted coefficients of the linear terms in these expansions lie
within 22\% of experiment with statistical uncertainties of up to 22\%.
The coefficients of the linear terms are obtained from small differences
between the masses of hadrons with different quark compositions and as a
result have relatively large statistical errors.

Following Refs.~\cite{Fermilab,Lepage}, we also determine the coupling
constant $g^{(0)}_{\overline{MS}}$ from the lattice coupling constant
$g_{lat}$. From $g^{(0)}_{\overline{MS}}$, by the two-loop
Callan-Symanzik equation, we determine $\Lambda^{(0)}_{\overline{MS}} a$
in units of lattice spacing $a$.  The values we find for the ratio
$(\Lambda^{(0)}_{\overline{MS}} a) / (m_{\rho} a)$ are constant within
statistical errors. Thus $m_{\rho} a$ depends on $a$ according to
asymptotic scaling to within statistical errors.  This result tends to
support the reliability of our extrapolation of masses to the continuum
limit.  The infinite volume, continuum limit of
$\Lambda^{(0)}_{\overline{MS}}/ m_{\rho}$ combined with the value of
$m_{\rho}$ in physical units permits a calculation of
$\Lambda^{(0)}_{\overline{MS}}$ in physical units. The result we obtain
agrees, to within 4\% statistical errors, with a value of the infinite
volume, continuum limit $\Lambda^{(0)}_{\overline{MS}}$ obtained from a
charmonium mass splitting \cite{Fermilab}.

An evaluation of meson decay constants using the data set from which
masses are extracted here is discussed in detail in
Ref.~\cite{Butler94}.

The calculations described here were done on the GF11 parallel computer
at IBM Research \cite{Weingar90} and took approximately one year to
complete.  The machine was used in configurations ranging from 384 to
480 processors, with sustained speeds ranging from 5 Gflops to 7 Gflops.
With the present set of improved algorithms and 480 processors, these
calculations could be repeated in about four months.

\section{COULOMB GAUGE HADRON OPERATORS} \label{sec:ops}

For all but one choice of lattice size and $\beta$, we construct hadron
propagators with nonlocal source and sink operators.  The operators which
we use are specified in lattice Coulomb gauge.  A transformation to
lattice Coulomb gauge is defined to give a local maximum of the sum over
all sites and space direction links
\begin{eqnarray}
\sum_{x, i = 1, 2, 3} Re Tr [ U_i(x)].
\end{eqnarray}
A transformation which produces a local maximum of this
sum is found by a method qualitatively similar to the
Cabbibo-Marinari-Okawa Monte Carlo algorithm. The lattice is swept
repeatedly, and at each site the target function is maximized first by a
gauge transformation in the $SU(2)$ subgroup of $SU(3)$ acting only on
gauge index values 1 and 2, then by a gauge transformation in the
$SU(2)$ subgroup acting only on index values 2 and 3, then by a gauge
transformation in the subgroup acting only on index values 1 and 3.
Maximizing the target function over $SU(2)$ subgroups is easier to
program than a direct maximization over all of $SU(3)$.  On the other
hand, it is not clear that maximizing each site over $SU(3)$ would
significantly accelerate the full transformation to Coulomb gauge.  A
local maximum is reached when at each site the quantity $R( x)$ vanishes
where
\begin{eqnarray}
\label{defqr}
Q( x) & = & \sum_i [ U_i( x) - U_i^{\dagger}( x - \hat{i}) ], \nonumber \\
R( x) & = & Q( x) - Q^{\dagger}( x) - \frac{2}{3} Im Tr[ Q( x)]. 
\end{eqnarray}
The vector $\hat{i}$ is a unit lattice vector in the positive $i$
direction.  We stop the iteration process when the sum over the lattice
of the quantity $Tr [ R^{\dagger}( x) R( x)] $ becames smaller than a
convergence parameter $c$.

In Coulomb gauge, a smeared quark field $\phi_r(\vec{x},t)$ is then
constructed from the local quark field $\psi(\vec{x},t)$ by
\begin{eqnarray}
\label{defphi}
\phi_r(\vec{x},t) & = & \sum_{\vec{y}} G_r(\vec{x} - \vec{y})
\psi(\vec{y},t), \nonumber \\ 
G_r(\vec{z}) & = & (\sqrt{\pi}r)^{-3} exp( - \frac{|\vec{z}|^2}{ r^2}).
\end{eqnarray}
Spin, flavor and color indices have been suppressed in
Eq.~(\ref{defphi}).  The field $\overline{\phi}_r(\vec{x},t)$ is defined
by a corresponding smearing of $\overline{\psi}(\vec{x},t)$.  We take
the smeared fields $\phi_0(x)$ and $\overline{\phi}_0(x)$ to be
$\psi(x)$ and $\overline{\psi}(x)$, respectively.

Smeared hadron fields can then be formed from local products of
smeared quark and antiquark fields.  The fields for a charged pion
and a changed rho are
\begin{eqnarray}
\label{defpirho}
\pi^+_r(x) & = & \overline{\phi}^d_r(x) \gamma^5 \phi^u_r(x) \nonumber \\
\rho^{+ i}_r(x) & = & \overline{\phi}^d_r(x) \gamma^i \phi^u_r(x).
\end{eqnarray}
For a proton, antiproton, $\Delta^{++}$ or $\overline{\Delta^{++}}$ with
with z-component of spin given by $s$
we have the nonrelativistic operators
\begin{eqnarray}
\label{defPDel}
P^s_r(x) & = &  
\phi^u_{a i r}(x) \phi^u_{b j r}(x) \phi^d_{c k r}(x)
\epsilon_{a b c} \Gamma^{P s}_{i j k} \nonumber \\
\overline{P}^s_r(x) & = &  
\overline{\phi}^u_{a i r}(x) \overline{\phi}^u_{b j r}(x) 
\overline{\phi}^d_{c k r}(x)
\epsilon_{a b c} \Gamma^{\overline{P} s}_{i j k} \nonumber \\
\Delta^s_r(x) & = &  
\phi^u_{a i r}(x) \phi^u_{b j r}(x) \phi^u_{c k r}(x)
\epsilon_{a b c} \Gamma^{\Delta s}_{i j k} \\
\overline{\Delta}^s_r(x) & = &  
\overline{\phi}^u_{a i r}(x) \overline{\phi}^u_{b j r}(x) 
\overline{\phi}^u_{c k r}(x)
\epsilon_{a b c} \Gamma^{\overline{\Delta} s}_{i j k} \nonumber 
\end{eqnarray}
where $a$, $b$ and $c$ are color indices, $i$, $j$, and $k$ are
spinor indices, $\epsilon_{a b c}$ is the alternating index, and
repeated indices are summed over.
For gamma matrices which are the same as the Bjorken and Drell
convention, but with space direction matrices multiplied by
$i$, the nonzero components of the baryon spin wave functions 
with maximum spin in the z-direction are
\begin{eqnarray}
\label{defgamma}
\Gamma^{P 1/2}_{1 1 2} & = & -\Gamma^{P 1/2}_{1 2 1} = 1, \nonumber \\
\Gamma^{\overline{P} 1/2}_{4 4 3} & = & -\Gamma^{\overline{P} 1/2}_{4 3 4} = 1, 
\nonumber \\
\Gamma^{\Delta 3/2}_{1 1 1} & = & 1, \\
\Gamma^{\overline{\Delta} 3/2}_{4 4 4} & = & 1. \nonumber 
\end{eqnarray}
Corresponding smeared fields can be defined for other nucleon and delta
charge and spin sates, and for other pseudoscalar mesons, vector mesons,
and spin 1/2 and 3/2 baryons.

Hadron field operators obtained
from local products of gaussian smeared fields, when fourier
transformed, are equivalent to operators formed from local fields with 
nonlocal gaussian relative wave functions.  For a pion at rest, for
example, we have
\begin{eqnarray}
\sum_{\vec{x}} \pi^{+}_ r(\vec{x},t) = \sum_{\vec{y} \vec{z}} G_s(\vec{z})
\overline{\psi}^u (\vec{y}, t) \gamma^5
\psi^d (\vec{y} + \vec{z}, t), 
\end{eqnarray}
where $s$ is $\sqrt{2} r$.

From Coulomb gauge smeared fields for each hadron we constructed zero momentum,
spin summed hadron propagators for a collection of source and sink sizes.
For a hadron $h$, the propagator is
\begin{eqnarray}
\label{defc}
c^h_{r r'}( t) & = & \sum_{\vec{x} s} < [h^s_r(\vec{x},t)]^\dagger 
h^s_{r'}( 0, 0) > 
\end{eqnarray}
The invariance of lattice QCD under parity
and charge conjugation transformations gives a variety of equalities
among various pairs of propagators. 
Propagators which should be equal
can be added together to decrease statistical fluctuations. 
For a
lattice with time direction period $T$, our final
zero-momentum propagators are then
\begin{eqnarray}
\label{defC}
C^{\pi}_{r r'}( t) & = & c^{\pi}_{r r'}( t) + c^{\pi}_{r r'}( T - t),
\nonumber \\ 
C^{\rho}_{r r'}( t) & = & c^{\rho}_{r r'}( t) + c^{\rho}_{r r'}( T - t),
\nonumber \\
C^P_{r r'}( t) & = & c^P_{r r'}( t) + c^{\overline{P}}_{r r'}(t), \\
C^{\Delta}_{r r'}( t) & = & c^{\Delta}_{r r'}( t) + c^{\overline{\Delta}}_{r r'}(t), 
\nonumber
\end{eqnarray}
with corresponding definitions for other mesons and baryons.

For a lattice with a sufficiently large space direction periodicity
$S$, statistical fluctuations in hadron propagators can, in principle,
be further decreased by introducing several
gaussian sources, each multiplied
by a random cube root of 1 to cancel cross terms between the propagation
of different sources for both baryon and meson propagators. A closely
related idea was first proposed in Ref.~\cite{wall}.
If $S$ is
large and even, for example, a quark field with eight sources is
\begin{eqnarray}
\label{defphiprime}
\phi'_r(t)  =  \sum_{\vec{y}} \xi_{\vec{y}} \phi_r(\vec{y},t),
\end{eqnarray}
for $\phi_r$ of Eq.~(\ref{defphi}).  Each component of the eight
$\vec{y}$ included in the sum in Eq.~(\ref{defphiprime}) is either 0 or
$S/2$, and the $\xi_{\vec{y}}$ for each different $\vec{y}$ and each
different gauge configuration used to find quark
propagators are independent random cube roots of 1. Using $\phi'_r$ of 
Eq.~(\ref{defphiprime}) to define hadron source operators but with
$\phi_r$ of Eq.~(\ref{defphi}) still used to define sink operators,
a new set of hadron propagators 
$C'^h_{r r'}( t)$, $h = \pi, \rho, P, \Delta$,
can be constructed similar to those in
Eq.~(\ref{defC}). If $S$ is sufficiently large, however, the
$C'^h_{r r'}( t)$ found from any ensemble of gauge fields will behave 
as though they have been averaged over eight times as many independent
gauge configurations as corresponding $C^h_{r r'}( t)$

\section{PROPAGATORS} \label{sec:props}

Table~\ref{tab:lattices} lists the lattice sizes and parameter values
for which quark and hadron propagators were evaluated. Up and down quark
masses were taken to be equal in all propagators. We therefore obtained
degenerate masses for each isospin multiplet of hadrons. This
approximation leads to almost no loss in useful results since since the
observed values of mass splittings in isospin multiplets in the real
world are generally somewhat smaller than our statistical errors.  We
also did no calculations directly including strange quarks.  Masses for
hadrons including strange quarks were found by an extrapolation in quark
mass to be discussed in Section~\ref{sec:strange}. Direct calculations
including strange quarks would have required additional programming but
would have increased our required computer time by only a small
fraction.

We chose periodic boundary conditions in all directions for both gauge
fields and quark fields.  Gauge configurations were generated using a
version of the Cabbibo-Marinari-Okawa~\cite{CMO} algorithm adapted for
parallel computers by Ding~\cite{Ding}. The number of sweeps skipped
between configurations and the total count of configurations for each
set of lattice parameters are given in the last two columns of the
table.  Values of the correlation between hadron propagators on
successive pairs of gauge configurations we found to be statistically
consistent with zero.  Thus the number of sweeps between configurations
in each case was sufficient to produce statistically independent hadron
propagators on all gauge configurations used in the calculation of
propagators.

For the $8^3 \times 32$ lattice at $\beta$ of 5.7, we used point sources
and sinks in the quark propagators. For all other lattices we used
gaussian sources with $r$ of 2, and four different gaussian sinks, with
$r$ of 0, 1, 2 and 3. For all lattices except $24^3 \times 32$ at
$\beta$ of 5.7, we found propagators for the source fields of
Eq.~(\ref{defphi}) including a single source.  For $24^3 \times 32$ at
$\beta$ of 5.7 we found propagators for the source fields of
Eq.~(\ref{defphiprime}) including eight sources.

For all the lattices listed in Table~\ref{tab:lattices}, the gauge
transformation convergence parameter $c$ defined following
Eq.~(\ref{defqr}) was set to $10^{-5}$.  The average number of sweeps
required for convergence to this accuracy ranged from 1525 on the
lattice $24^3 \times 36$ at $\beta$ of 5.93, to 2270 on the lattice
$16^3 \times 32$ at $\beta$ of 5.7.

Quark propagators were constructed using the conjugate gradient
algorithm for the $8^3 \times 32$ lattice at $\beta$ of 5.7, using
red-black preconditioned conjugate gradient for the other lattices at
$\beta$ of 5.7 and 5.93, and using a red-black preconditioned minimum
residual algorithm at $\beta$ of 6.17~\cite{Algs}.  At the largest
hopping constant values at $\beta$ of 5.7 and 5.93, preconditioning the
conjugate gradient algorithm increased its speed by a factor of 3, and
at the largest hopping constant values at $\beta$ of 6.17, the change
from conjugate gradient to the minimum residual algorithm yielded an
additional factor of 2 in speed.  Table~\ref{tab:sweeps} gives
the average number of lattice sweeps
needed to find quark propagators for all of the lattice sizes and parameter
values which we use to find infinite volume continuum limit masses.
The number of sweeps in all cases was chosen
large enough to insure that effective pion, rho, nucleon and delta
masses evaluated within the time interval used for final fits are 
within 0.2\% of their values obtained on propagators run to machine
precision. The number of sweeps required to obtain this precision was
found by calculating propagators with several different convergence
criteria on small ensembles of configurations.

\begin{figure}
\epsfxsize=\textwidth \epsfbox{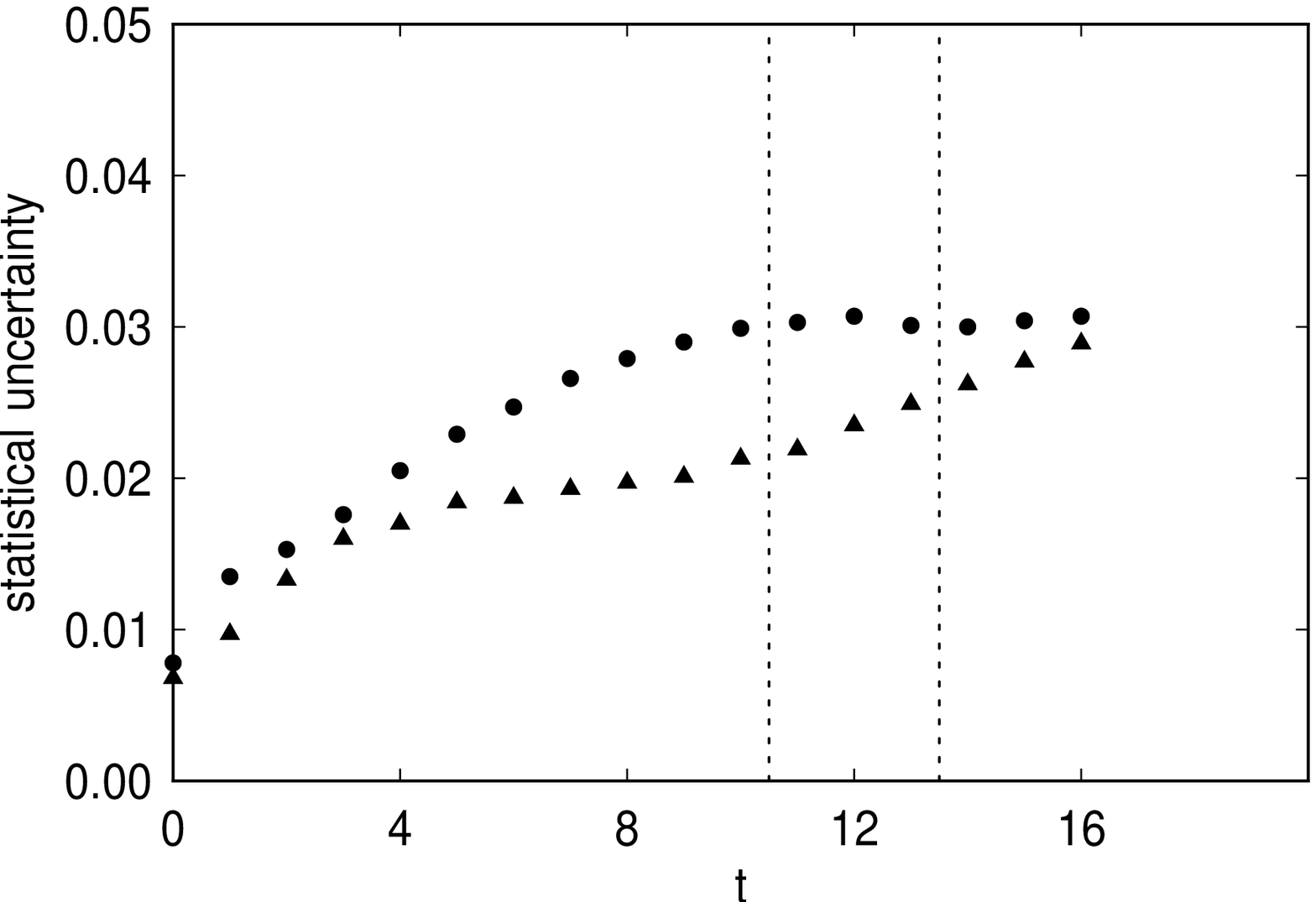}
\caption{
Fractional uncertainty in the pion propagator obtained from one source (circles)
and from eight sources (triangles) on the lattice $16^3 \times 32$ at $\beta = 5.70$
and $k = 0.1650$.}
\label{fig:pik165x16src}
\end{figure}

\begin{figure}
\epsfxsize=\textwidth \epsfbox{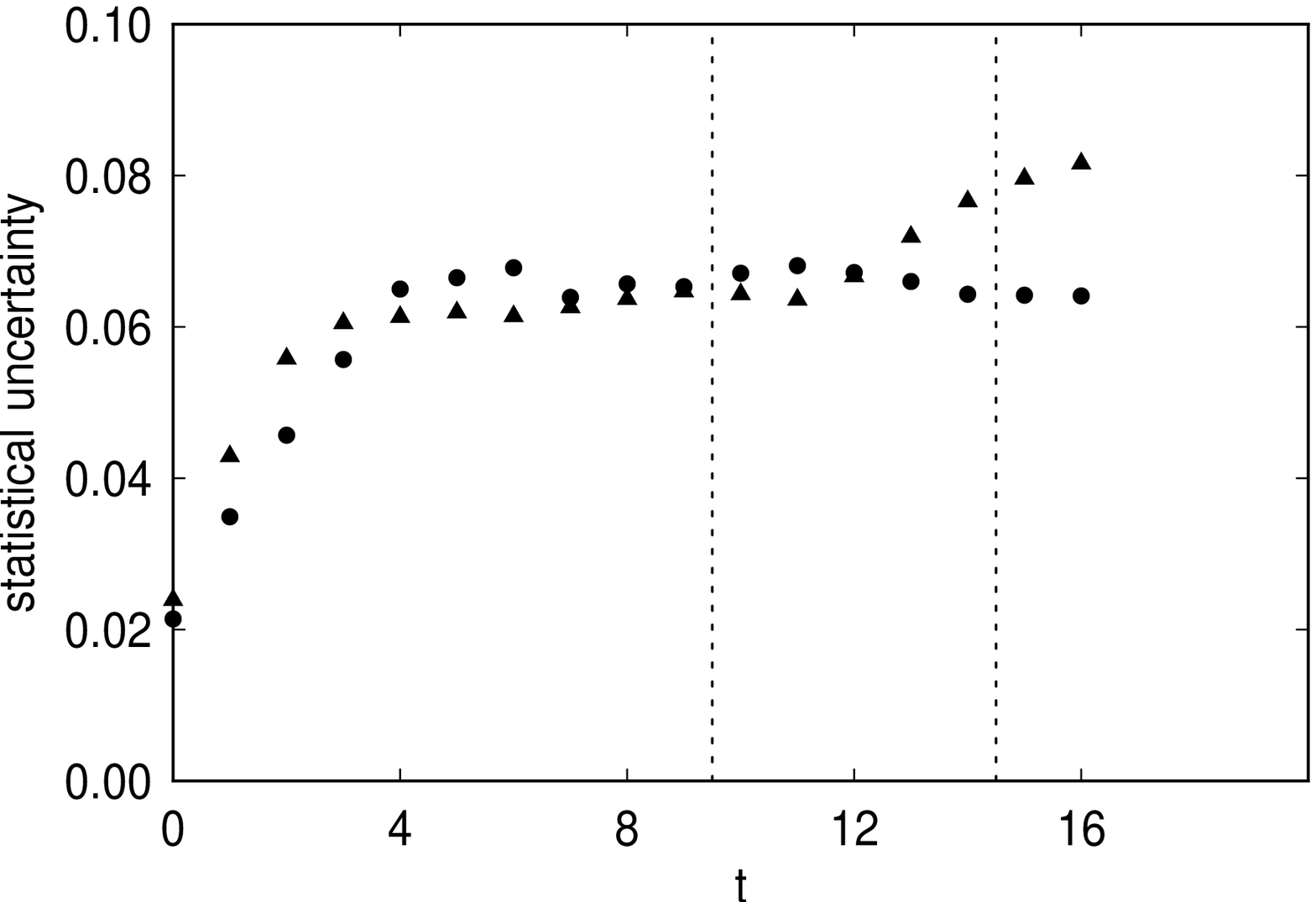}
\caption{
Fractional uncertainty in the pion propagator obtained from one source (circles)
and from eight sources (triangles) on the lattice $16^3 \times 32$ at $\beta = 5.70$
and $k = 0.1675$.}
\label{fig:pik167x16src}
\end{figure}

\begin{figure}
\epsfxsize=\textwidth \epsfbox{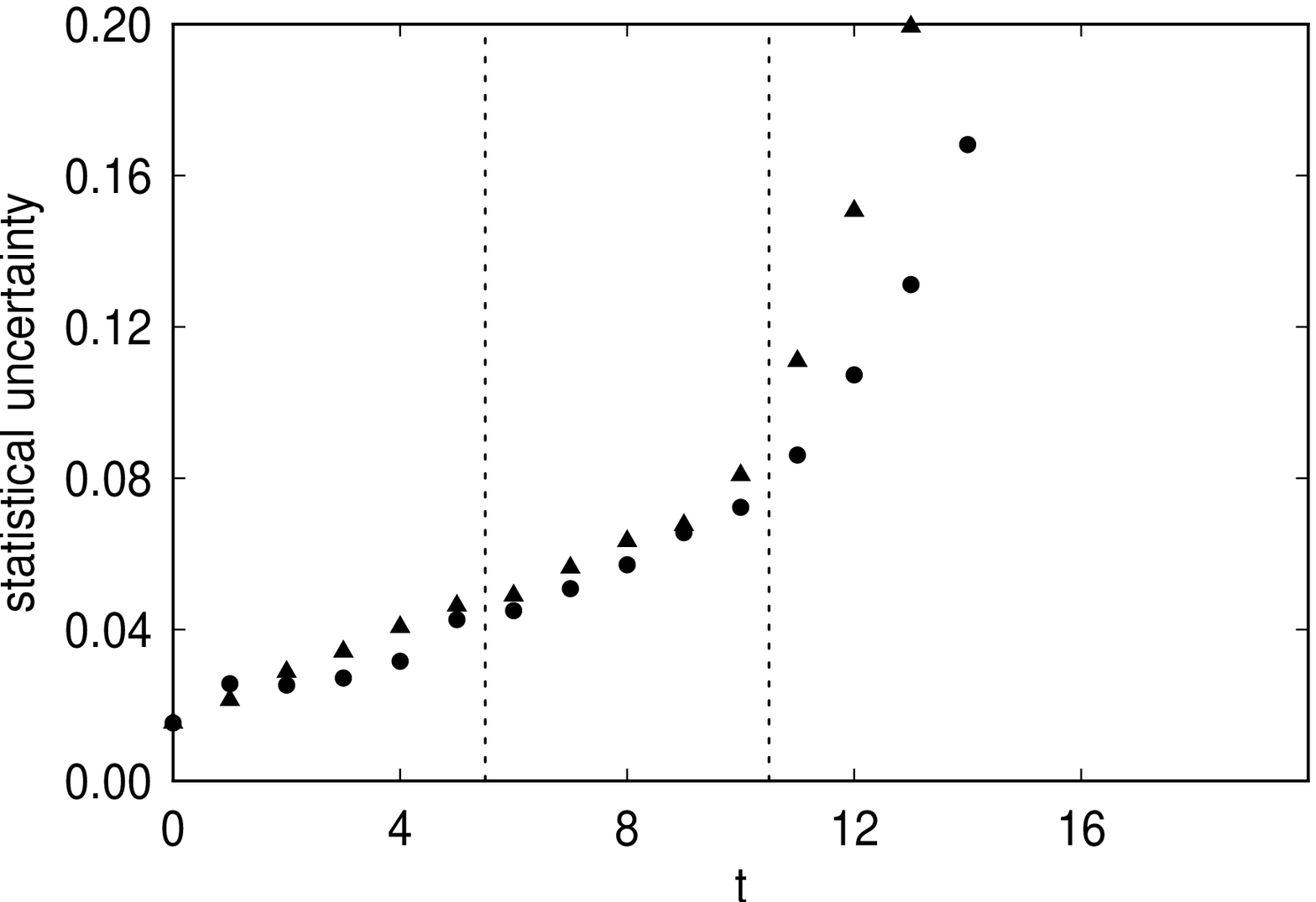}
\caption{
Fractional uncertainty in the nucleon propagator obtained from one
source (circles)
and from eight sources (triangles) on the lattice $16^3 \times 32$ at $\beta = 5.70$
and $k = 0.1650$.}
\label{fig:prk165x16src}
\end{figure}

\begin{figure}
\epsfxsize=\textwidth \epsfbox{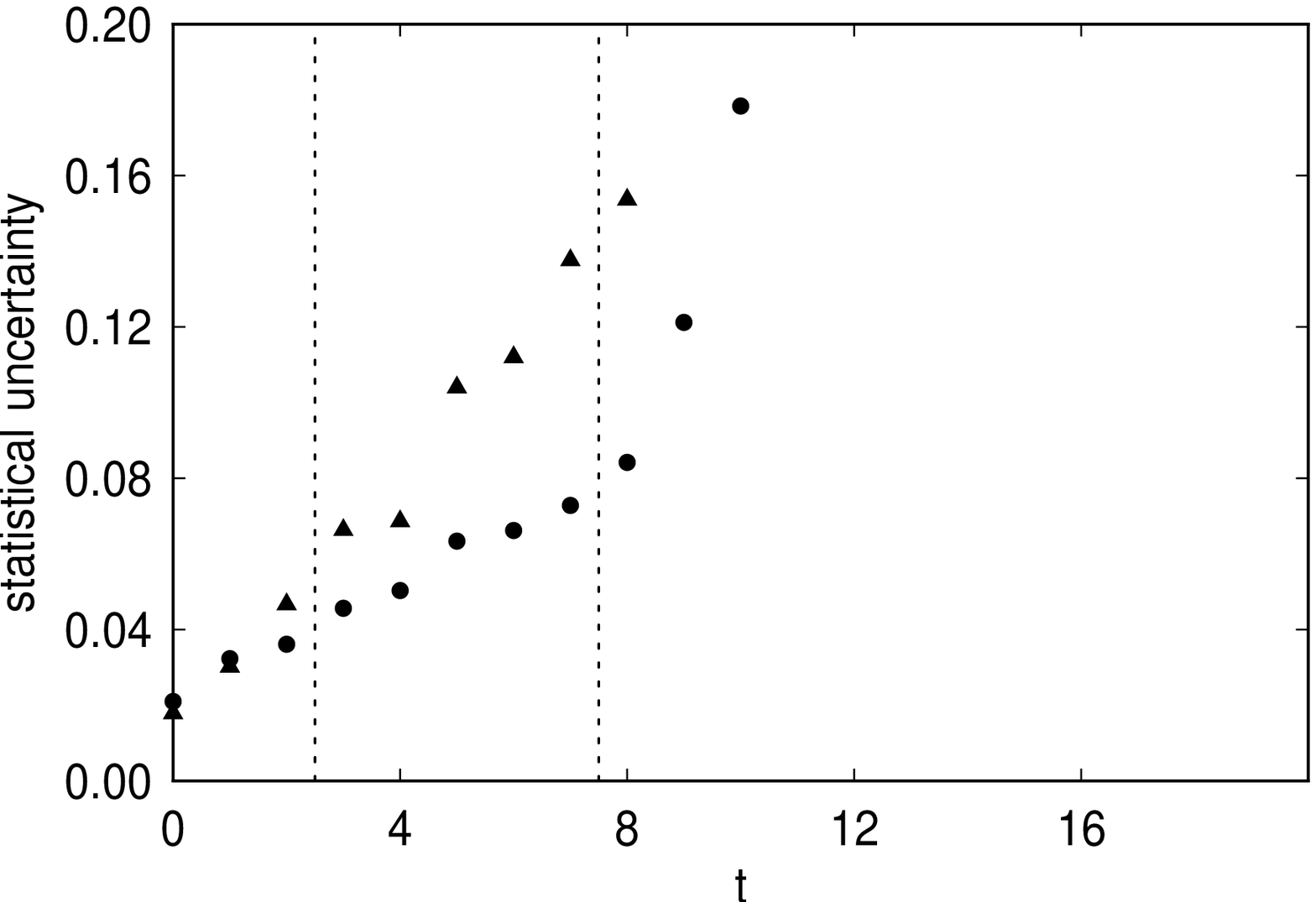}
\caption{
Fractional uncertainty in the nucleon propagator obtained from one
source (circles)
and from eight sources (triangles) on the lattice $16^3 \times 32$ at $\beta = 5.70$
and $k = 0.1675$.}
\label{fig:prk167x16src}
\end{figure}

\begin{figure}
\epsfxsize=\textwidth \epsfbox{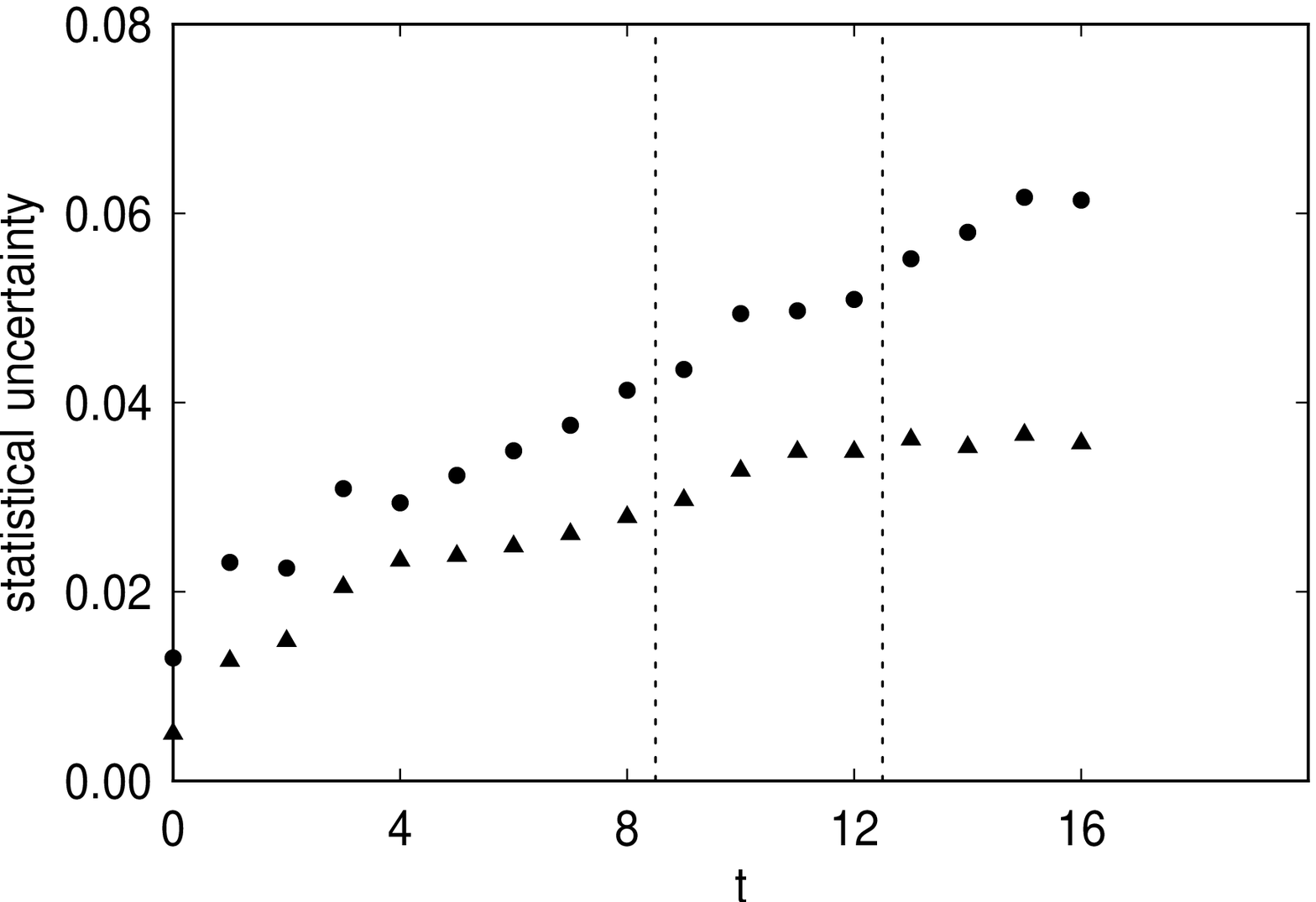}
\caption{
Fractional uncertainty in the pion propagator obtained from one source (circles)
and from eight sources (triangles) on the lattice $24^3 \times 32$ at $\beta = 5.70$
and $k = 0.1650$.}
\label{fig:pik165x24src}
\end{figure}

\begin{figure}
\epsfxsize=\textwidth \epsfbox{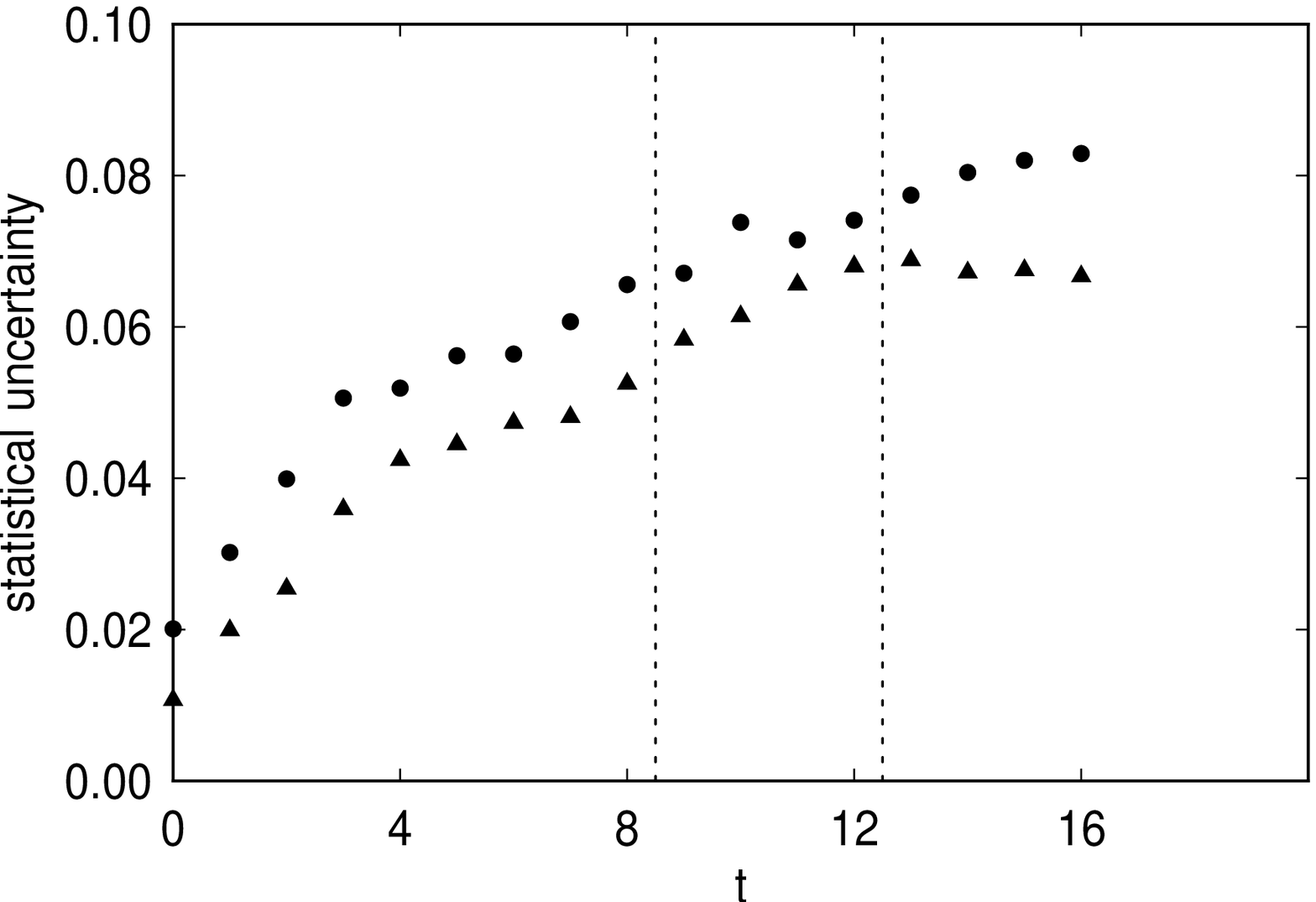}
\caption{
Fractional uncertainty in the pion propagator obtained from one source (circles)
and from eight sources (triangles) on the lattice $24^3 \times 32$ at $\beta = 5.70$
and $k = 0.1675$.}
\label{fig:pik167x24src}
\end{figure}

\begin{figure}
\epsfxsize=\textwidth \epsfbox{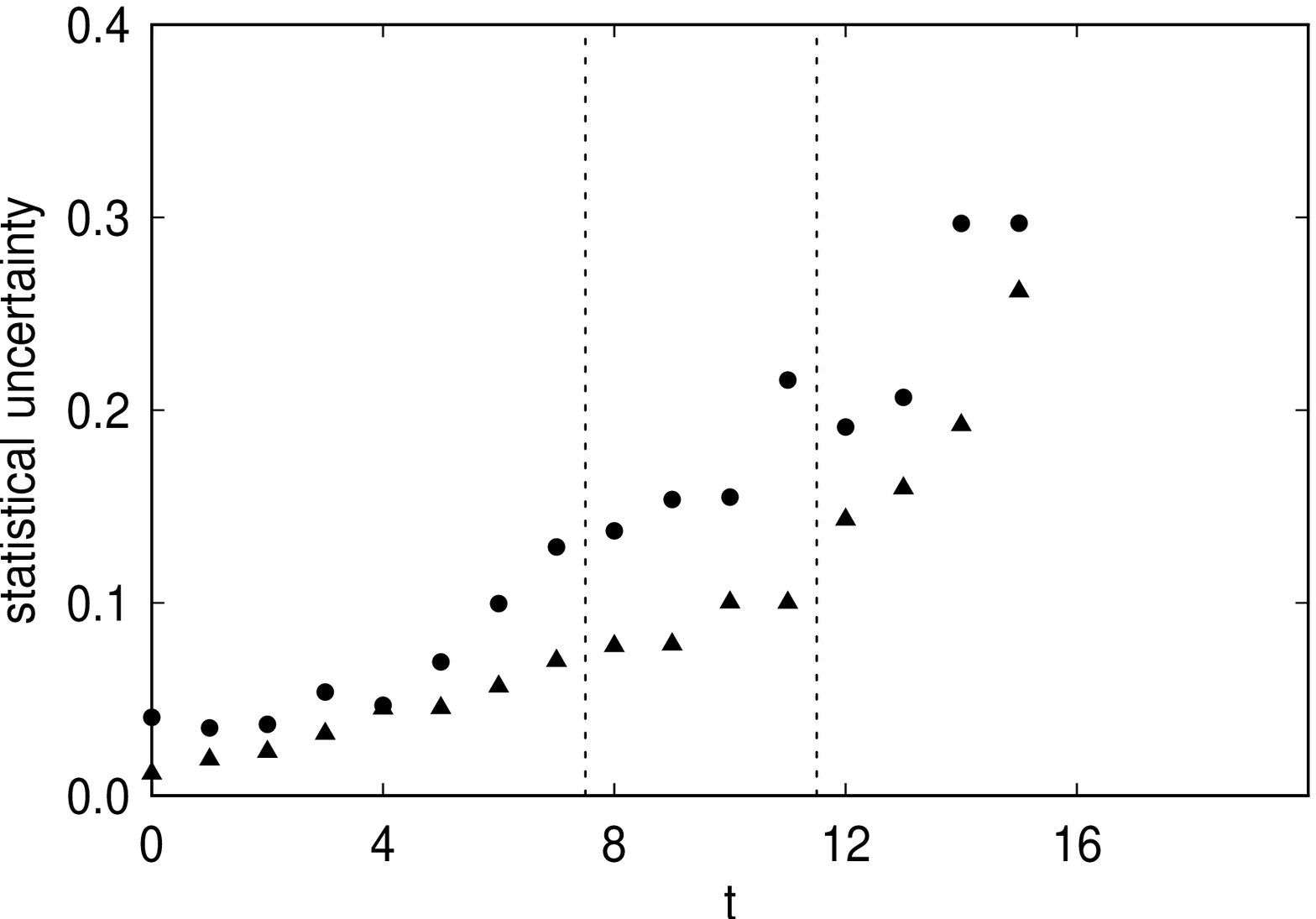}
\caption{
Fractional uncertainty in the nucleon propagator obtained from one
source (circles)
and from eight sources (triangles) on the lattice $24^3 \times 32$ at $\beta = 5.70$
and $k = 0.1650$.}
\label{fig:prk165x24src}
\end{figure}

\begin{figure}
\epsfxsize=\textwidth \epsfbox{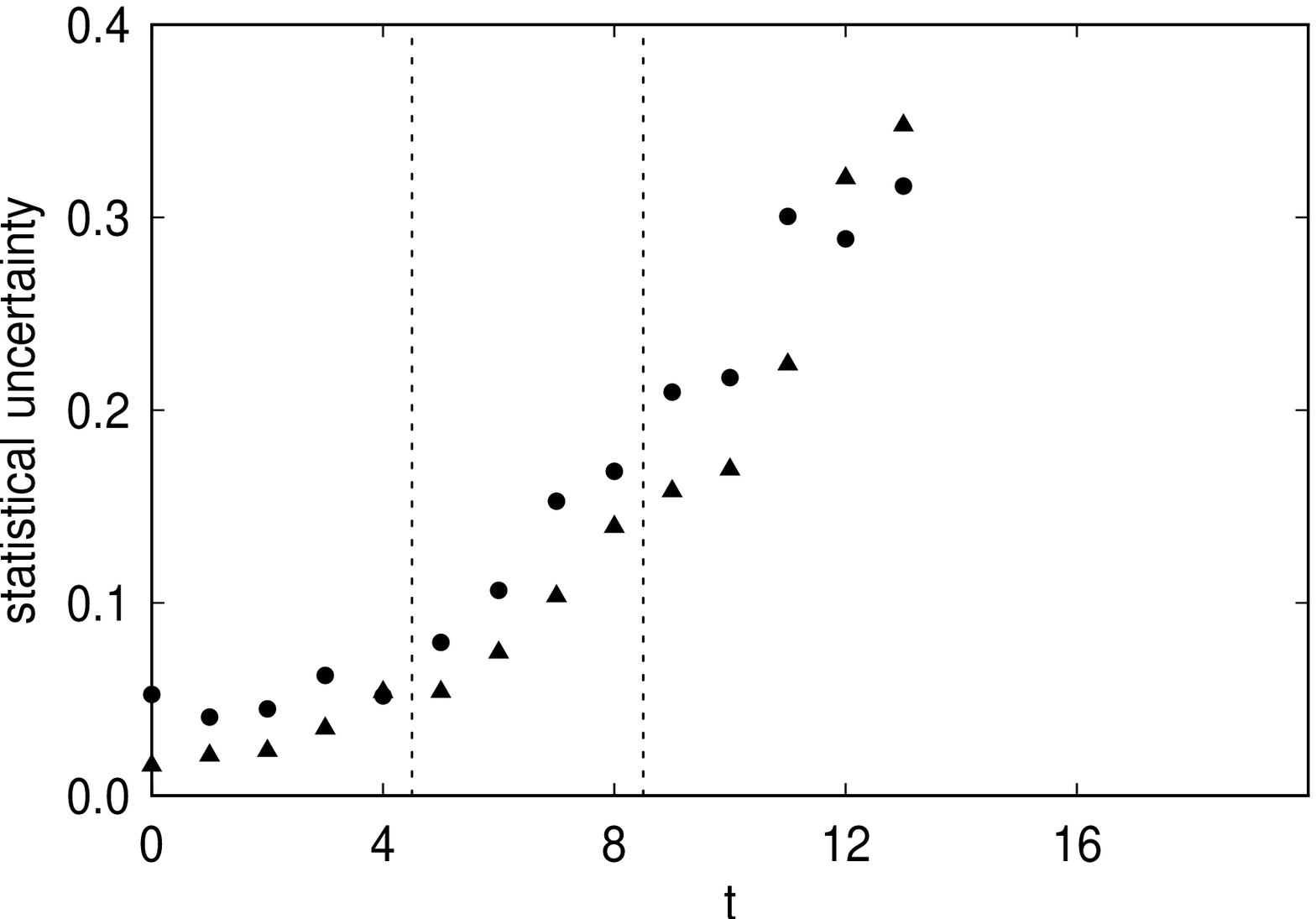}
\caption{
Fractional uncertainty in the nucleon propagator obtained from one
source (circles)
and from eight sources (triangles) on the lattice $24^3 \times 32$ at $\beta = 5.70$
and $k = 0.1675$.}
\label{fig:prk167x24src}
\end{figure}

At $\beta$ of 5.7 we also found propagators for the lattice $16^3 \times
32$ using a source field including eight sources for an ensemble of 81
configurations.  For the lattice $24^3 \times 32$ additional propagators
were found using a source field including only a single source for an
ensemble of 18 configurations. Figures \ref{fig:pik165x16src} -
\ref{fig:prk167x24src} show propagator statistical dispersions divided by
propagators for various combinations of sources and lattice sizes.  For
the lattice $16^3 \times 32$, Figures~\ref{fig:pik165x16src} and
\ref{fig:pik167x16src} compare the dispersion in the pion
propagators found from a single source and found from eight sources, in
both cases using 81 gauge field configurations.
Figures~\ref{fig:prk165x16src} and
\ref{fig:prk167x16src} show this comparison for the proton propagator.
Figures~\ref{fig:pik165x16src} - \ref{fig:prk167x16src} compare single
source and eight source propagators for the lattice $24^3 \times 32$
using 18 gauge field configurations.
The range parameter $r$ of Eq.~(\ref{defphi}) has the value 2 for all
sources and sinks in these figures.  The vertical dashed lines in each
figure mark the fitting intervals, to be discussed later, which we found
to be optimal for determining the corresponding hadron mass.  Figures
\ref{fig:pik165x16src} - \ref{fig:prk167x24src} show that except for
the pion at $k$ of 0.1650, the smallest errors within the fitting
intervals for propagators on the lattice $16^3 \times 32$ are obtained
with a single source.
For the lattice $24^3 \times 32$, on the other hand,
eight sources give significantly smaller statistical errors within the
fitting interval at $k$ of 0.1650 and at least somewhat smaller errors
at $k$ of 0.1675.  
Similar results to those shown in Figures \ref{fig:pik165x16src} -
\ref{fig:prk167x24src} for pion and nucleon propagators are also
obtained for rho and delta baryon propagators.

Overall it appears that single sources are the most efficient choice for
hadron propagators for the lattice $16^3 \times 32$ at $\beta$ of 5.7
and therefore also for $24^3 \times 36$ at $\beta$ of 5.93 and $30
\times 32^2 \times 40$ at 6.17, which have nearly the same volume in physical
units as $16^3 \times 32$ at $\beta$ of 5.7.  For the larger physical
volume of $24^3 \times 32$ at $\beta$ of 5.7, however, it appears that
eight sources are a more efficient choice.

\section{HADRON MASSES} \label{sec:mass}

Hadron masses were determined by fitting hadron propagators to their
asymptotic forms, for large values of $t$ and the lattice time period
$T$.  For any meson $M$ and any baryon $B$,
\begin{eqnarray}
\label{masym}
C^M_{r r'}( t) & \rightarrow & Z^M_{r r'} \{
exp( -m_M t) + exp[ -m_M ( T - t)]\}, \\ 
\label{basym}
C^B_{r r'}( t) & \rightarrow & Z^B_{r r'} 
exp( -m_B t).
\end{eqnarray}
Here $r$ and $r'$ are the smearing parameters for propagator sink and
source, respectively. In the asymptotic form for meson propagators we
include a contribution from backward propagation across the lattice's
periodic boundary but omit this term in baryon propagators.  Backward
propagating baryons arising from the operators of Eqs.~(\ref{defPDel}) are
opposite parity excitations with a larger mass than the forward
propagating ground state.  For large enough $T$, with $t < T$, the
omitted term is negligible. Including the backward term, on the other
hand, would require fitting four parameters to baryon propagators rather
than two.

To determine the range of time separations to fit to the asymptotic
forms of Eqs.~(\ref{masym}) and \ref{basym}, we evaluated for each
hadron effective masses $m( t)$ defined to be the result of fitting the
corresponding propagators to Eqs.~(\ref{masym}) and \ref{basym} at the
pair of successive time values $t$ and $t+1$.  The largest interval at
large $t$ showing an approximate plateau in an effective mass was chosen
as the initial trial fitting range for the corresponding propagator. In
all cases the initial trial fitting range included more than four values
of $t$.  An automatic fitting program was then used to choose the final
fitting range within the initial trial range.  For each possible
interval including at least four values of $t$ within the trial fitting
range, the program chose the parameters in Eqs.~(\ref{masym}) and
\ref{basym} which minimize the full correlated $\chi^2$ of the fit to
the data.  The interval giving the smallest value of $\chi^2$ per fitted
degree of freedom and the corresponding mass were then chosen as the
final fitting range and mass prediction.

Statistical uncertainties of parameters obtained from fits and of any
function of these parameters were determined by the bootstrap method
\cite{Efron}.  From each ensemble of N gauge configurations, 100
bootstrap ensembles were generated. Each bootstrap ensemble consists of
a set of N gauge configurations randomly selected from the underlying N
member ensemble allowing repeats. For each bootstrap ensemble the entire
fit was repeated, including a possibly new choice of the final fitting
interval.  The collection of 100 bootstrap ensembles thus yields a
collection of 100 values of any fitted parameter or any function of any
fitted parameter.  The statistical uncertainty of any parameter is taken
to be half the difference between a value which is higher than all but
15.9\% of the bootstrap values and a value which is lower than all but
15.9\% of the bootstrap values. In the limit of large N, the collection
of bootstrap values of a parameter $p$ approaches a gaussian
distribution and the definition we use for statistical uncertainty
approaches the dispersion $\sqrt{< p^2 > - < p >^2}$.

In the absence of some independent method for determing the predictions
of QCD, it appears inevitable that the choice of $t$ interval on which
to fit data to a large $t$ asymptotic form must be made by some
procedure which depends on the Monte Carlo data itself.  Thus the
statistical uncertainties in the data lead to a corresponding
uncertainty in the choice of fitting interval which, in turn, could lead
to an additional uncertainty in the fitted result.  Another advantage of
our procedure for choosing the fitting interval combined with bootstrap
evaluation of statistical uncertainties is that the values we obtain for
statistical uncertainties include the uncertainty arising from the
choice of fitting interval.  A comparison of the error bars found for
our final fits with the error bars found using the same fitting range
held fixed across the bootstrap ensemble shows that typically about 10\%
of the final statistical uncertainty comes from fluctuations over the
bootstrap ensemble of the fitting range itself.

\begin{figure}
\epsfxsize=\textwidth \epsfbox{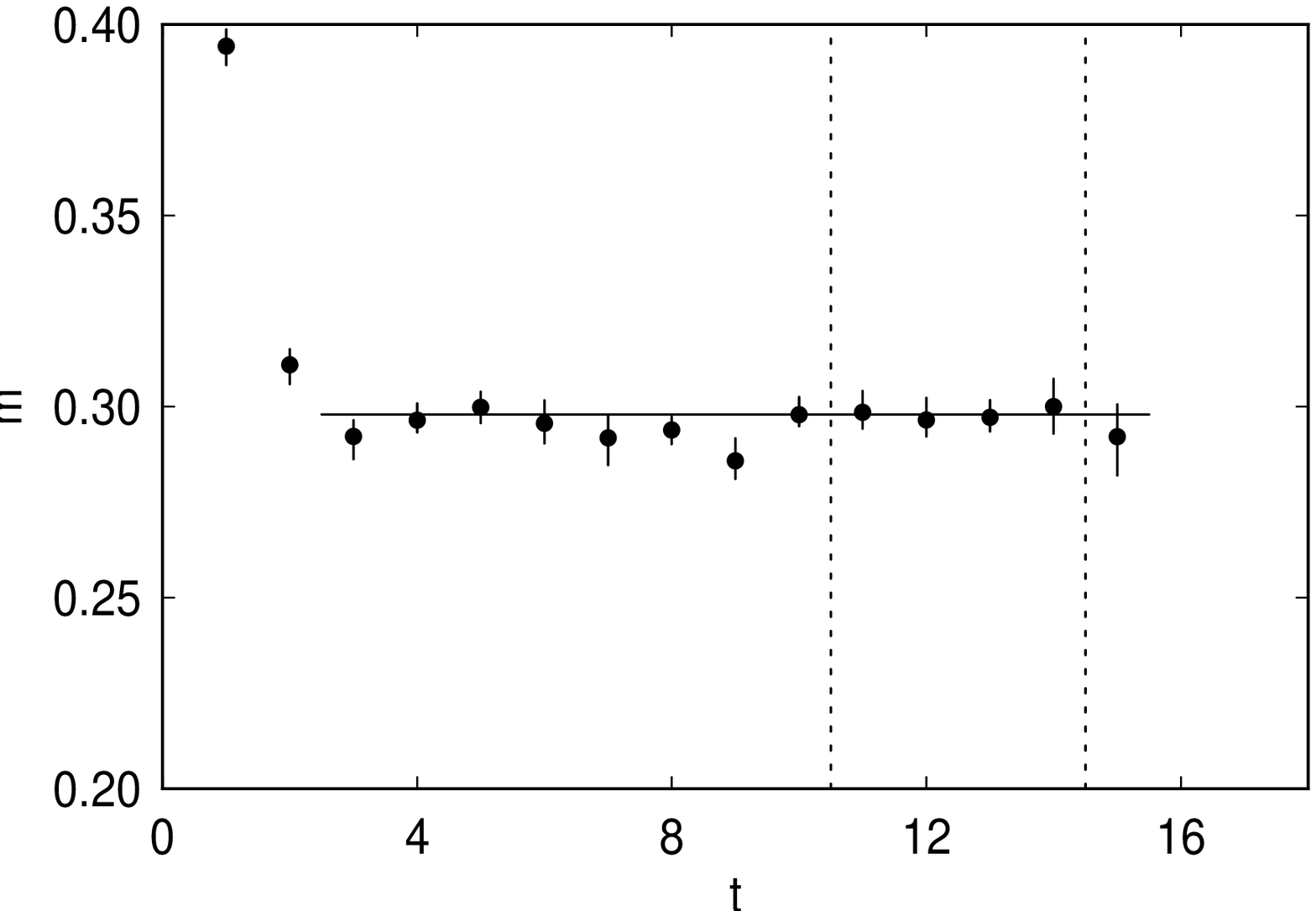}
\caption{ 
Effective masses, final fitting range and fitted mass for the
pseudoscalar propagator with sink size 2 on the lattice $16^3 \times 32$
at at $\beta = 5.70$ and $k = 0.1675$}
\label{fig:pimeffs2x16}
\end{figure}

\begin{figure}
\epsfxsize=\textwidth \epsfbox{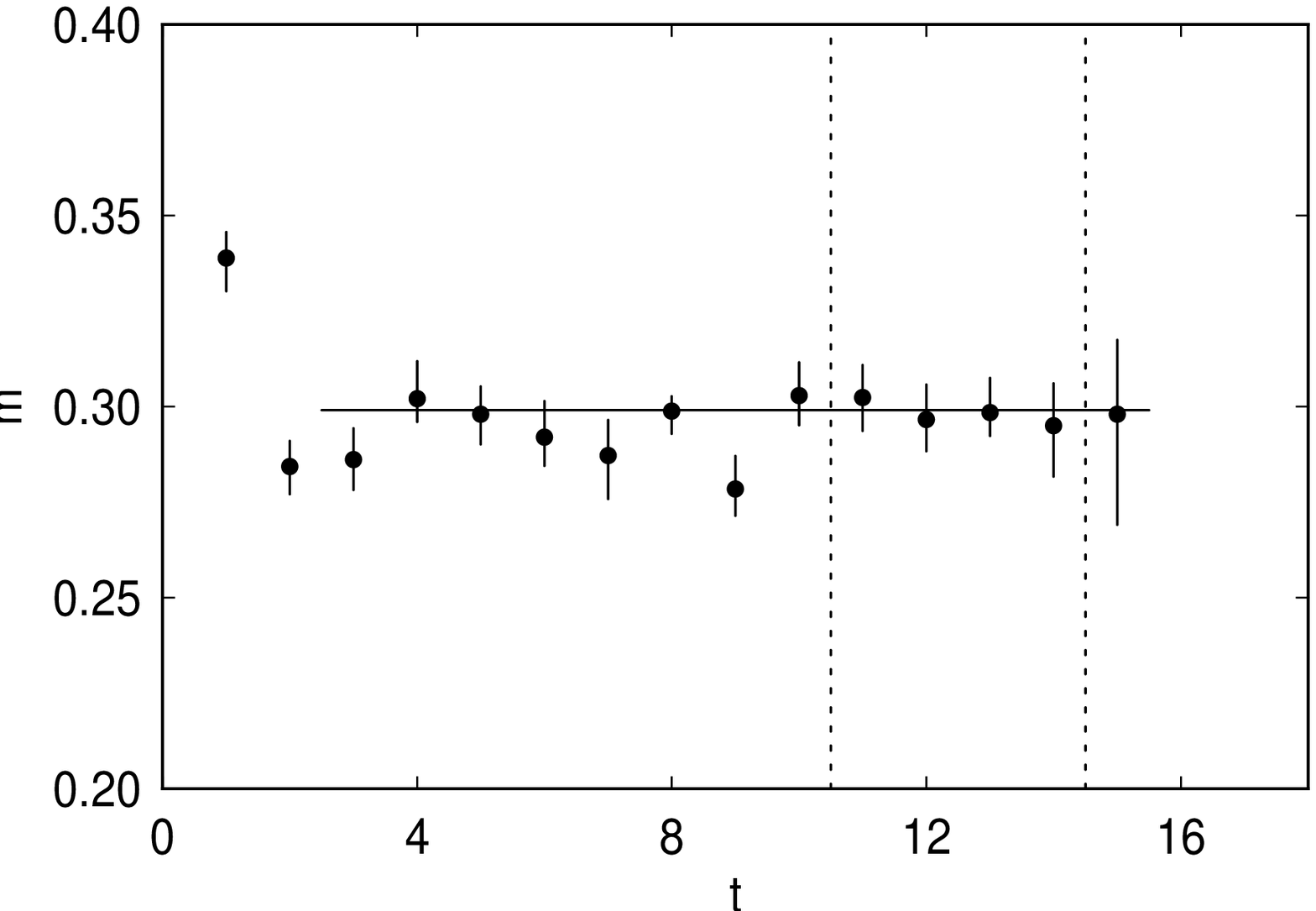}
\caption{ 
Effective masses, final fitting range and fitted mass for the
pseudoscalar propagator with sink size 4 on the lattice $16^3 \times 32$
at at $\beta = 5.70$ and $k = 0.1675$}
\label{fig:pimeffs4x16}
\end{figure}

\begin{figure}
\epsfxsize=\textwidth \epsfbox{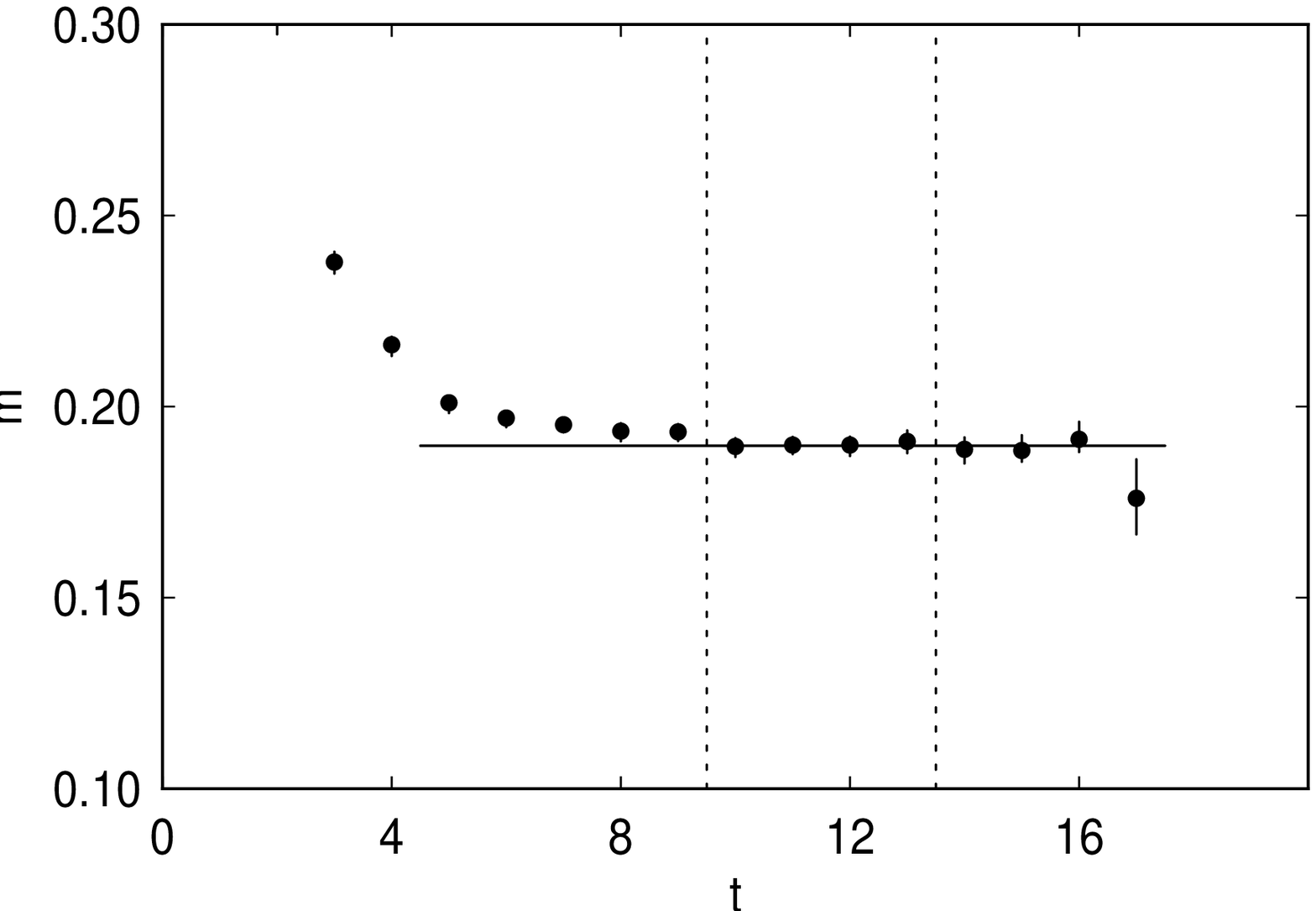}
\caption{ 
Effective masses, final fitting range and fitted mass for the
pseudoscalar propagator with sink size 2 on the lattice $24^3 \times 36$
at at $\beta = 5.93$ and $k = 0.1581$}
\label{fig:pimeffs2x24}
\end{figure}

\begin{figure}
\epsfxsize=\textwidth \epsfbox{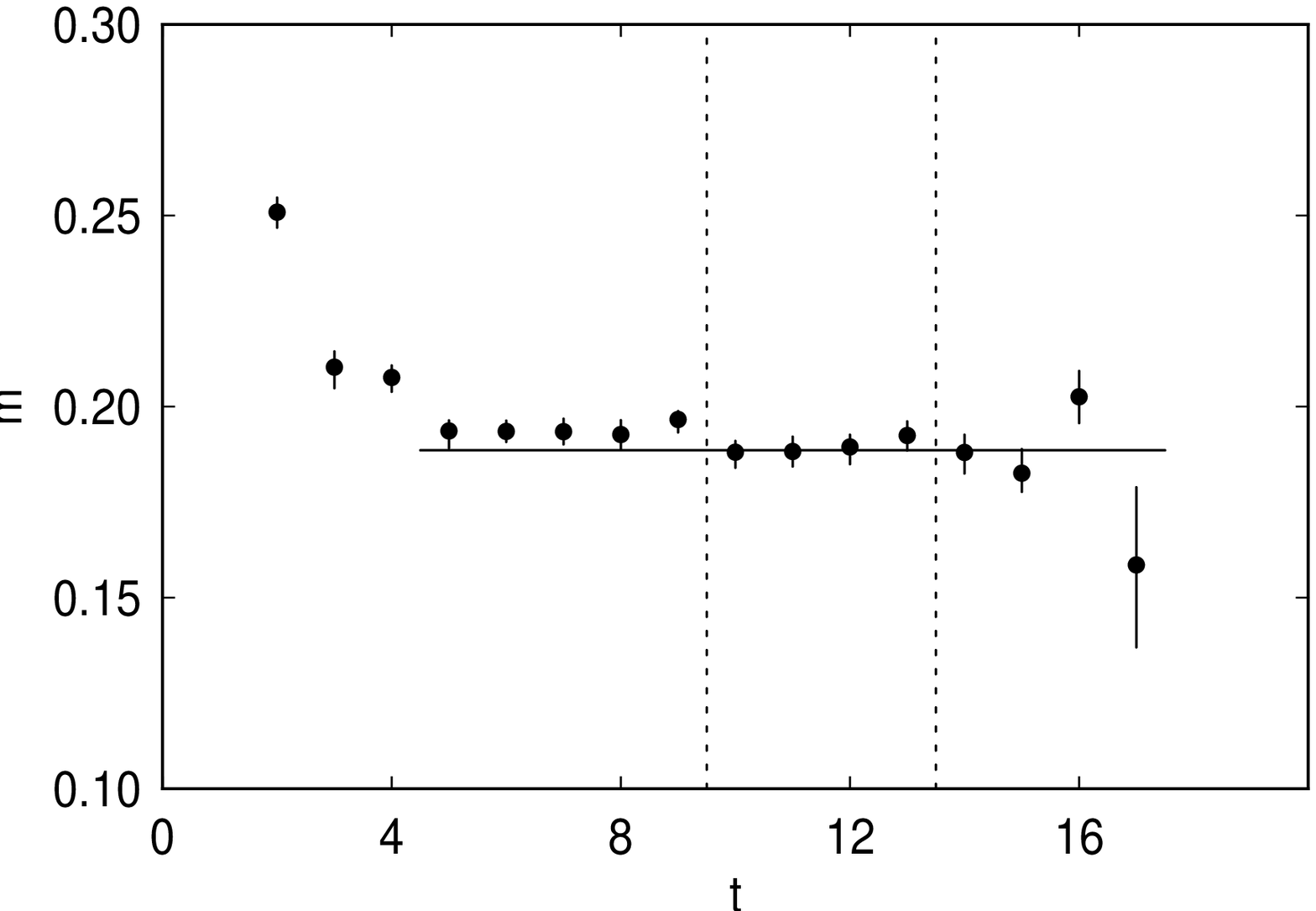}
\caption{ 
Effective masses, final fitting range and fitted mass for the
pseudoscalar propagator with sink size 4 on the lattice $24^3 \times 36$
at at $\beta = 5.93$ and $k = 0.1581$}
\label{fig:pimeffs4x24}
\end{figure}

\begin{figure}
\epsfxsize=\textwidth \epsfbox{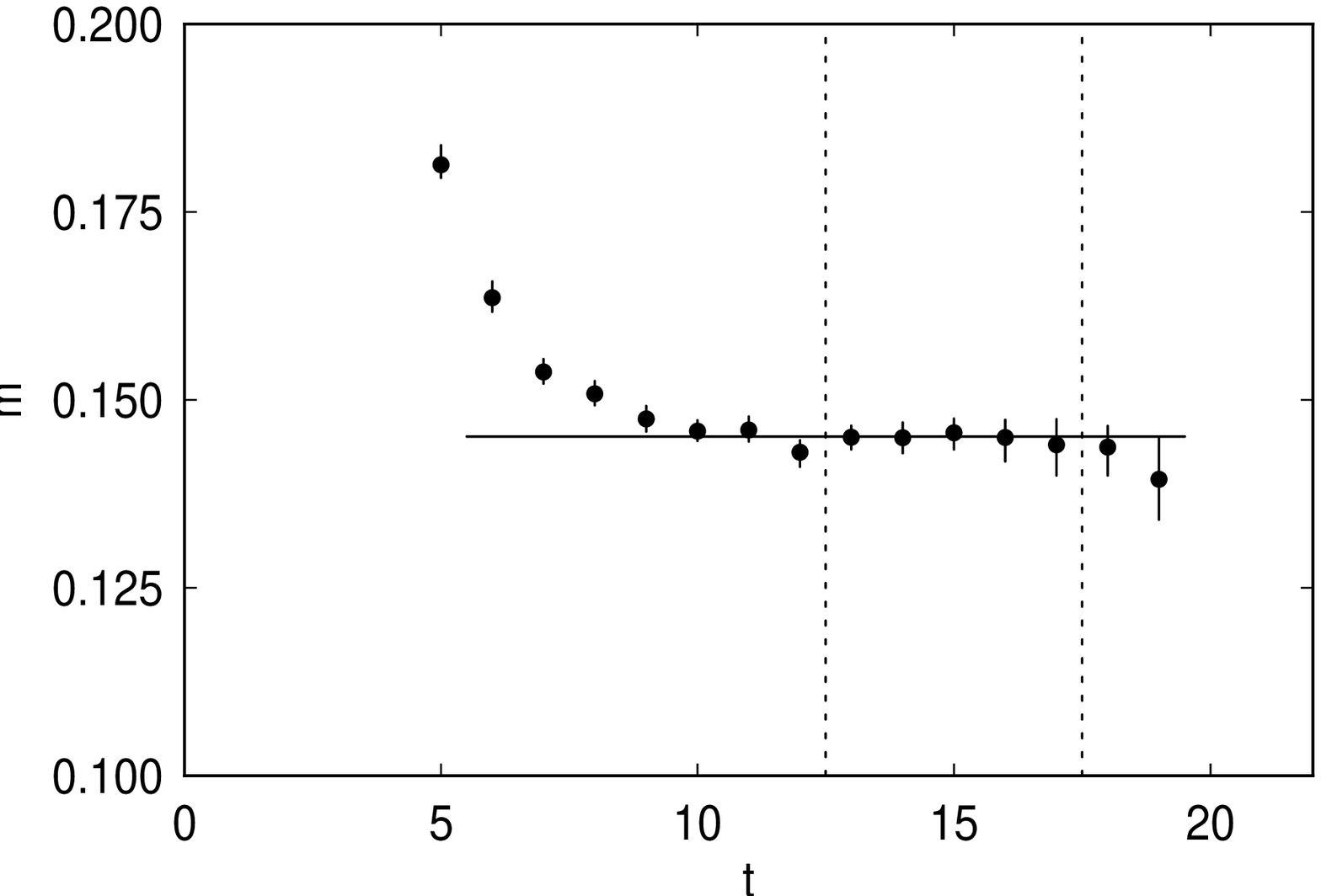}
\caption{ 
Effective masses, final fitting range and fitted mass for the
pseudoscalar propagator with sink size 2 on the lattice $30 \times 32^2
\times 40$ at at $\beta = 6.17$ and $k = 0.1532$}
\label{fig:pimeffs2x32}
\end{figure}

\begin{figure}
\epsfxsize=\textwidth \epsfbox{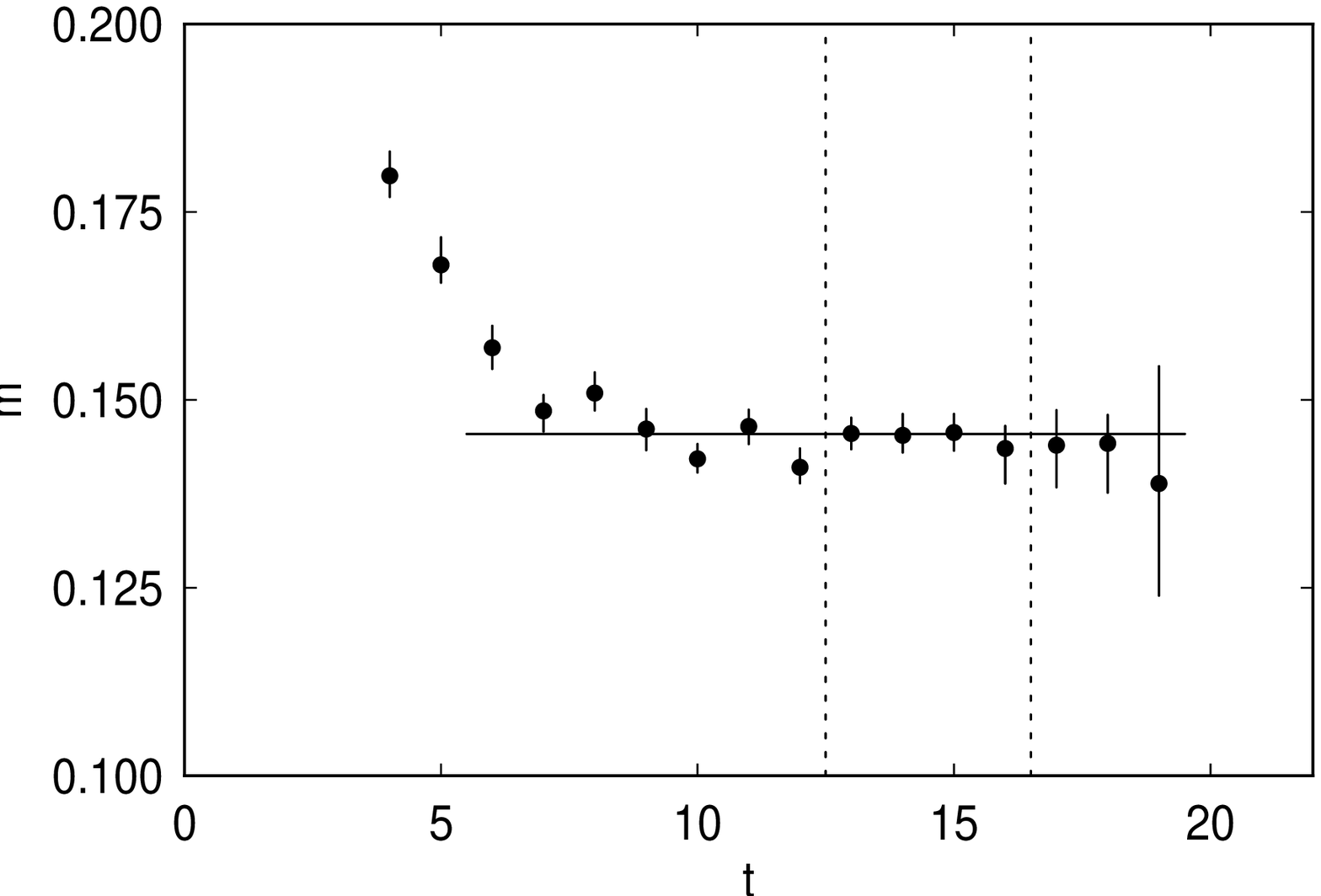}
\caption{ 
Effective masses, final fitting range and fitted mass for the
pseudoscalar propagator with sink size 4 on the lattice $30 \times 32^2
\times 40$ at at $\beta = 6.17$ and $k = 0.1532$}
\label{fig:pimeffs4x32}
\end{figure}

\begin{figure}
\epsfxsize=\textwidth \epsfbox{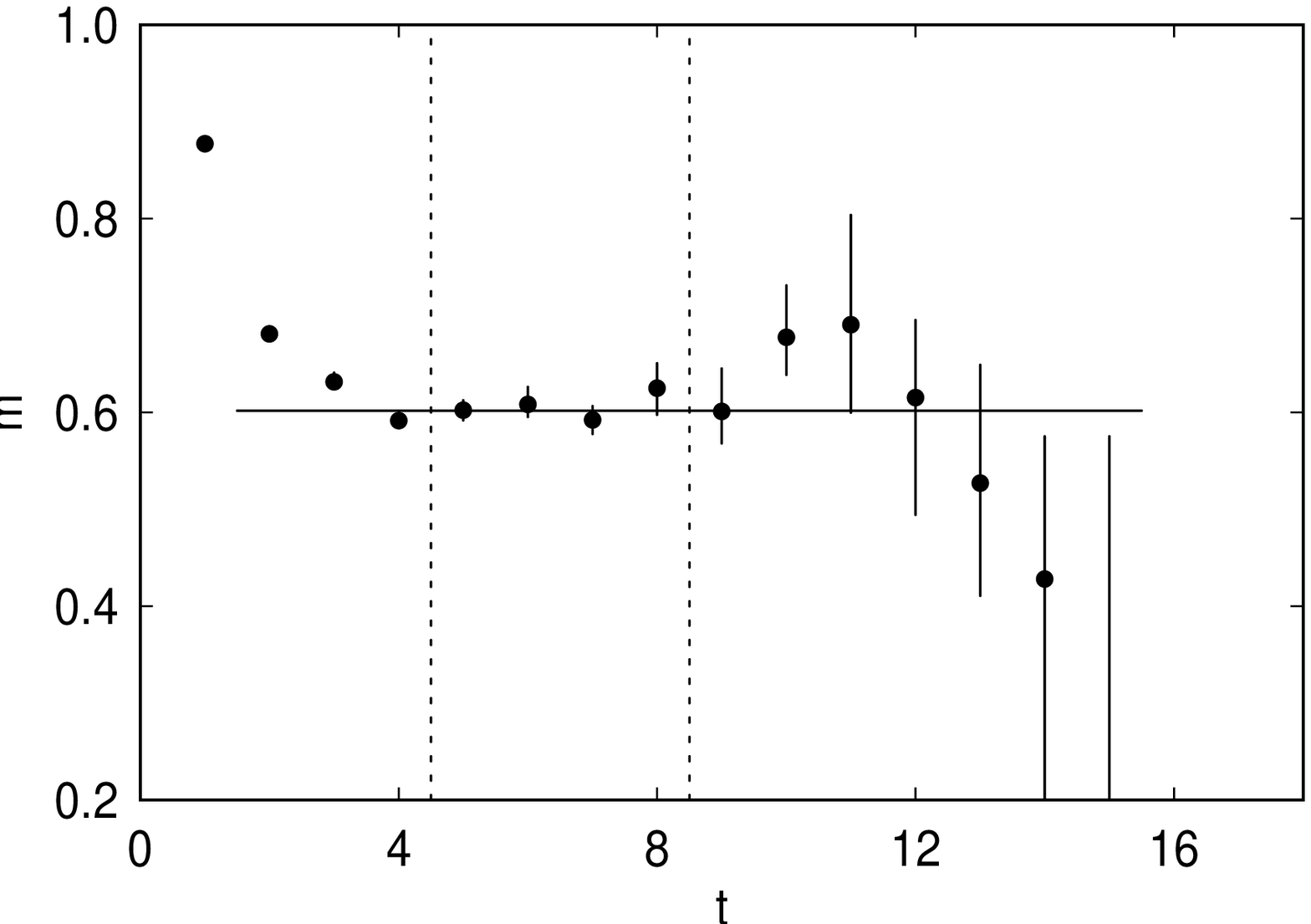}
\caption{ 
Effective masses, final fitting range and fitted mass for the vector
propagator with sink size 2 on the lattice $16^3 \times 32$ at at $\beta
= 5.70$ and $k = 0.1675$}
\label{fig:rhmeffs2x16}
\end{figure}

\begin{figure}
\epsfxsize=\textwidth \epsfbox{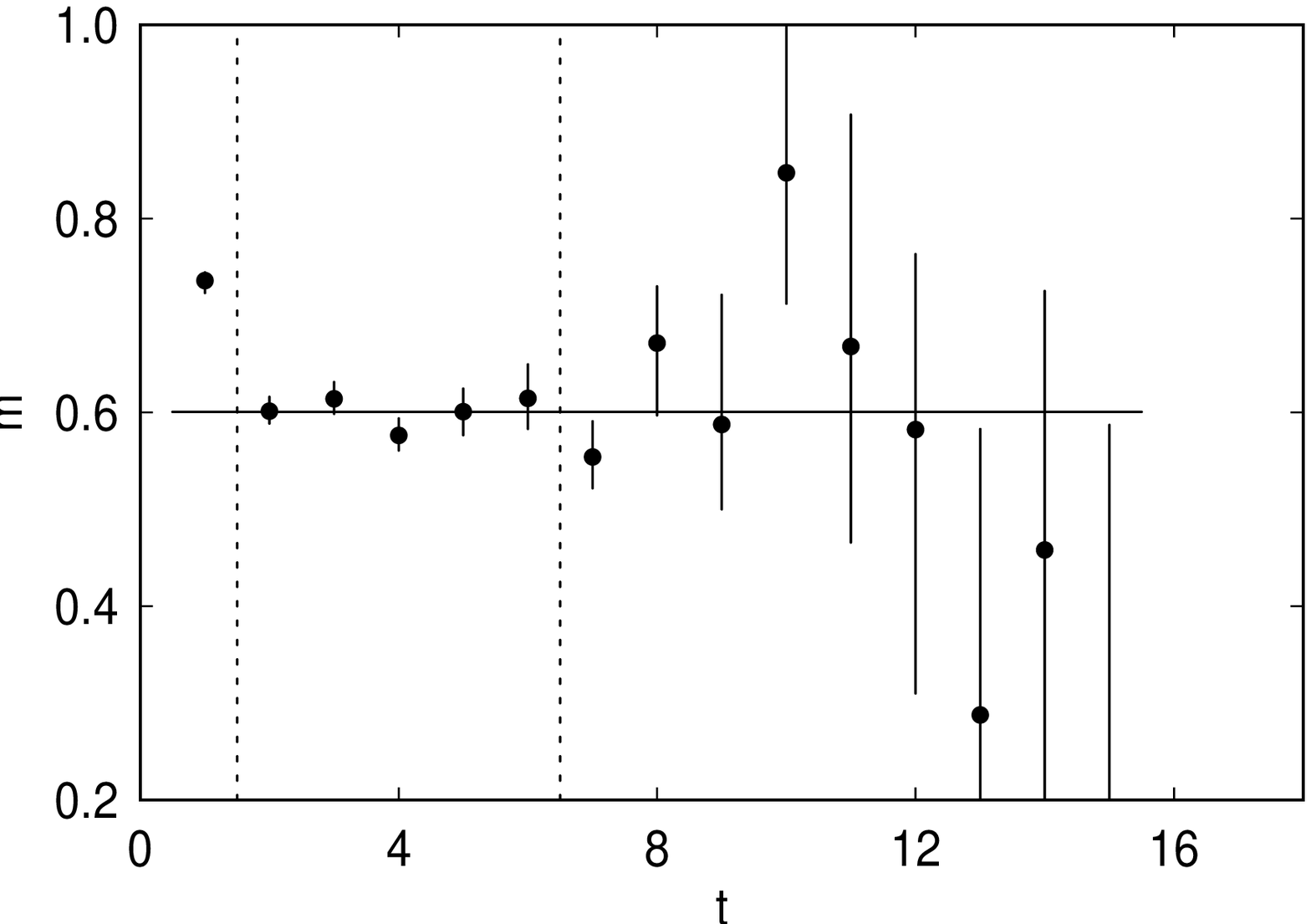}
\caption{ 
Effective masses, final fitting range and fitted mass for the vector
propagator with sink size 4 on the lattice $16^3 \times 32$ at at $\beta
= 5.70$ and $k = 0.1675$}
\label{fig:rhmeffs4x16}
\end{figure}

\begin{figure}
\epsfxsize=\textwidth \epsfbox{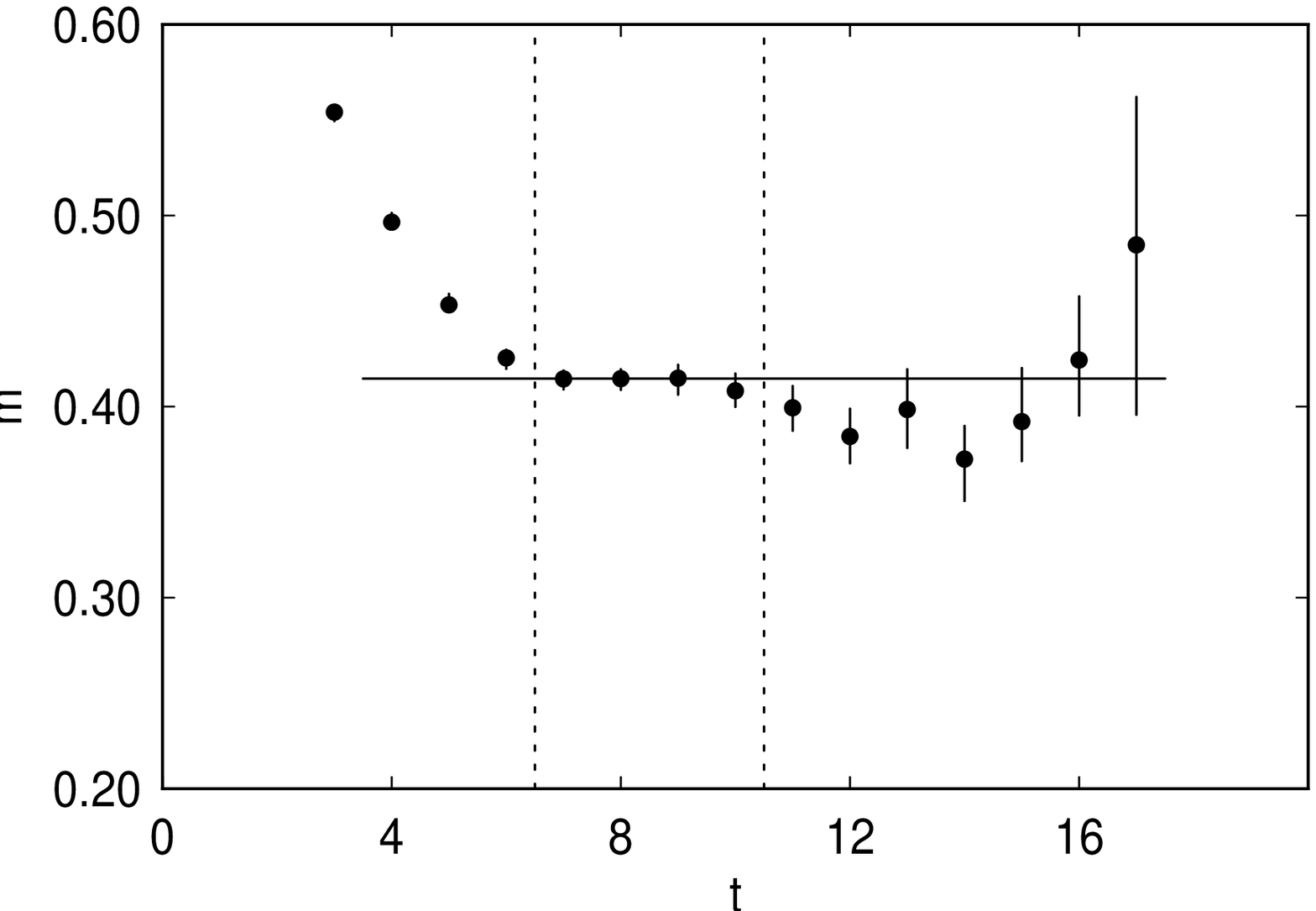}
\caption{ 
Effective masses, final fitting range and fitted mass for the vector
propagator with sink size 2 on the lattice $24^3 \times 36$ at at $\beta
= 5.93$ and $k = 0.1581$}
\label{fig:rhmeffs2x24}
\end{figure}

\begin{figure}
\epsfxsize=\textwidth \epsfbox{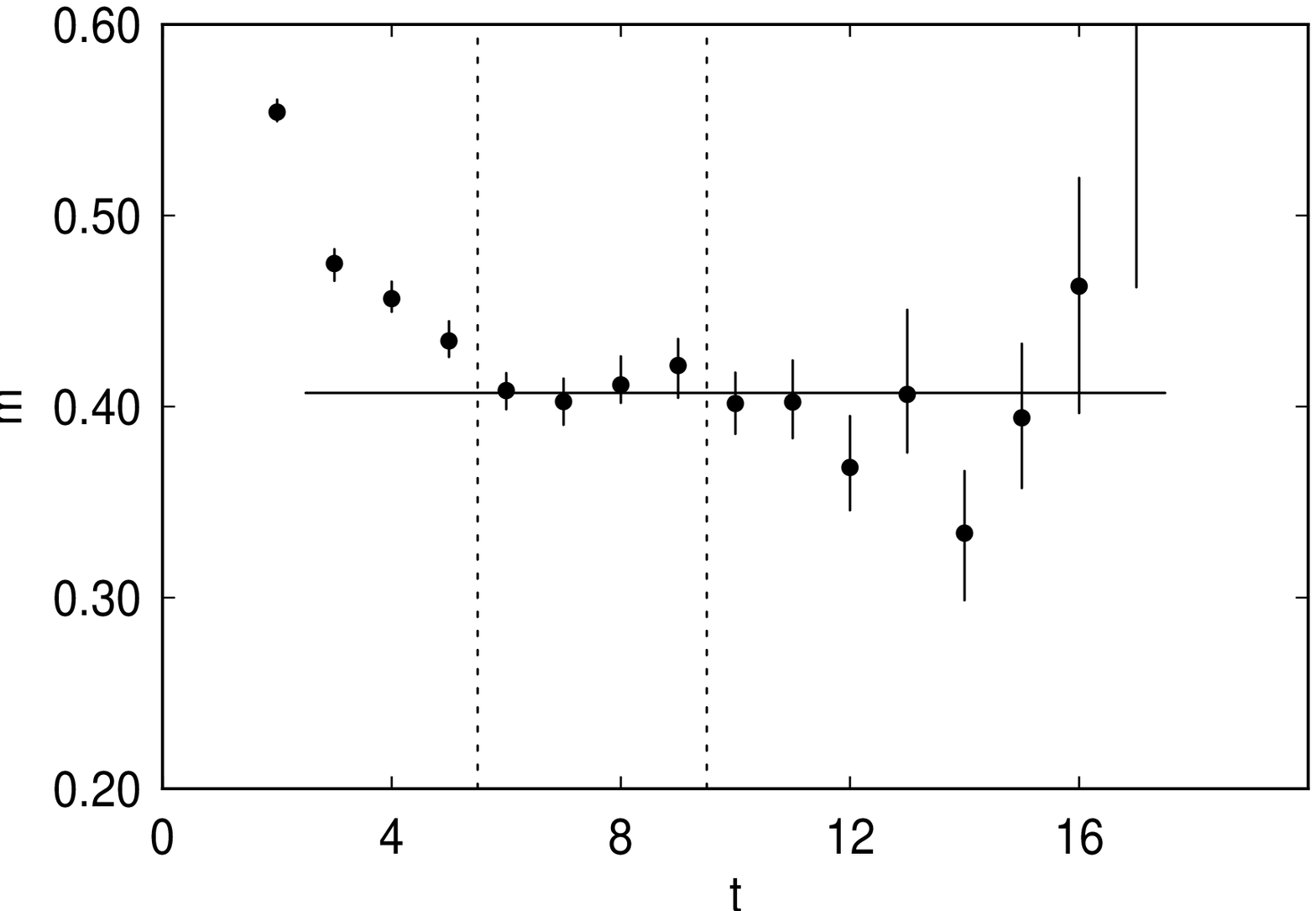}
\caption{ 
Effective masses, final fitting range and fitted mass for the vector
propagator with sink size 4 on the lattice $24^3 \times 36$ at at $\beta
= 5.93$ and $k = 0.1581$}
\label{fig:rhmeffs4x24}
\end{figure}

\begin{figure}
\epsfxsize=\textwidth \epsfbox{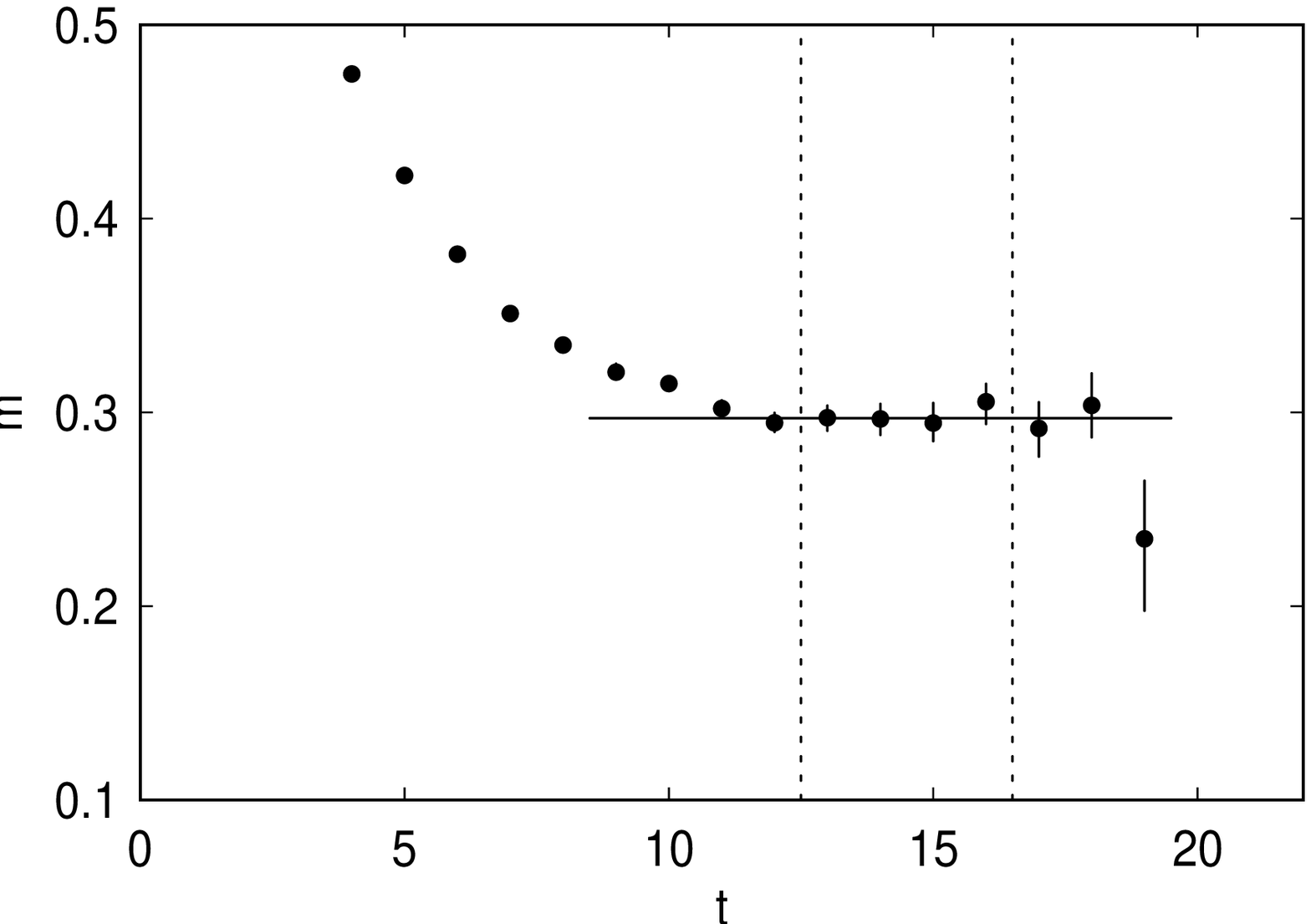}
\caption{ 
Effective masses, final fitting range and fitted mass for the vector
propagator with sink size 2 on the lattice $30 \times 32^2 \times 40$ at
at $\beta = 6.17$ and $k = 0.1532$}
\label{fig:rhmeffs2x32}
\end{figure}

\begin{figure}
\epsfxsize=\textwidth \epsfbox{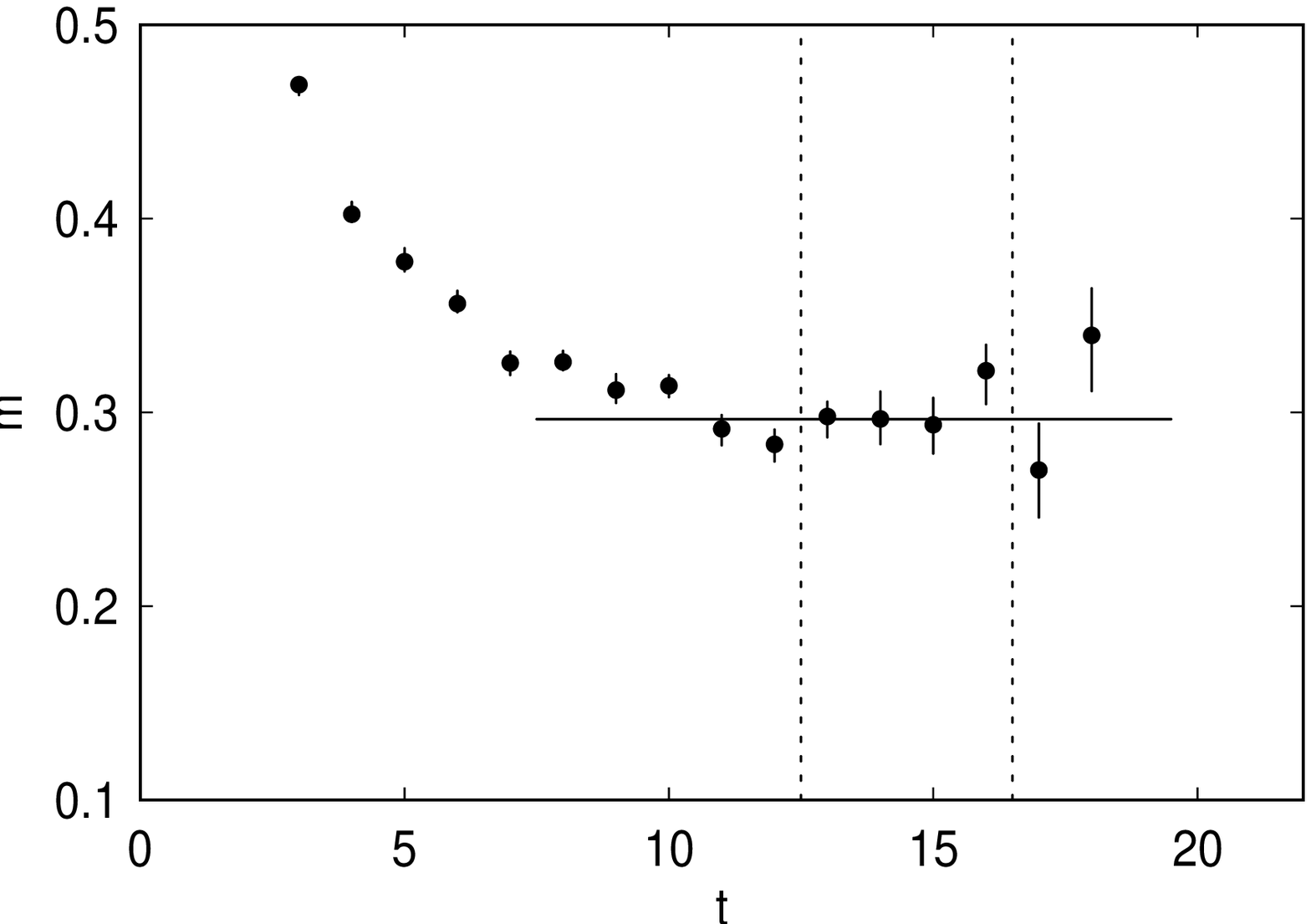}
\caption{ 
Effective masses, final fitting range and fitted mass for the vector
propagator with sink size 4 on the lattice $30 \times 32^2 \times 40$ at
at $\beta = 6.17$ and $k = 0.1532$}
\label{fig:rhmeffs4x32}
\end{figure}

\begin{figure}
\epsfxsize=\textwidth \epsfbox{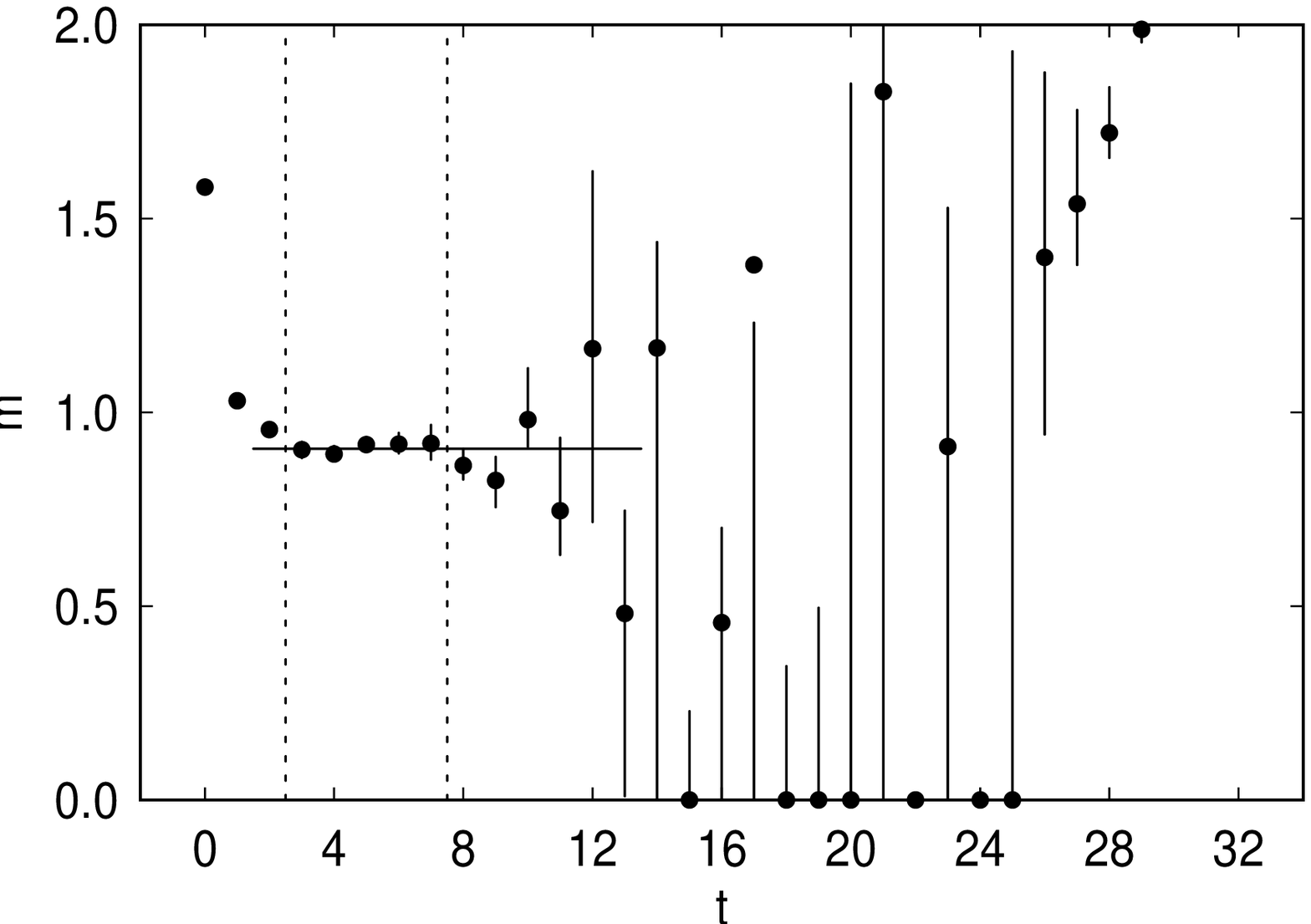}
\caption{ 
Effective masses, final fitting range and fitted mass for the nucleon
propagator with sink size 2 on the lattice $16^3 \times 32$ at at $\beta
= 5.70$ and $k = 0.1675$}
\label{fig:numeffs2x16}
\end{figure}

\begin{figure}
\epsfxsize=\textwidth \epsfbox{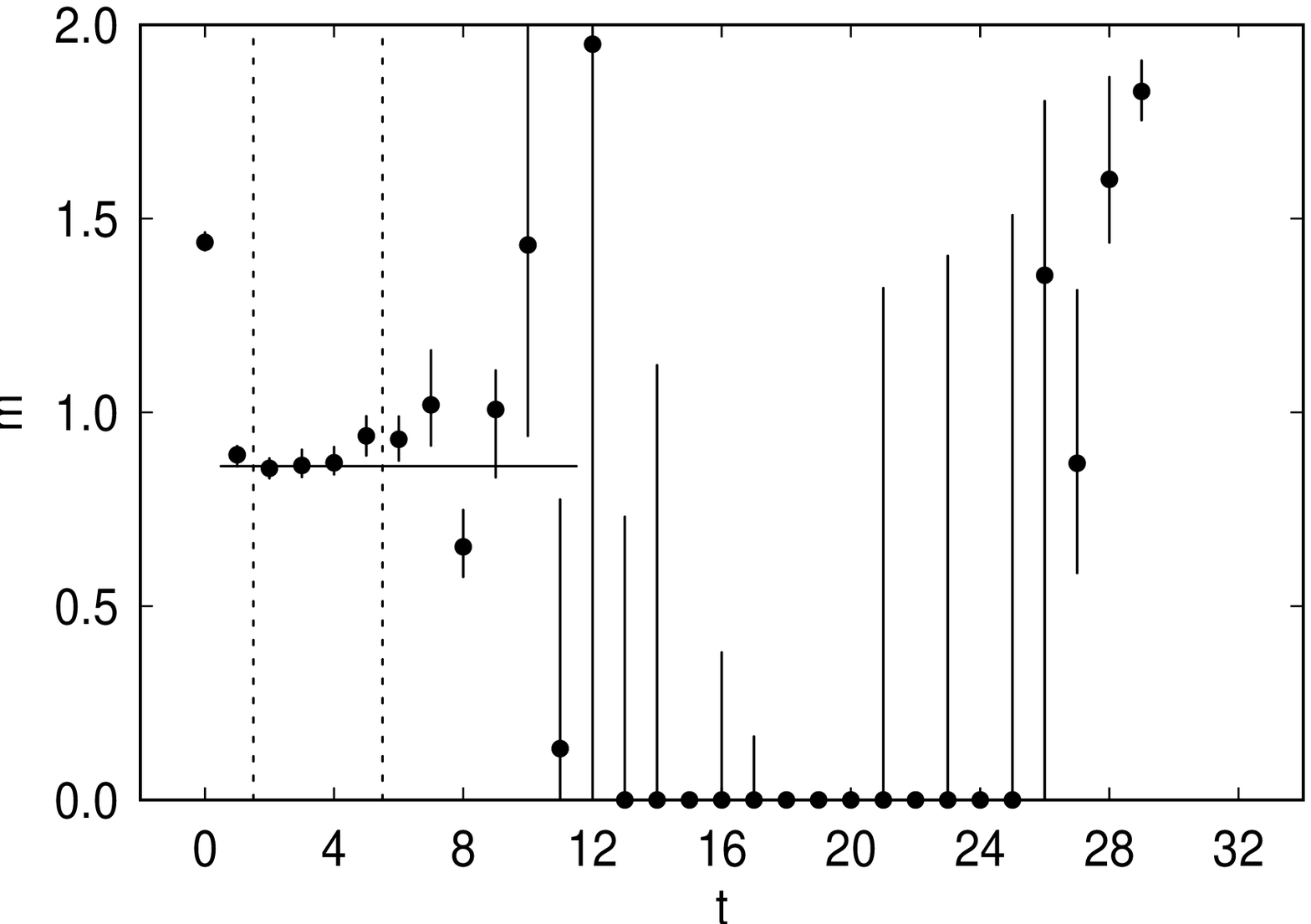}
\caption{ 
Effective masses, final fitting range and fitted mass for the nucleon
propagator with sink size 4 on the lattice $16^3 \times 32$ at at $\beta
= 5.70$ and $k = 0.1675$}
\label{fig:numeffs4x16}
\end{figure}

\begin{figure}
\epsfxsize=\textwidth \epsfbox{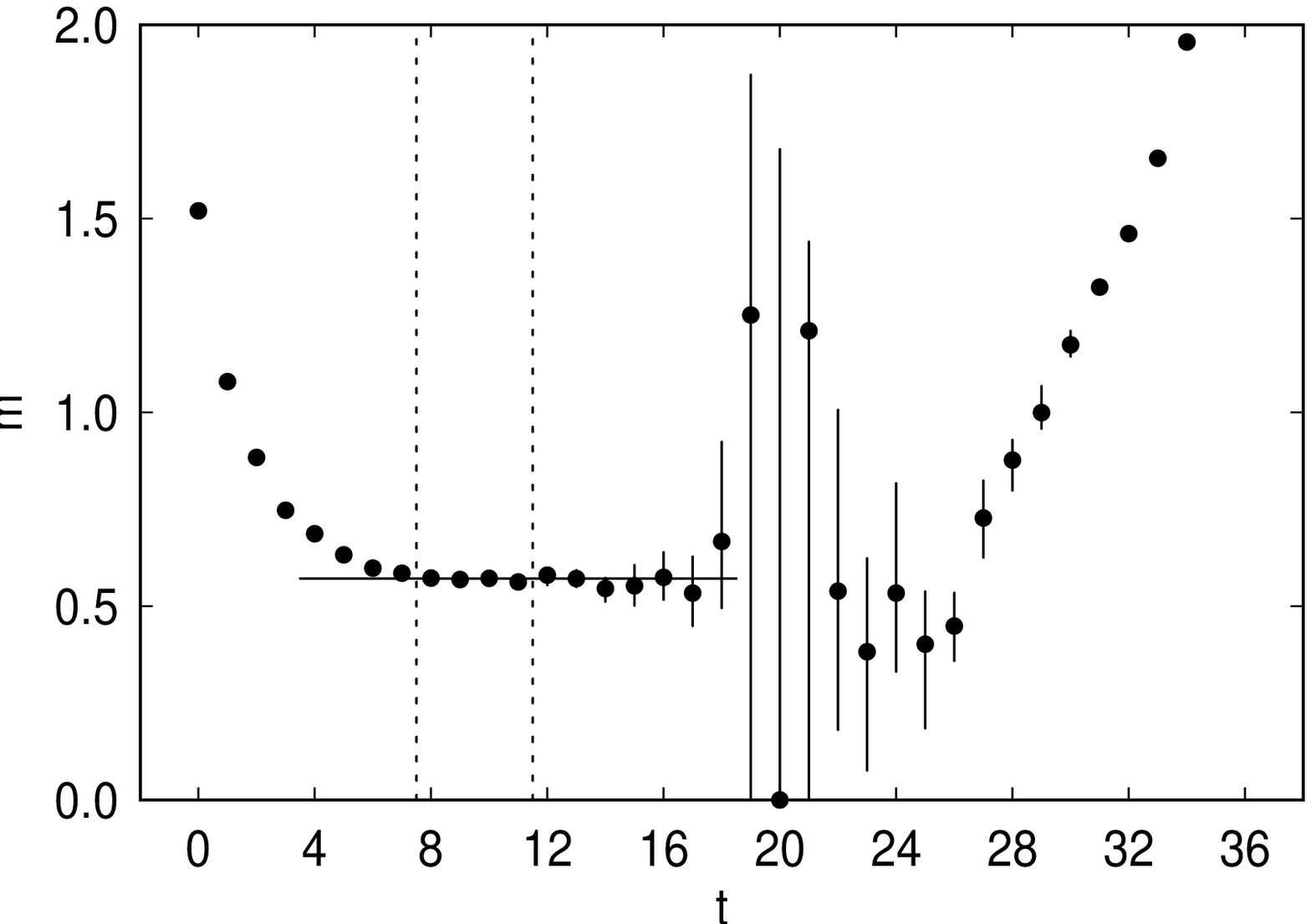}
\caption{ 
Effective masses, final fitting range and fitted mass for the nucleon
propagator with sink size 2 on the lattice $24^3 \times 36$ at at $\beta
= 5.93$ and $k = 0.1581$}
\label{fig:numeffs2x24}
\end{figure}

\begin{figure}
\epsfxsize=\textwidth \epsfbox{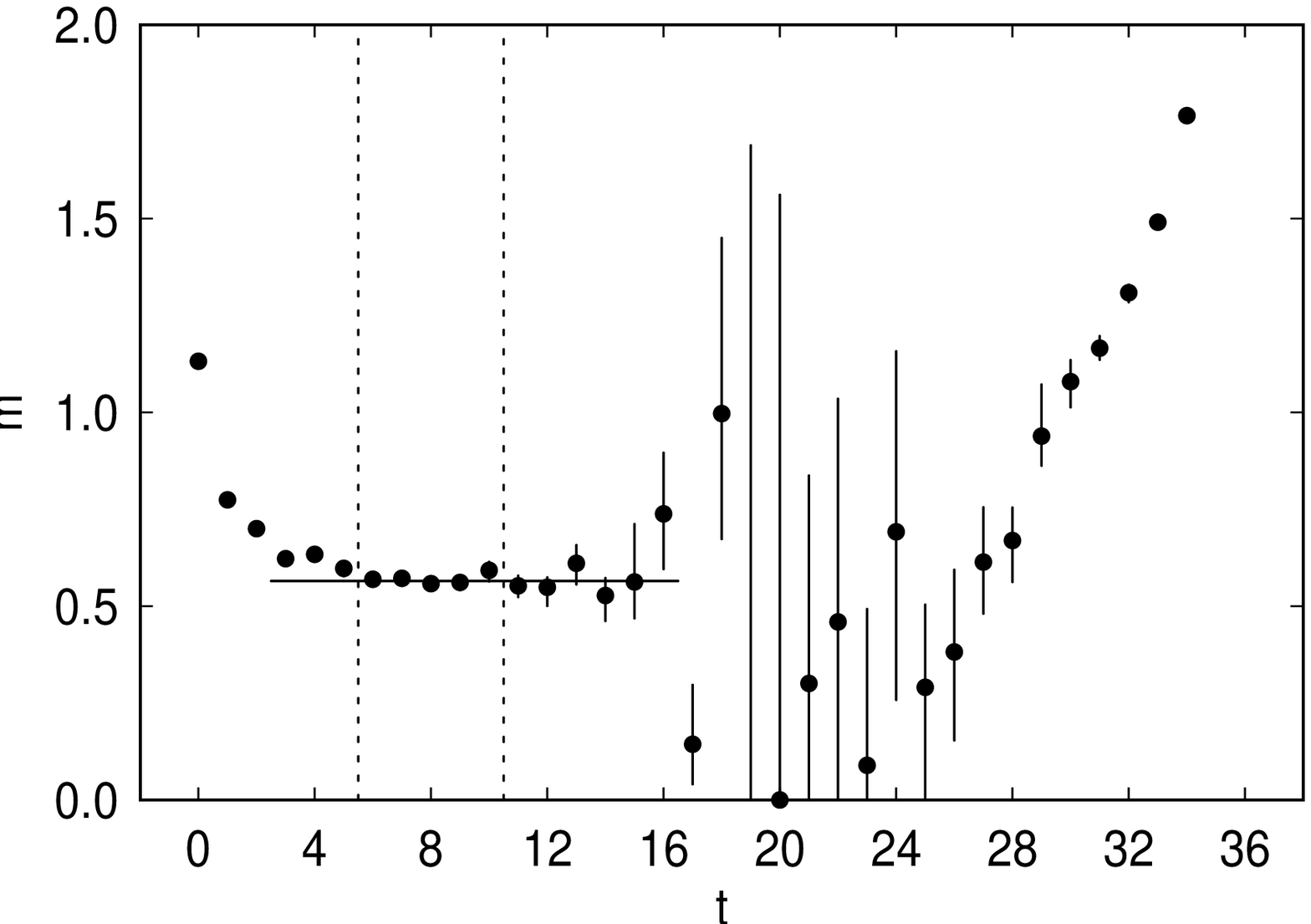}
\caption{ 
Effective masses, final fitting range and fitted mass for the nucleon
propagator with sink size 4 on the lattice $24^3 \times 36$ at at $\beta
= 5.93$ and $k = 0.1581$}
\label{fig:numeffs4x24}
\end{figure}

\begin{figure}
\epsfxsize=\textwidth \epsfbox{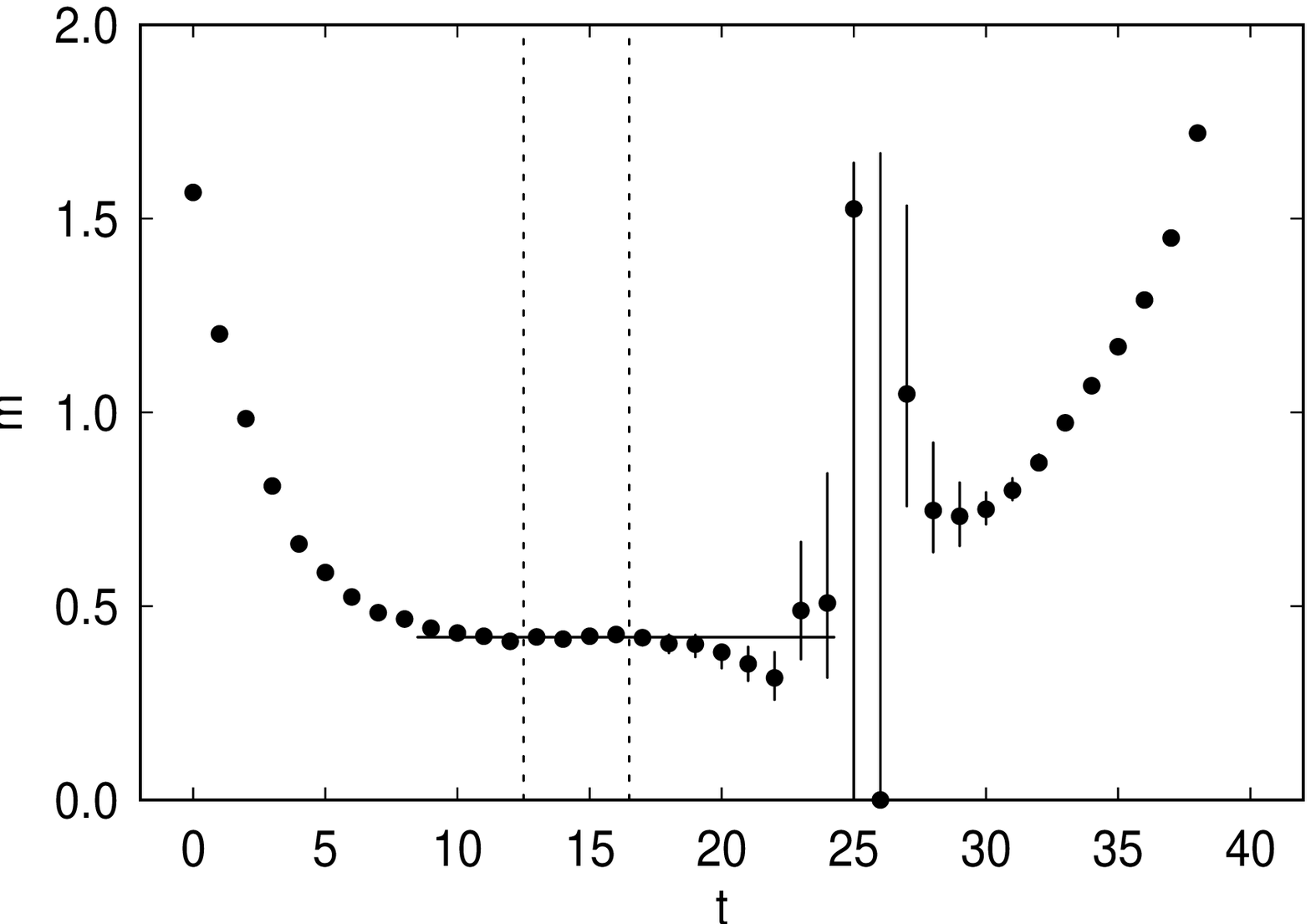}
\caption{ 
Effective masses, final fitting range and fitted mass for the nucleon
propagator with sink size 2 on the lattice $30 \times 32^2 \times 40$ at
at $\beta = 6.17$ and $k = 0.1532$}
\label{fig:numeffs2x32}
\end{figure}

\begin{figure}
\epsfxsize=\textwidth \epsfbox{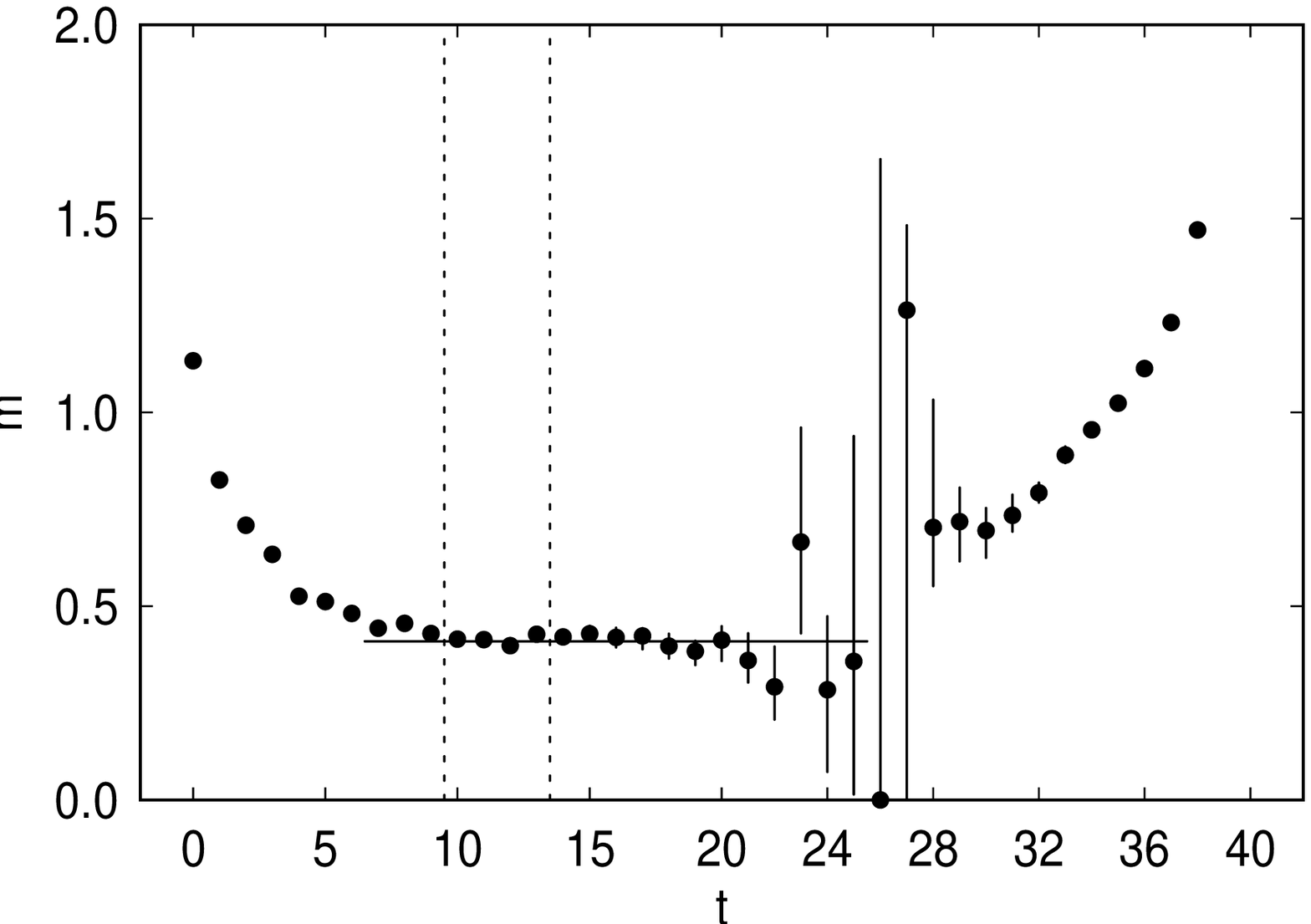}
\caption{ 
Effective masses, final fitting range and fitted mass for the nucleon
propagator with sink size 4 on the lattice $30 \times 32^2 \times 40$ at
at $\beta = 6.17$ and $k = 0.1532$}
\label{fig:numeffs4x32}
\end{figure}

\begin{figure}
\epsfxsize=\textwidth \epsfbox{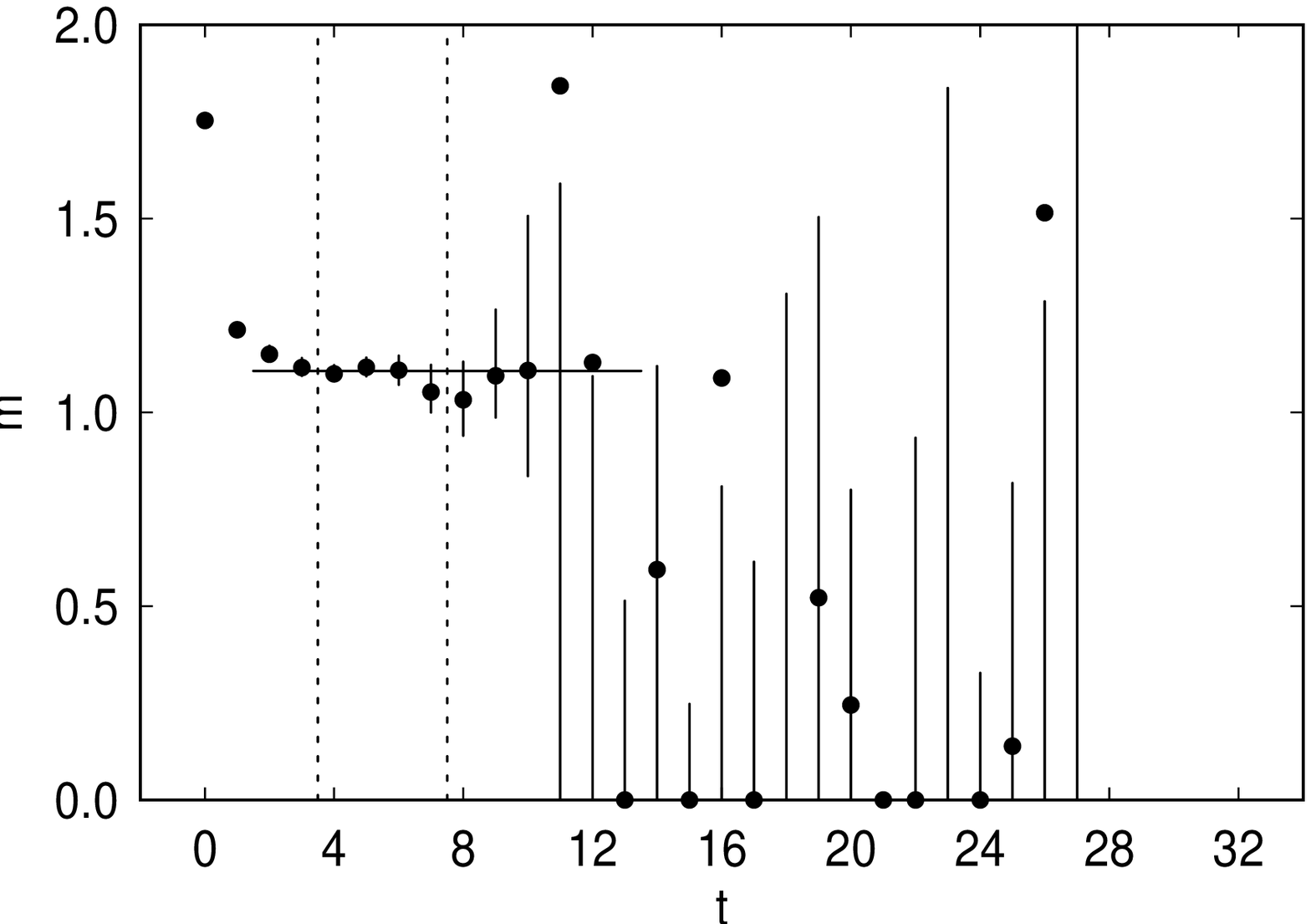}
\caption{ 
Effective masses, final fitting range and fitted mass for the delta
baryon propagator with sink size 2 on the lattice $16^3 \times 32$ at at
$\beta = 5.70$ and $k = 0.1675$}
\label{fig:demeffs2x16}
\end{figure}

\begin{figure}
\epsfxsize=\textwidth \epsfbox{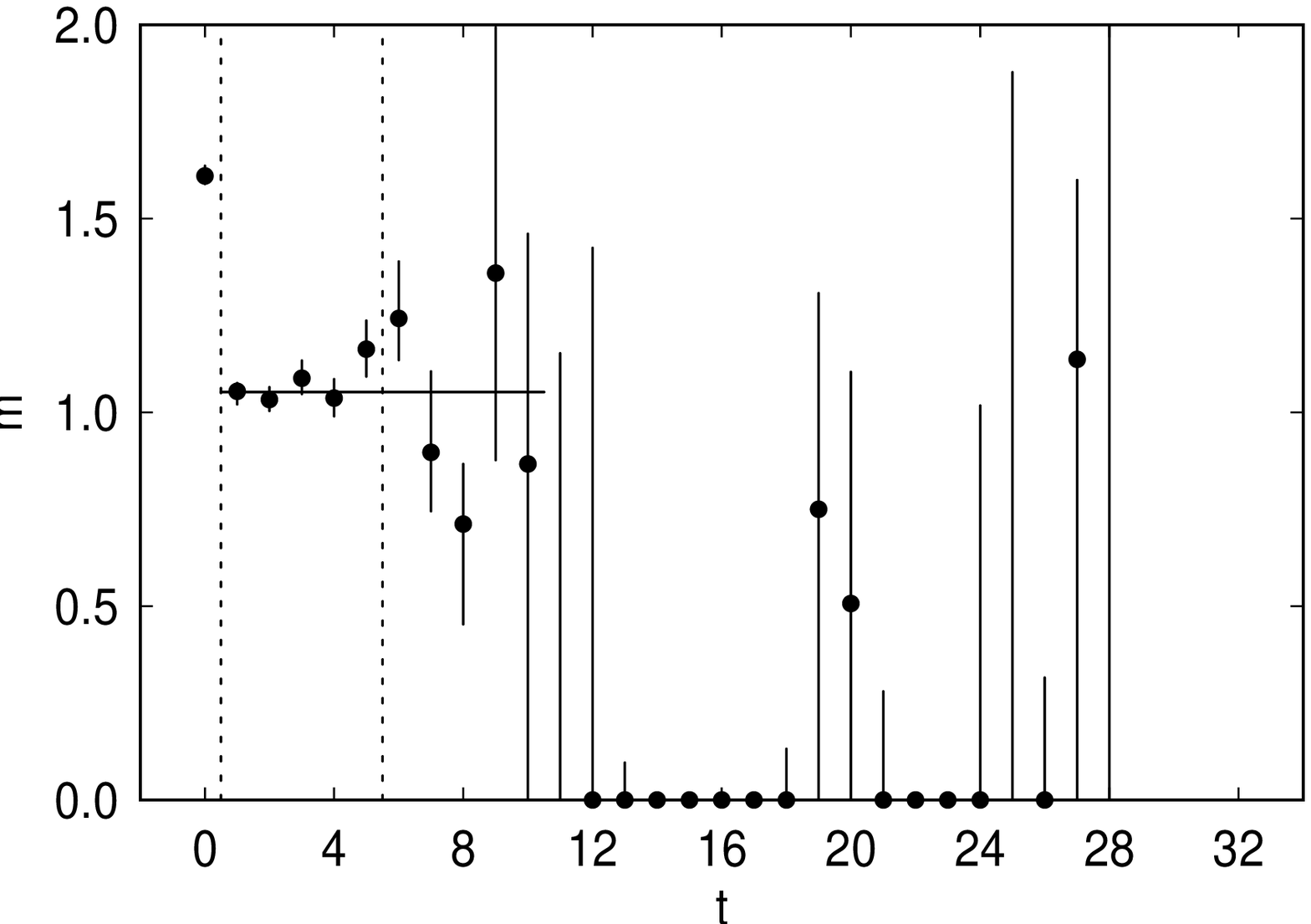}
\caption{ 
Effective masses, final fitting range and fitted mass for the delta
baryon propagator with sink size 4 on the lattice $16^3 \times 32$ at at
$\beta = 5.70$ and $k = 0.1675$}
\label{fig:demeffs4x16}
\end{figure}

\begin{figure}
\epsfxsize=\textwidth \epsfbox{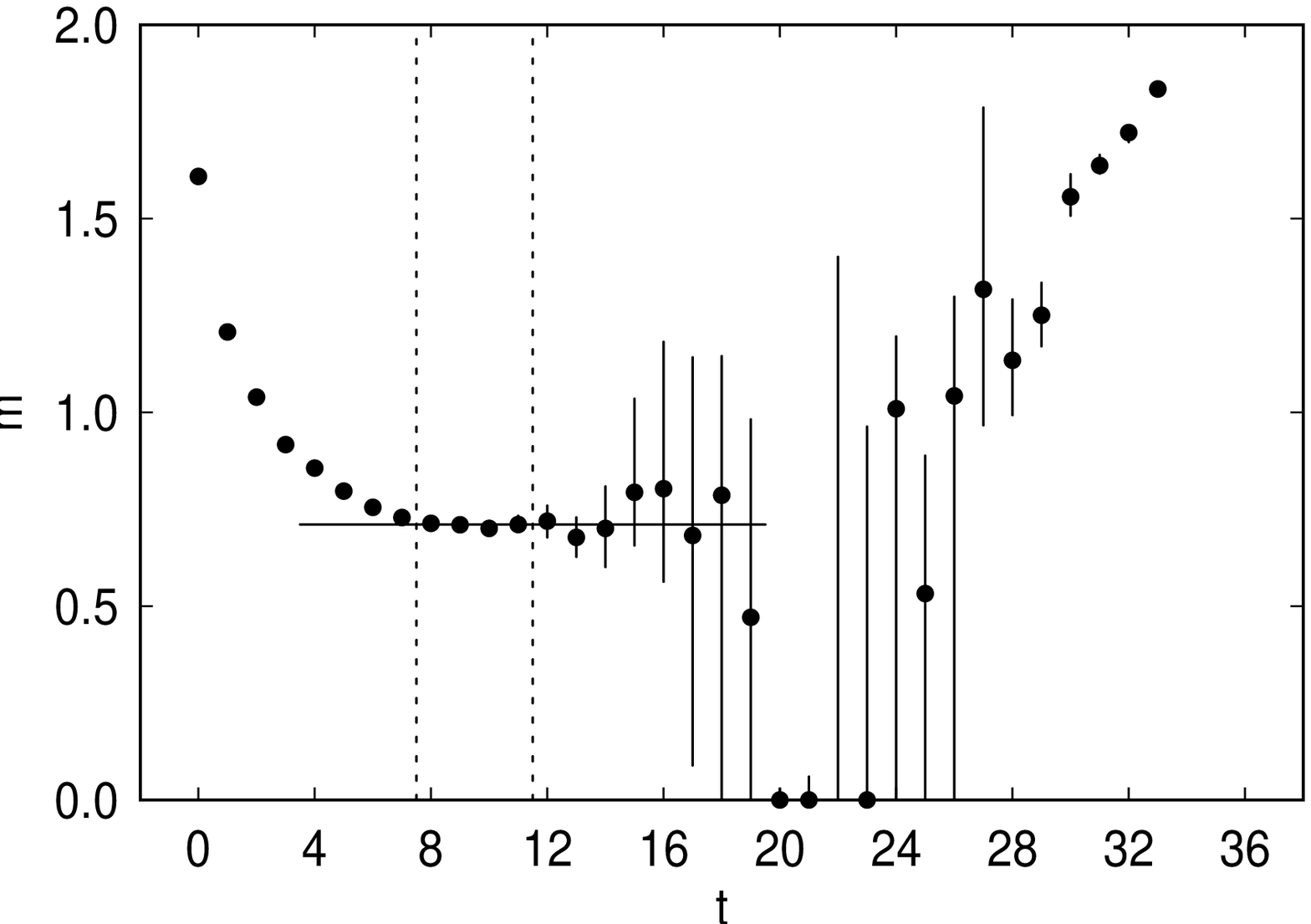}
\caption{ 
Effective masses, final fitting range and fitted mass for the delta
baryon propagator with sink size 2 on the lattice $24^3 \times 36$ at at
$\beta = 5.93$ and $k = 0.1581$}
\label{fig:demeffs2x24}
\end{figure}

\begin{figure}
\epsfxsize=\textwidth \epsfbox{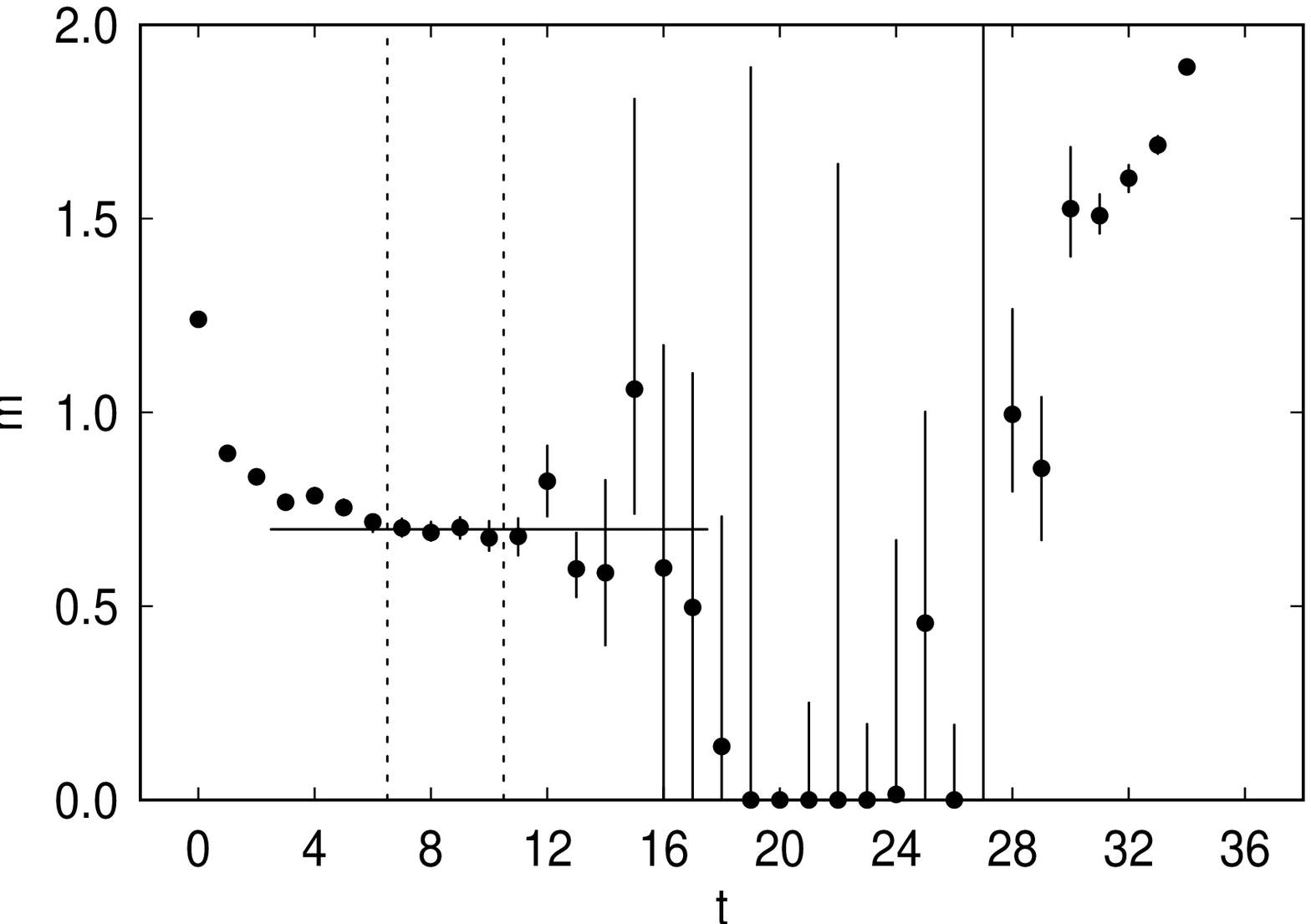}
\caption{ 
Effective masses, final fitting range and fitted mass for the delta
baryon propagator with sink size 4 on the lattice $24^3 \times 36$ at at
$\beta = 5.93$ and $k = 0.1581$}
\label{fig:demeffs4x24}
\end{figure}

\begin{figure}
\epsfxsize=\textwidth \epsfbox{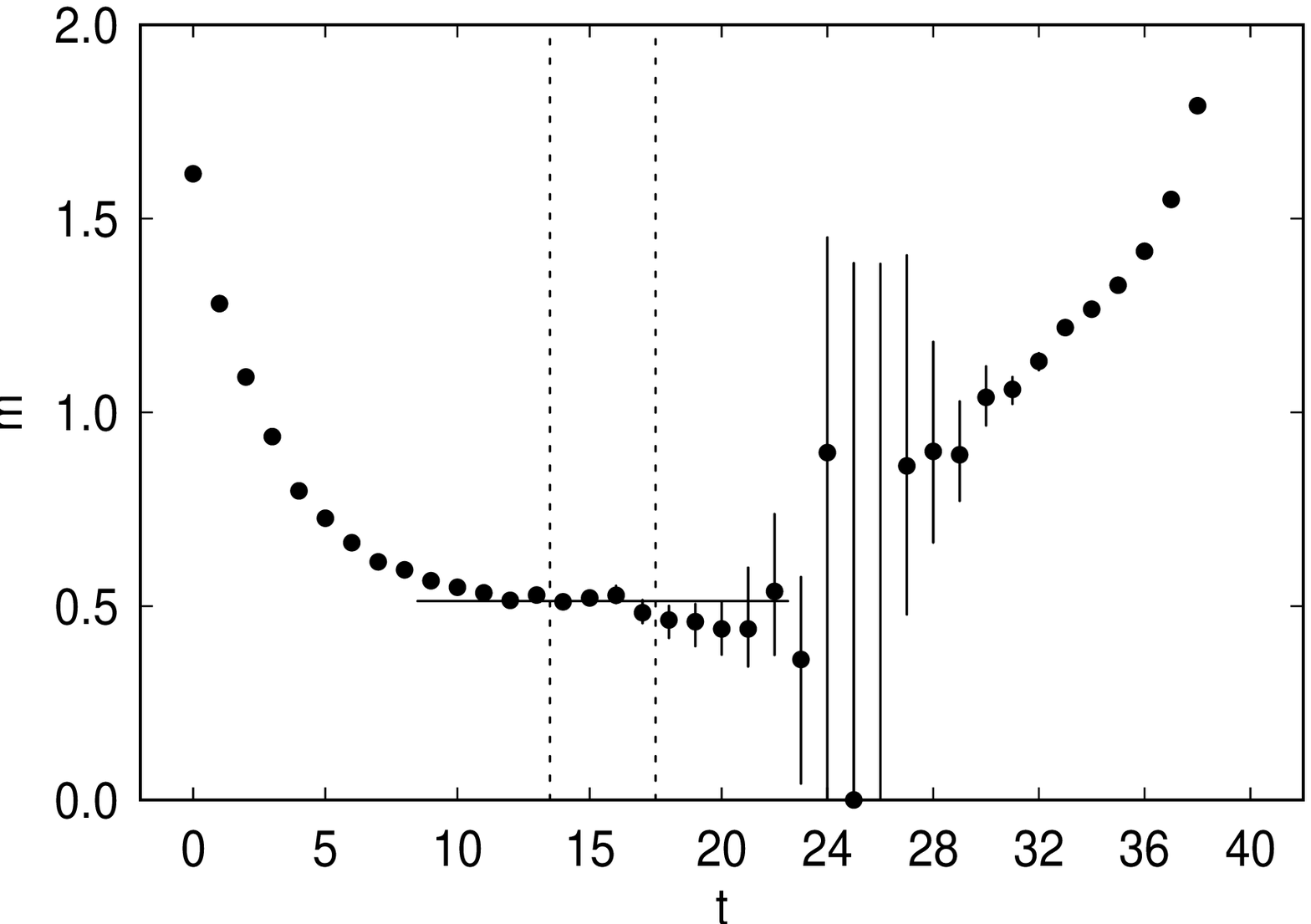}
\caption{ 
Effective masses, final fitting range and fitted mass for the delta
baryon propagator with sink size 2 on the lattice $30 \times 32^2 \times
40$ at at $\beta = 6.17$ and $k = 0.1532$}
\label{fig:demeffs2x32}
\end{figure}

\begin{figure}
\epsfxsize=\textwidth \epsfbox{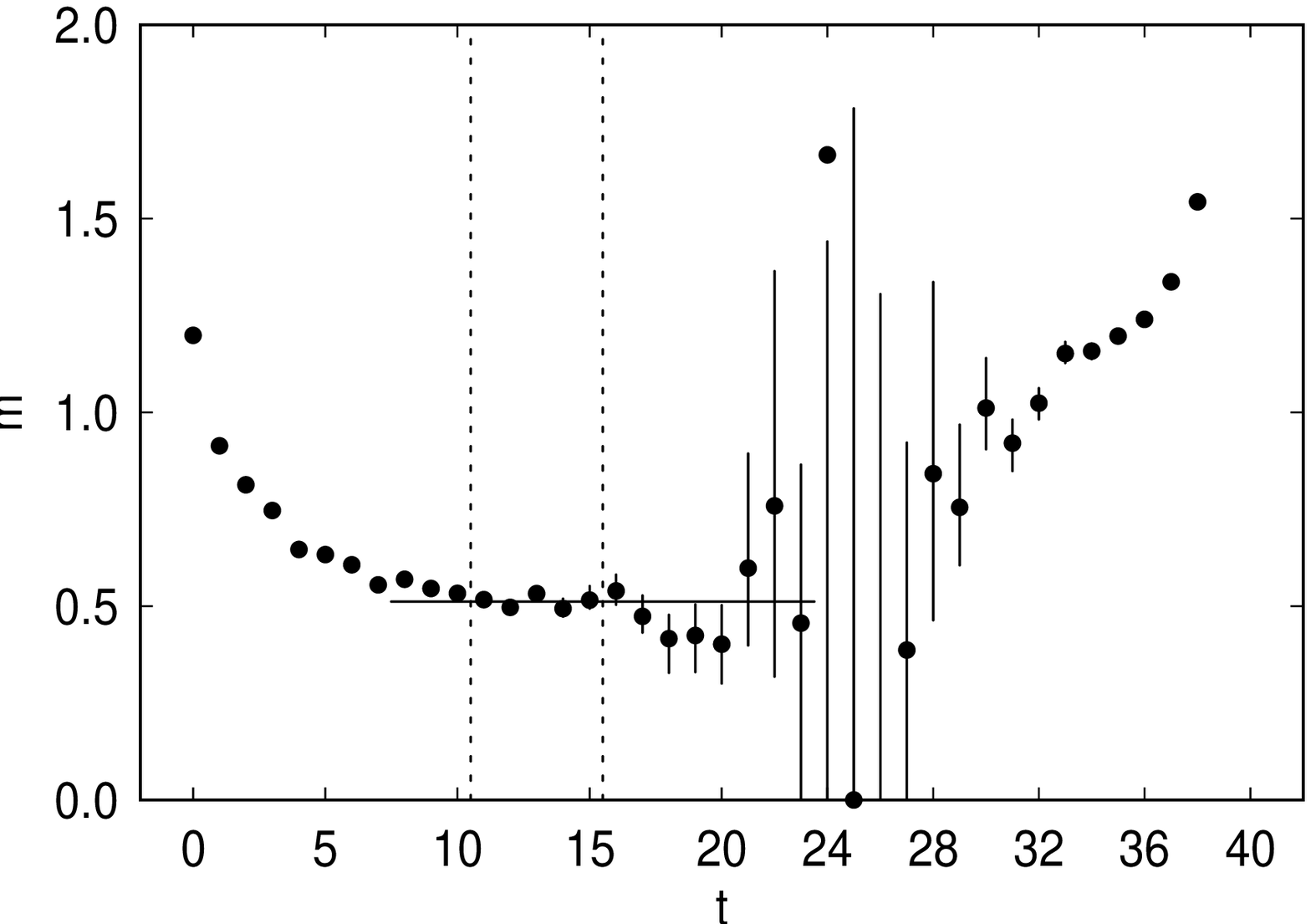}
\caption{ 
Effective masses, final fitting range and fitted mass for the delta
baryon propagator with sink size 4 on the lattice $30 \times 32^2 \times
40$ at at $\beta = 6.17$ and $k = 0.1532$}
\label{fig:demeffs4x32}
\end{figure}

Hadron masses were calculated from all propagators discussed in
Section~\ref{sec:props}.  Figures~\ref{fig:pimeffs2x16} -
\ref{fig:demeffs4x32} show effective hadron masses, fitted hadron masses
and final fitting ranges for pion, rho, nucleon and delta propagators,
for the lattices $16^3 \times 32$, $24^3 \times 36$ and $30 \times 32^2
\times 40$ which will be used in 
Section~\ref{sec:contlim} to obtain continuum limits. Data is shown
for the lightest quark mass used on each lattice both for sinks with $r$
of 2 and of 4. Final fitted mass values, fitting ranges and $\chi^2$ per
degree of freedom are listed in 
Tables~\ref{tab:allmx8} - \ref{tab:demx32}.
For all but the lattice $8^3 \times 32$, and the lattice $16^3 \times
32$ at small $k$, these tables list results for
$r$ ranging from 0 to 4 along with masses obtained by fitting
simultaneously data for sinks with $r$ of 0, 1 and 2. 

The masses determined simultaneously from sinks 0, 1 and 2 we chose as
our overall best values. These numbers generally showed somewhat smaller
error bars than single sink fits. They also provided an unbiased
resolution of the small disagreements between fits to single sinks of
different sizes.  We tried simultaneous fits to more than three sinks
but found that excessive computer time was required and the minimal
$\chi^2$ was hard to find reliably. For the pion we found no noticeably
improvement in errors with multiple sink fits and arbitrarily chose the
masses obtained from sinks of size 0 as our best values. 

In nearly all cases, the effective masses in the figures show plateaus
extending over four or more values of $t$. Since each effective mass is
found from a propagator at a pair of values of $t$, it follows that 
in nearly all cases the propagators shown fall off with a consistent
mass over five or more values of $t$.  Propagators with $r$ of 4 show
wider plateaus beginning at smaller $t$ than propagators with $r$ of 2.
Thus, as expected, hadron operators with $r$ of 4 couple more strongly
to ground state hadrons and less strongly to excited states than do
operators with $r$ of 2.  The consistency, within statistical errors, of
mass values shown in Tables~\ref{tab:allmx8} -
\ref{tab:demx32} for varying choices of $r$ shows that errors in ground
state masses which might arise from contamination by excited states 
are within the statistical errors in each mass. 

For larger values of $r$, the propagators in Figures~\ref{fig:pimeffs2x16}
- \ref{fig:demeffs4x32} and the masses in Tables~\ref{tab:allmx8} -
\ref{tab:demx32} show larger statistical errors. These larger errors
arise because hadron sink operators with largyer $r$ couple significantly
to a larger collection of gauge link operators, those between the
postions of the sink quark and antiquark fields, and thus are more
sensitive to fluctuations in the gauge field.

In subsequent sections of this paper, extrapolations to obtain physical
predictions for hadron masses will be presented both for mass values
obtained from simultaneous fits to sinks 0, 1 and 2, and for masses
solely from fits to sink size 4. The consistency of these two sets of
predicitions, within statistical errors, will serve as further evidence
that our results are not biased by the contamination of ground state
masses with masses from excited states.

\section{EXTRAPOLATION TO SMALL QUARK MASS} \label{sec:mextrap}

As the hopping constant $k$ is made larger, the amount of computer time
required to evaluate a quark propagator grows, as does the size of the
ensemble of gauge field configurations required to find hadron
propagators to within a fixed statistical error. As a consequence of
these effects, particularly the second, we were unable to evaluate
hadron masses at large enough $k$ to give the physical value of $m_{\pi}
/ m_{\rho}$.  Thus to determine the masses of hadrons composed only of
up and down quarks we extrapolated to the physical value of $k$ data
found at smaller $k$ on the lattice $16^3 \times 32$, $24^3 \times 32$,
$24^3 \times 36$ and $30 \times 32^2 \times 40$.  For the lattice $8^3
\times 32$ we were not able to run at $k$ large enough to make this
extrapolation reliably.

\begin{figure}
\epsfxsize=\textwidth \epsfbox{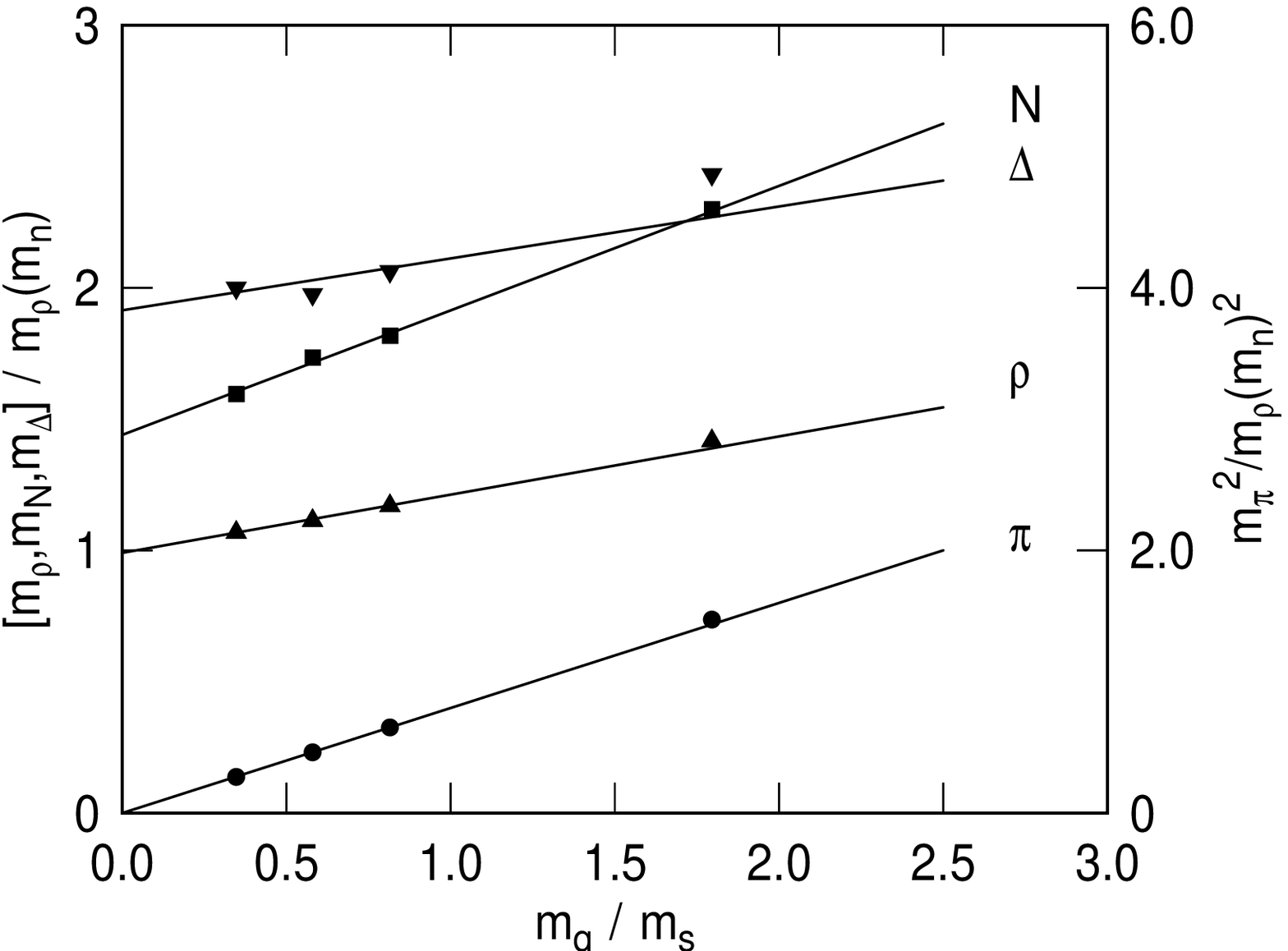}
\caption{ 
For a $16^3 \times 32$ lattice at $\beta$ of 5.70 combining sinks of
sizes 0, 1 and 2, $m_{\pi}^2$, $m_{\rho}$, $m_N$ and $m_{\Delta}$, in
units of the physical rho mass $m_{\rho}(m_n)$, as functions of the
quark mass $m_q$, in units of the strange quark mass $m_s$. The symbol
at each point is larger than the error bars.}
\label{fig:mexx16s012}
\end{figure}

\begin{figure}
\epsfxsize=\textwidth \epsfbox{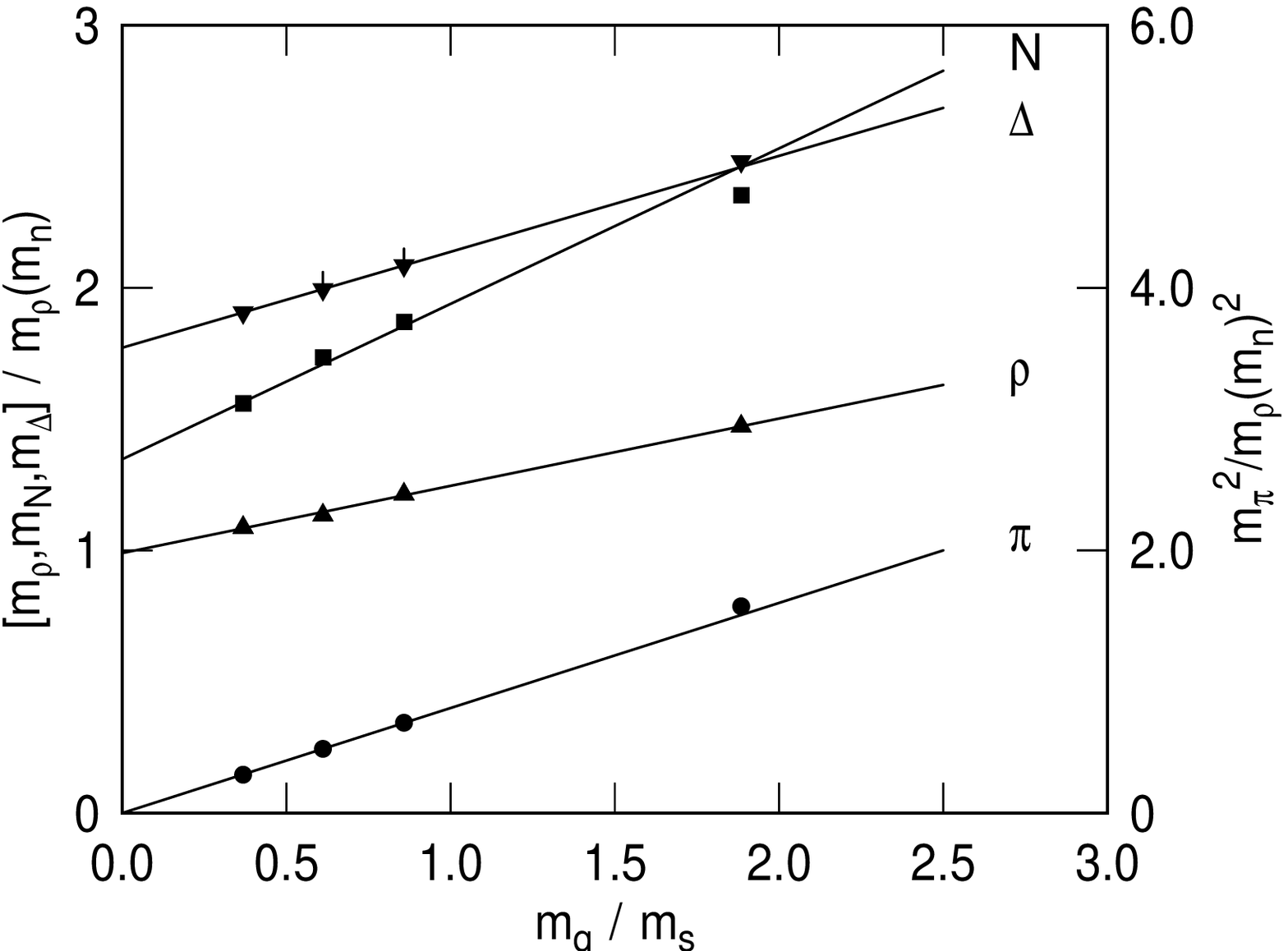}
\caption{ 
For a $16^3 \times 32$ lattice at $\beta$ of 5.70 with a sink of
size 4, $m_{\pi}^2$, $m_{\rho}$, $m_N$ and $m_{\Delta}$, in
units of the physical rho mass $m_{\rho}(m_n)$, as functions of the
quark mass $m_q$, in units of the strange quark mass $m_s$. The symbols
at most points are larger than the error bars.}
\label{fig:mexx16s4}
\end{figure}

\begin{figure}
\epsfxsize=\textwidth \epsfbox{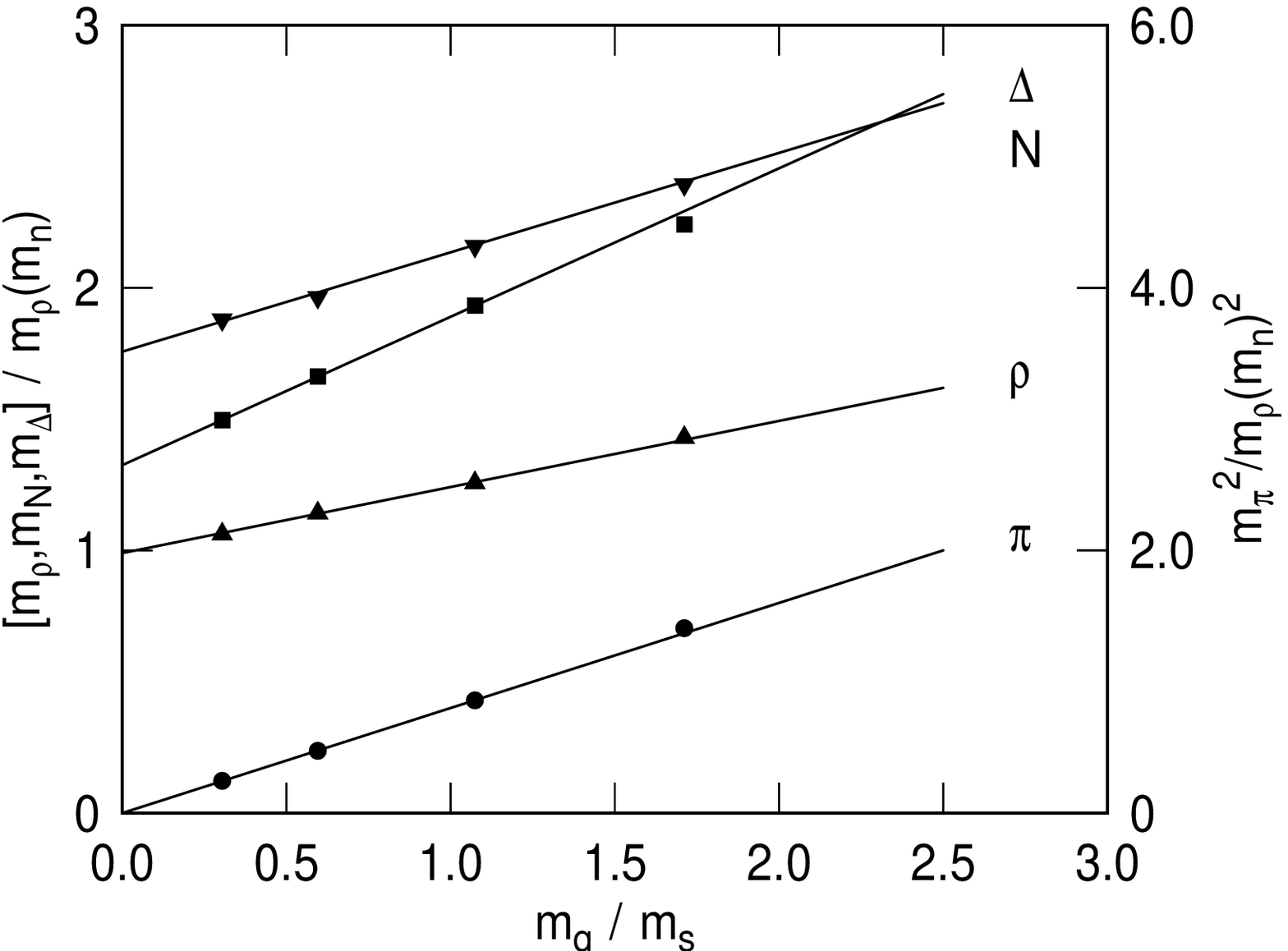}
\caption{ 
For a $24^3 \times 36$ lattice at $\beta$ of 5.93 combining sinks of
sizes 0, 1 and 2, $m_{\pi}^2$, $m_{\rho}$, $m_N$ and $m_{\Delta}$, in
units of the physical rho mass $m_{\rho}(m_n)$, as functions of the
quark mass $m_q$, in units of the strange quark mass $m_s$. The symbol
at each point is larger than the error bars.}
\label{fig:mexx24s012}
\end{figure}

\begin{figure}
\epsfxsize=\textwidth \epsfbox{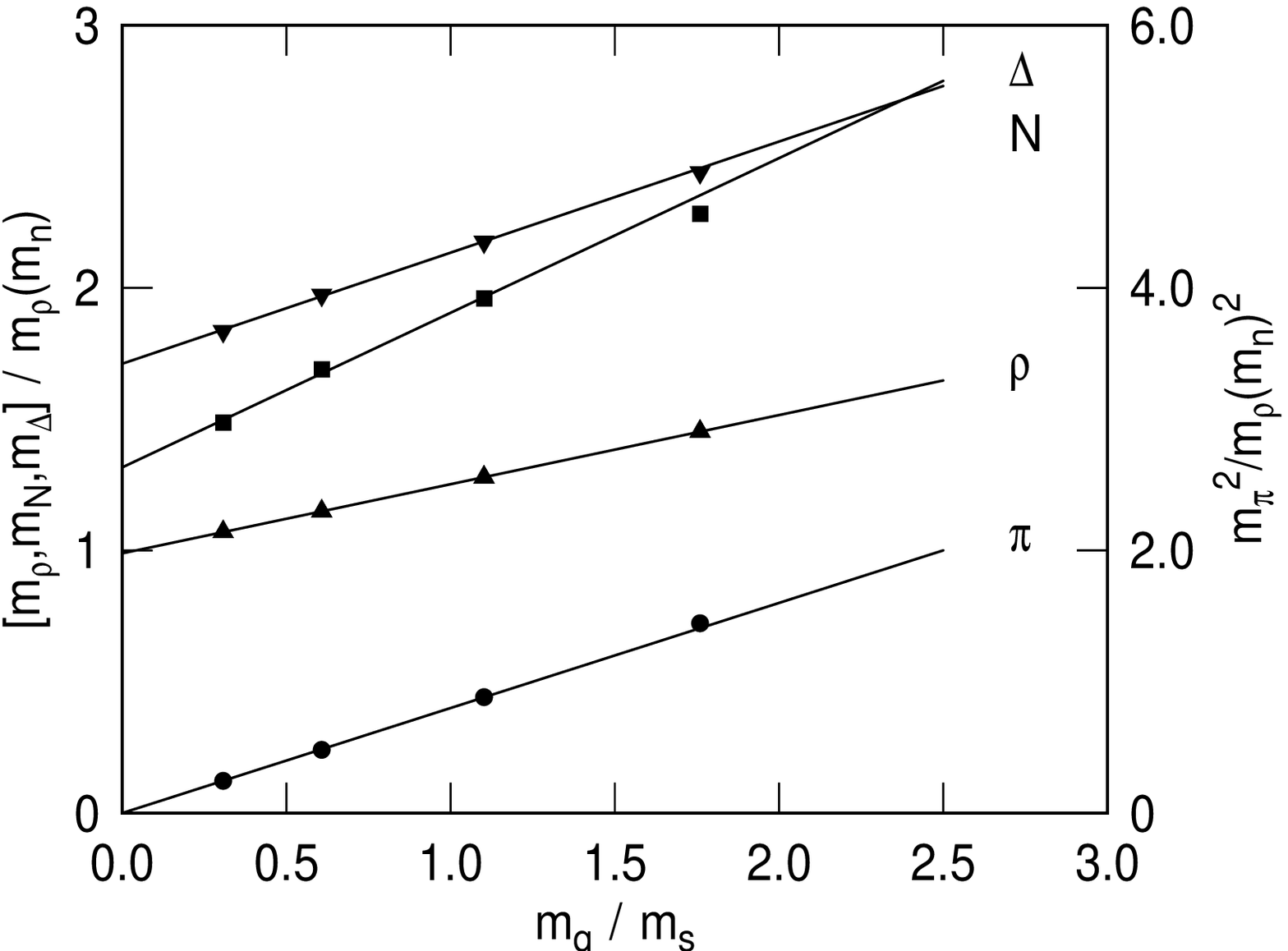}
\caption{ 
For a $24^3 \times 32$ lattice at $\beta$ of 5.93 with a sink of
size 4, $m_{\pi}^2$, $m_{\rho}$, $m_N$ and $m_{\Delta}$, in
units of the physical rho mass $m_{\rho}(m_n)$, as functions of the
quark mass $m_q$, in units of the strange quark mass $m_s$. The symbol
at each point is larger than the error bars.}
\label{fig:mexx24s4}
\end{figure}

\begin{figure}
\epsfxsize=\textwidth \epsfbox{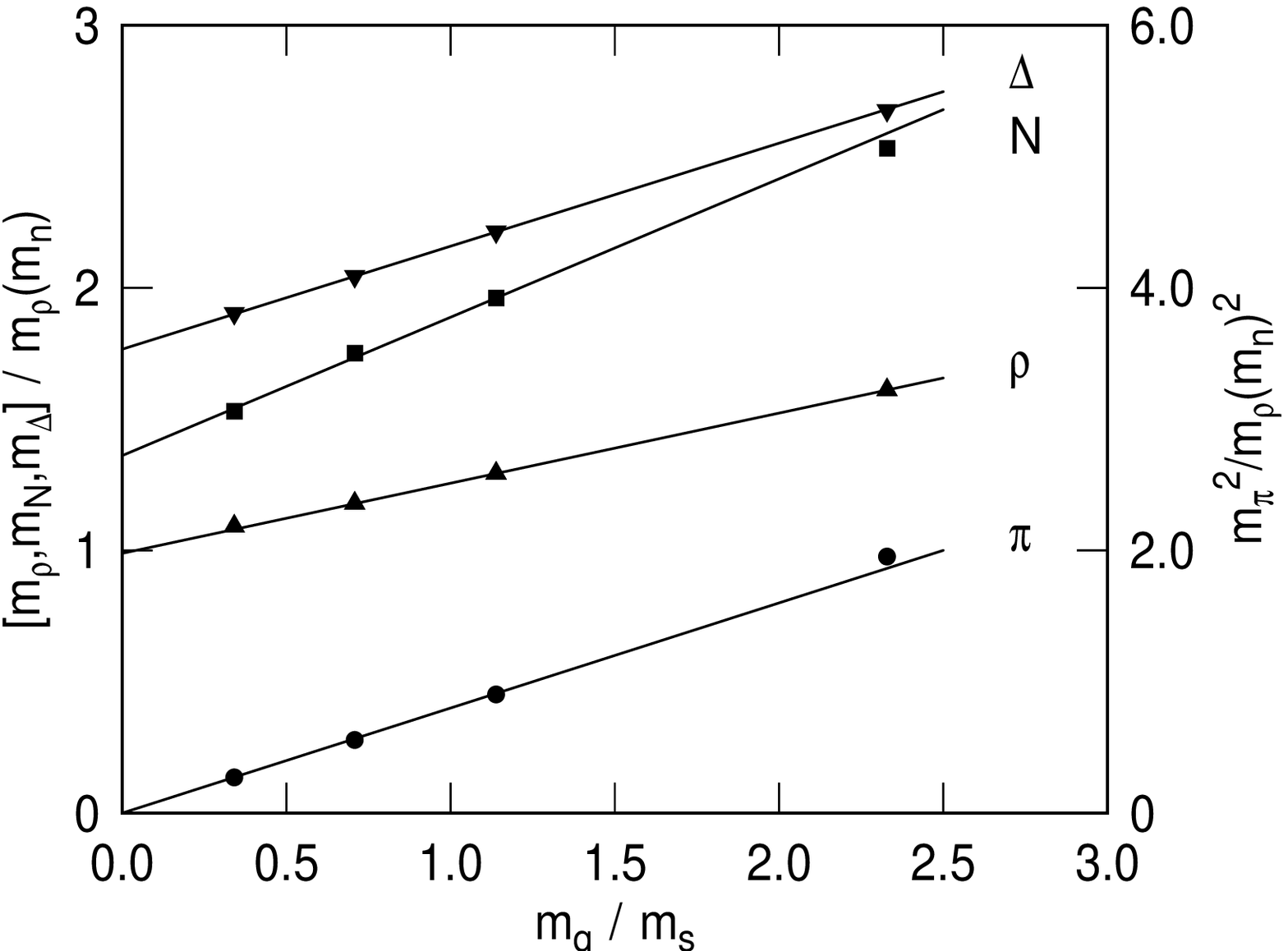}
\caption{ 
For a $30 \times 32^2 \times 40$ lattice at $\beta$ of 6.17 combining sinks of
sizes 0, 1 and 2, $m_{\pi}^2$, $m_{\rho}$, $m_N$ and $m_{\Delta}$, in
units of the physical rho mass $m_{\rho}(m_n)$, as functions of the
quark mass $m_q$, in units of the strange quark mass $m_s$. The symbol
at each point is larger than the error bars.}
\label{fig:mexx32s012}
\end{figure}

\begin{figure}
\epsfxsize=\textwidth \epsfbox{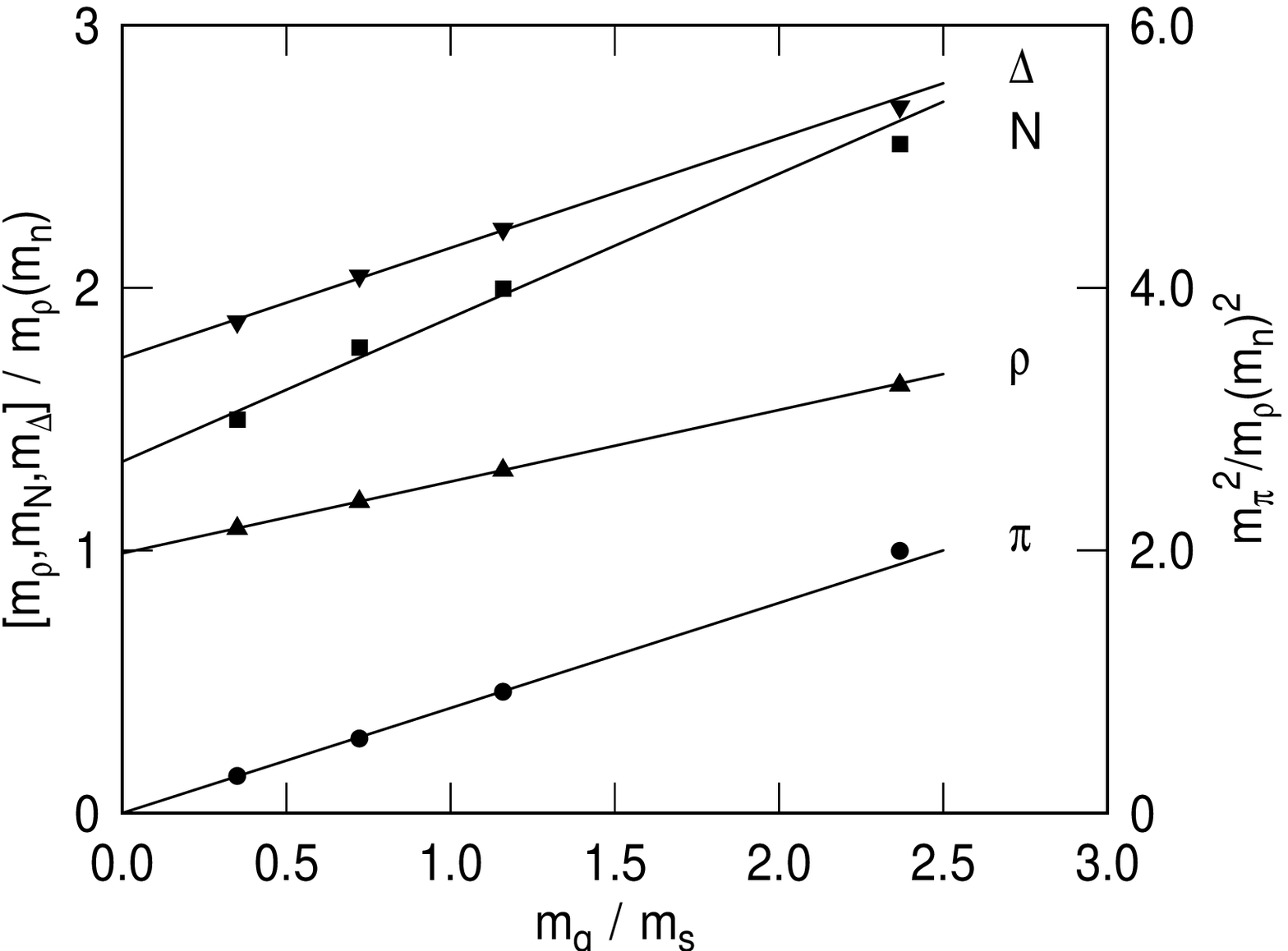}
\caption{ 
For a $30 \times 32^2 \times 40$ lattice at $\beta$ of 6.17 with a sink of
size 4, $m_{\pi}^2$, $m_{\rho}$, $m_N$ and $m_{\Delta}$, in
units of the physical rho mass $m_{\rho}(m_n)$, as functions of the
quark mass $m_q$, in units of the strange quark mass $m_s$. The symbol
at each point is larger than the error bars.}
\label{fig:mexx32s4}
\end{figure}

The quark mass in lattice units $m_q a$ can be defined as
\begin{eqnarray}
\label{defmq}
m_q a = \frac{1}{2 k} - \frac{1}{2 k_c},
\end{eqnarray}
where $k_c$ is the critical hopping constant at which $m_{\pi}$ becomes
zero. A naive application of the hypothesis that the pion is the
pseudo-Goldstone boson of spontaneously broken chiral symmetry implies
that $m_{\pi}^2$ should be linear in $m_q$ if $m_q$ is made small
enough.  Eq.~(\ref{defmq}) then yields linearity in $1/(2k)$ near
$1/(2k_c)$.  On the lattices $16^3 \times 32$, $24^3 \times 32$, $24^3
\times 36$ and $30 \times 32^2 \times 40$, we found $m_{\pi}^2$ 
to be close to a linear function of $1/(2k)$ at all the values of $k$ in
Table~\ref{tab:lattices}, and statistically consistent with exact
linearity at the three largest values of $k$. The best fit at the three
largest values of $k$ in each case was found by minimizing the fit's
full correlated $\chi^2$. Values of $k_c$ were taken from these fits.
Eq.~(\ref{defmq}) then gives a translation from $k$ to $m_q$.  For
$m_{\rho}$, $m_N$, and $m_{\Delta}$, a simple
application of perturbation theory suggests linearity in $m_q$ at small
enough $m_q$. These quantities we found to be close to linear in $m_q$
at all $k$ and statistically consistent with exact linearity at the
three largest $k$.  Figures~\ref{fig:mexx16s012} - \ref{fig:mexx32s4}
show these fits and extrapolations for masses from sinks 0, 1 and 2, and
for masses from sink 4 for the lattices $16^3 \times 32$, $24^3 \times
32$, $24^3 \times 36$ and $30 \times 32^2
\times 40$. 
Hadron masses in Figures~\ref{fig:mexx16s012} - \ref{fig:mexx32s4} are
shown in units of $m_{\rho}$ extrapolated to physical quark mass, and
the quark mass is shown in units of the strange quark mass $m_s$.  How
we determine $m_s$ will be discussed in Section~\ref{sec:strange}.

On each lattice, the fits of $m_{\pi}^2$ and $m_{\rho}$ to linear
functions of $m_q a$ we used to determine the value of $m_q a$ which
gives the physical value of $m_{\pi} / m_{\rho}$.  This choice of $m_q
a$ we define to be the ``normal quark'' mass $m_n a$. The fits, for each
lattice, of $m_N a$ and $m_{\Delta} a$ to linear functions of $m_q a$ we
then evaluated at $m_n a$ to find $m_N a$ and $m_{\Delta} a$ at physical
quark mass.  These masses taken at $m_n a$ in all cases differ by less
than one standard deviation from their values extrapolated to zero
$m_q$.  Numbers for $m_n a$ and $m_{\rho}( m_n) a$ are given in
Table~\ref{tab:mq_mrho}.  Tables~\ref{tab:mexx16} - \ref{tab:mexx32}
give a variety of ratios of extrapolated hadron mass values both for
data obtained from simultaneous fits to sinks of size 0, 1 and 2 and for
data from fits to sinks of size 4. For each ratio in each table, the
last column gives the value of $\chi^2$ per degree of freedom of the
linear fit used to extrapolate the ratio's numerator to the correct
quark mass.

Following Refs.~\cite{Lepage,Fermilab} we have also calculated, for each
lattice, values of $g^{(0)}_{\overline{MS}}$. From these and the
two-loop beta function for the Callan-Symanzik equation, we have found
for each lattice $\Lambda^{(0)}_{\overline{MS}} a$.
Tables~\ref{tab:mexx16} - \ref{tab:mexx32} give these results measured
in units of $m_{\rho}( m_n) a$.

\section{STRANGE HADRON MASSES} \label{sec:strange}

The linear relations between hadron masses and quark and antiquark
masses which we have found to be approached at small enough quark and
antiquark mass for hadrons composed of a single mass of quark and
antiquark can be summarized
\begin{eqnarray}
\label{pilin}
m_{\pi}(m_q)^2 & = & a_{\pi} m_q,  \\
\label{rhlin}
m_{\rho}(m_q) & = & a_{\rho} m_q + c_{\rho},  \\
\label{bnlin}
m_B(m_q) & = & a_B m_q + c_B.
\end{eqnarray}
Here $m_q$ is a variable mass put in place of the normal quark and
antiquark mass, and $B$ is either the nucleon or the delta baryon. These
equations strongly suggest that for strange hadrons composed of quarks
and antiquarks with different masses we should have in addition
\begin{eqnarray}
\label{kalin}
m_K(m_{q 1}, m_{q 2})^2 & = & a_K m_{q 1} + b_K m_{q 2},  \\
\label{kslin}
m_{K^*}(m_{q 1}, m_{q 2}) & = & a_{K^*} m_{q 1} + b_{K^*} m_{q 2} + c_{K^*},  \\
\label{bslin}
m_B(m_{q 1}, m_{q 2}) & = & a_B m_{q 1} + b_B m_{q 2} + c_B, 
\end{eqnarray}
where $m_{q 1}$ and $m_{q 2}$ are variable masses put in place of the
normal and strange quark masses, respectively, and B is any strange
baryon.

Assuming Eqs.~(\ref{kalin}) - (\ref{bslin}), a set of equations can be
derived among the coefficients of related hadrons.  These equations
permit the masses of hadrons containing both strange quarks and normal
quarks to be obtained from the masses of related hadrons composed of a
single flavor of quark with quark mass some weighted average of normal
and strange quark masses. The equations among masses found in this way
are closely connected to the Gell-Mann-Okubo mass formulas.

For the pseudo-scalar and vector mesons, charge conjugation invariance
implies
\begin{eqnarray}
\label{abeqs}
a_K & = & b_K, \nonumber \\
a_{K^*} & = & b_{K^*}. 
\end{eqnarray}
By flavor SU(3) symmetry, pion and kaon masses become degenerate if
$m_{q 1}$ and $m_{q 2}$ are equal, and the rho and k-star masses become equal if
$m_{q 1}$ and $m_{q 2}$ are equal. Thus
\begin{eqnarray}
\label{aacceqs}
a_K & = & a_{\pi}, \nonumber \\
a_{K^*} & = & a_{\rho}, \nonumber \\
c_K & = & c_{\pi}, \nonumber \\
c_{K^*} & = & c_{\rho}.
\end{eqnarray}

We obtain for strange meson masses at physical values of the normal and
strange quark masses
\begin{eqnarray}
\label{kamasseq}
m_K( m_n, m_s) & = & m_{\pi}[ (m_n + m_s)/2], \\
\label{kstarmasseq}
m_{K^*}( m_n, m_s) & = & m_{\rho}[ (m_n + m_s)/2]. 
\end{eqnarray}
In addition, in the valence approximation, the phi vector meson does not
mix with the omega and is composed purely of a strange quark and
antiquark, giving for the physical phi mass
\begin{eqnarray}
\label{phimasseq} 
m_{\Phi}( m_s)  =  m_{\rho}( m_s).
\end{eqnarray}

For baryons, relations similar to Eqs.~(\ref{kamasseq}) - (\ref{phimasseq}) can
be derived by combining Eq.~(\ref{bnlin}) and (\ref{bslin})
with the asymptotic form Eq.~(\ref{basym}) and definitions
Eqs.~(\ref{defPDel}) - (\ref{defgamma}).  Differentiating Eq.~(\ref{basym})
with respect to a quark mass and taking the asymptotic behavior for
large $t$ gives
\begin{eqnarray}
\label{bdasym}
\frac{\partial}{\partial m_{q i}} C^B( t) & \rightarrow & 
- t \left[ \frac{\partial}{\partial m_{q i}} m_B \right] C^B( t).
\end{eqnarray}
Evaluating the left side of Eq.~(\ref{bdasym}) for $m_{q 1}$ equal to $m_{q 2}$
and using Eqs.~(\ref{defPDel}) - (\ref{defgamma}), a variety of linear
relations can be obtained among the derivatives of baryon masses with
respect to quark masses.  

For the spin 1/2 baryon multiplet we find
\begin{eqnarray}
\label{abNeq}
a_N & = & b_{\Sigma} + b_{\Xi}, \\
\label{bbNeq}
b_{\Xi} & = & \frac{1}{2} b_{\Sigma} + \frac{3}{2} b_{\Lambda},
\end{eqnarray}
and the coefficients $c_B$ are the same for all members of the
multiplet.  For the spin 3/2 baryon multiplet we find
\begin{eqnarray}
\label{baSeq}
b_{\Sigma^*} & = & \frac{1}{3} a_{\Delta}, \\
\label{baXeq}
b_{\Xi^*} & = & \frac{2}{3} a_{\Delta}, \\
\label{baOeq}
b_{\Omega} & = & a_{\Delta},
\end{eqnarray}
and the coefficients $c_B$ are, again, the same for all members of the
multiplet.

From Eqs.~(\ref{abNeq}) - (\ref{baOeq}) we obtain
\begin{eqnarray}
\label{sxmasseq}
m_{\Sigma}(m_n, m_s) + m_{\Xi}( m_n, m_s) - m_N( m_n) & = & m_N( m_s), \\
\label{simasseq}
m_{\Sigma^*}(m_n, m_s) & = & m_{\Delta}[ (2 m_n +  m_s)/3], \\
\label{xsmasseq}
m_{\Xi^*}(m_n, m_s) & = & m_{\Delta}[ ( m_n +  2 m_s)/3], \\
\label{ommasseq}
m_{\Omega}( m_s) & = & m_{\Delta}( m_s).
\end{eqnarray}

Eqs.~(\ref{kamasseq}) - (\ref{phimasseq}) and (\ref{sxmasseq}) - (\ref{ommasseq})
permit a variety of strange hadron masses to be obtained from the fits
in Section~\ref{sec:mextrap} of hadron masses to Eqs.~(\ref{pilin}) -
(\ref{bnlin}).  For each lattice, we determine the strange quark mass by
tuning $m_s$ in Eq.~(\ref{kamasseq}) to produce the physical value of
$m_K / m_{\rho}$. For the values of $m_n$ and $m_{\rho}$ needed in these
calculations we take the results of Section~\ref{sec:mextrap}. Values of
$m_s$ found in this way are listed in Table~\ref{tab:mq_mrho}. Strange
hadron masses determined from $m_n$ and $m_s$ are given in
Tables~\ref{tab:mexx16} -
\ref{tab:mexx32} both for simultaneous fits to sinks of size 0, 1 and 2
and for fits to sink size 4.  The hadron masses in these tables are all
measured in units of the extrapolated $m_{\rho}$ given in
Table~\ref{tab:mq_mrho}.
For each mass ratio in each table, the last column again gives $\chi^2$
per degree of freedom for the linear fits used to determine the ratios
numerator.

\section{TEST OF EXTRAPOLATION TO SMALL QUARK MASS} \label{sec:test}

A test of the linear relations Eqs.~(\ref{pilin}) - (\ref{bslin}) and of our
determination in Section~\ref{sec:mextrap} of the masses of light hadrons
by extrapolation can be made using observed values of hadron masses.

Eqs.~(\ref{kstarmasseq}) and (\ref{phimasseq}) combined with physical values
of the k-star and phi masses give values for the mass of a rho made
of heavy quarks. Eqs.~(\ref{simasseq}) - (\ref{ommasseq}) can be used similarly
to find the mass of a delta baryon composed of heavy quarks. For the
nucleon, define $m_{\Sigma \Lambda}$ to be
\begin{eqnarray}
\label{defsl}
m_{\Sigma \Lambda}( m_{q 1}, m_{q 2}) = 
\frac{1}{4} m_{\Sigma}( m_{q 1}, m_{q 2}) + 
\frac{3}{4} m_{\Lambda}( m_{q 1}, m_{q 2}).
\end{eqnarray}
Then we have
\begin{eqnarray}
\label{xi2masseq}
m_{\Xi}( m_n, m_s) & = & m_{\Sigma \Lambda}( m_n, 2 m_s), \\ 
\label{xi3masseq}
m_N( m_n) & = & m_{\Sigma \Lambda} ( m_n, m_n).
\end{eqnarray}
The first of these equations follows from Eq.~(\ref{bbNeq}), and the second
holds because the masses of the octet of spin 1/2 baryons becomes degenerate
if $m_s$ and $m_n$ are equal.

\begin{figure}
\epsfxsize=\textwidth \epsfbox{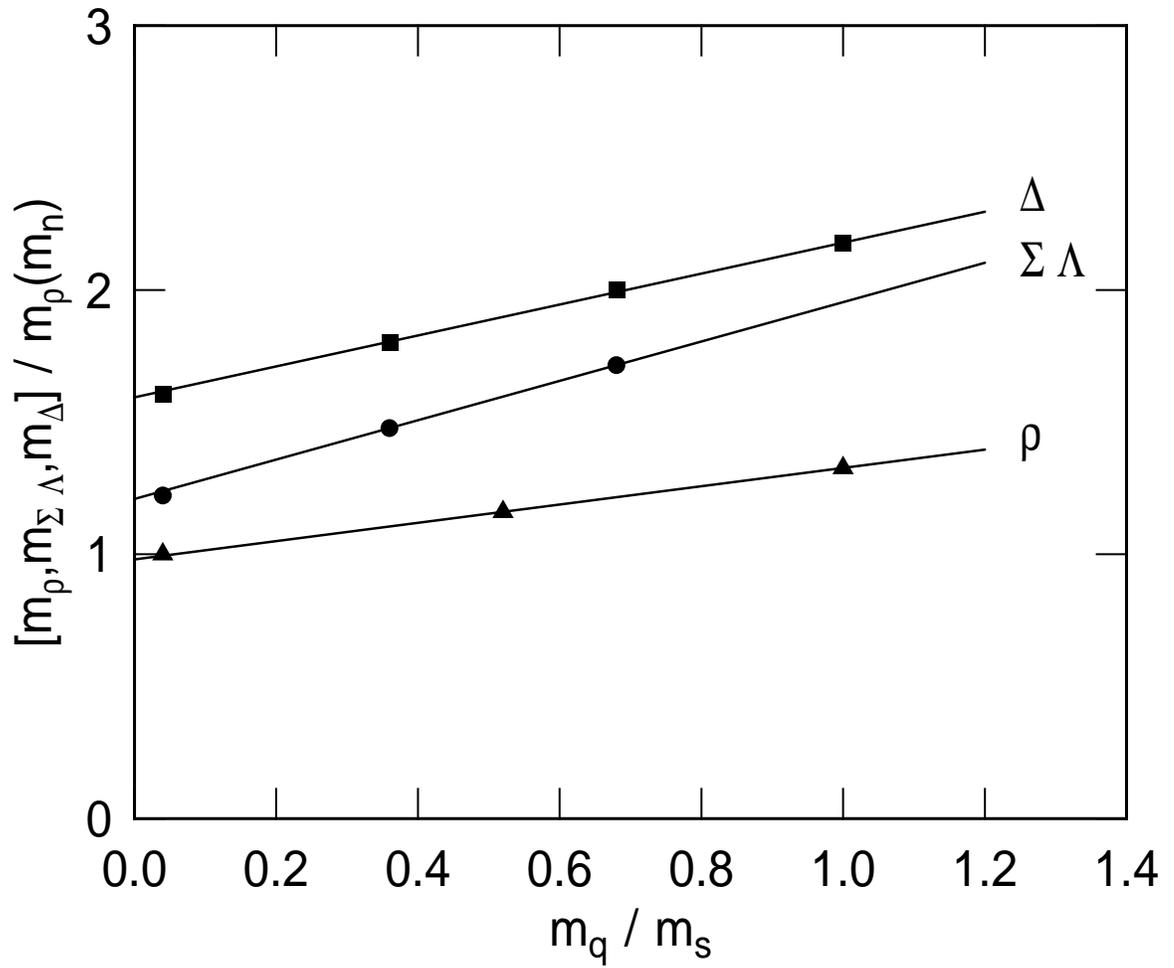}
\caption{ 
Linear extrapolation down to $m_q = m_n$ of synthetic values of
$m_{\rho}( m_q)$, $m_{\Sigma \Lambda}(m_n, m_q)$ and $m_{\Delta}(m_q)$
obtained from strange hadron masses.}
\label{fig:mexreal}
\end{figure}

From $m_{\rho}$, $m_{\Delta}$ and $m_{\Sigma \Lambda}$ for heavy quarks
we can now attempt to recover the physical values $m_{\rho}(m_n)$,
$m_{\Delta}(m_n)$ and $m_N(m_n)$ by linear extrapolation as done in
Section~\ref{sec:mextrap}.  Figure~\ref{fig:mexreal} shows linear
extrapolations in $m_q$ down to the value $m_n$ of $m_{\rho}( m_q)$,
$m_{\Sigma \Lambda}(m_n, m_q)$ and $m_{\Delta}(m_q)$. The scale for
$m_q$ is shown in units of the strange quark mass $m_s$ and the hadron
mass scale is shown in units of the physical rho mass $m_{\rho}( m_n)$. The
hadron mass at $m_q$ equal to $m_n$ compared with the $m_{\Sigma
\Lambda}$ extrapolation, following Eq.~(\ref{xi3masseq}), is the physical
nucleon mass $m_N( m_n)$. The extrapolated value for $m_{\rho}( m_n)$ is
low by 0.53\%, for $m_N(m_n)$ is high by 1.38\%, and for
$m_{\Delta}(m_n)$ is high by 0.81\%. The linear fit in each of these
extrapolations uses a range of $m_q / m_s$ contained within the range
used in our extrapolations in Section~\ref{sec:mextrap}.  Overall these
results support the accuracy of the linearity assumed in
Section~\ref{sec:strange} and the extrapolation to find hadron masses in
Section~\ref{sec:mextrap}.

\section{CONTINUUM LIMIT} \label{sec:contlim}

The value of $\beta$ for each of the three lattices $16^3 \times 32$,
$24^3 \times 36$ and $30 \times 32^2 \times 40$ was chosen so that the
physical volume in each case is nearly the same.  For lattice period $L$,
the quantity $m_{\rho} L$ is respectively, 9.08 $\pm$ 0.13, 9.24 $\pm$
0.19 and, averaged over three directions, 8.67 $\pm$ 0.12. Thus a
sequence of corresponding results on these three lattices gives each
predicition's behavior as the lattice spacing is made smaller with the
volume held fixed in physical units.

\begin{figure}
\epsfxsize=\textwidth \epsfbox{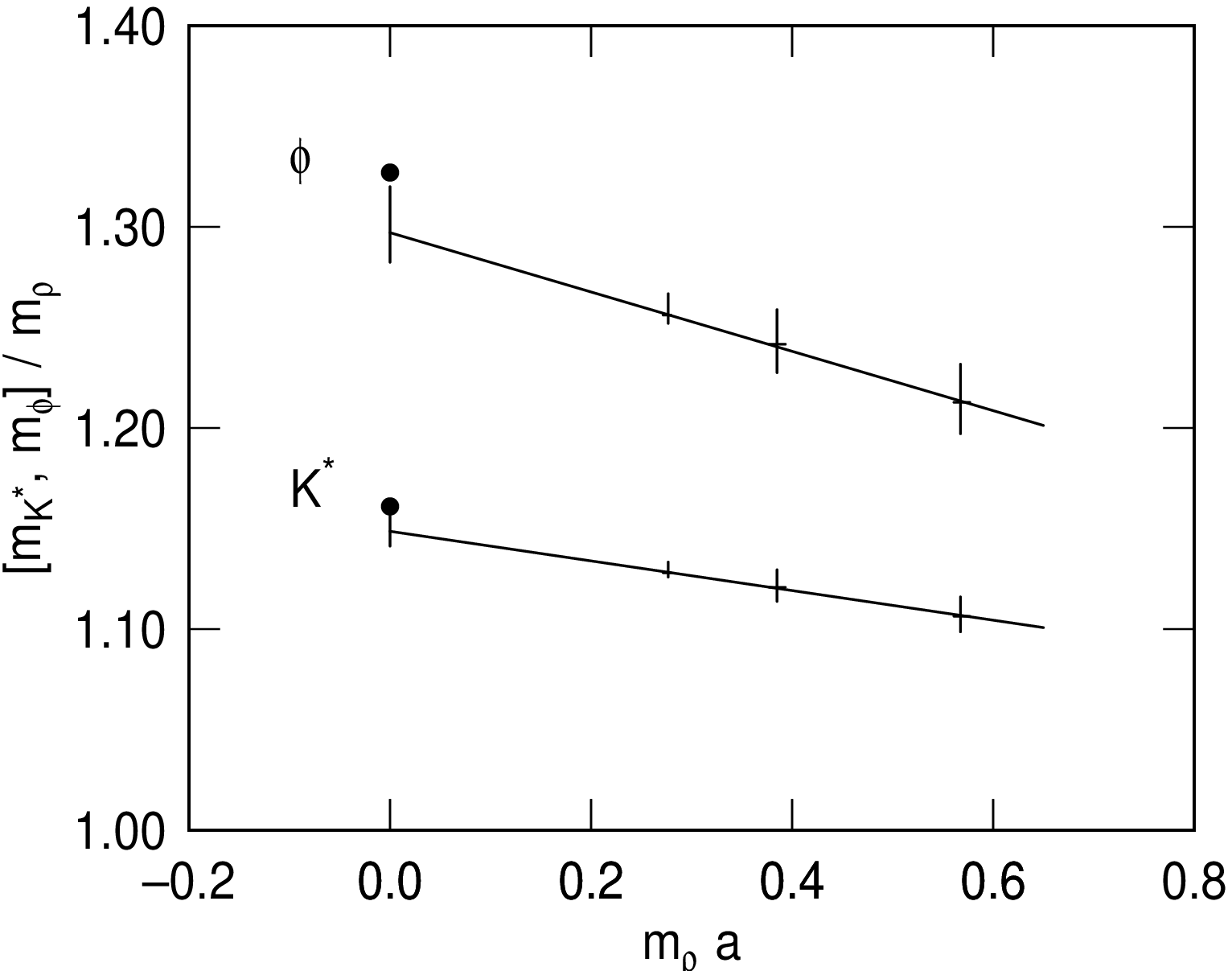}
\caption{ 
For sinks 0, 1 and 2 combined, $m_{K^{*}}/m_{\rho}$ and
$m_{\phi}/m_{\rho}$ as functions of the lattice spacing $a$, in units of
$1/m_{\rho}$.  The straight lines are extrapolations to zero lattice
spacing, the error bars at zero lattice spacing are uncertainties in the
extrapolated ratios, and the points at zero lattice spacing are observed
values.}
\label{fig:aexkstars012}
\end{figure}

\begin{figure}
\epsfxsize=\textwidth \epsfbox{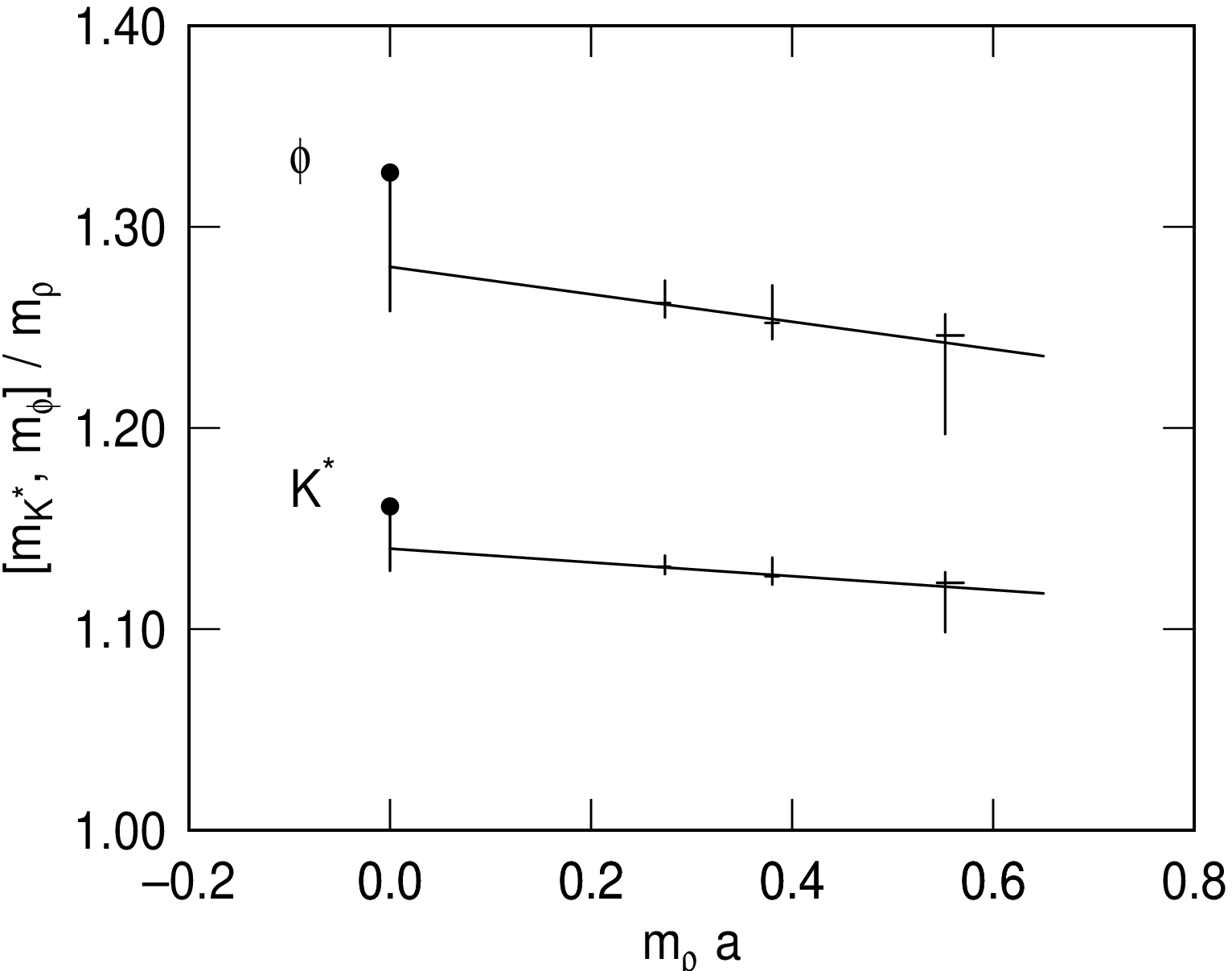}
\caption{ 
For sink 4, $m_{K^{*}}/m_{\rho}$ and $m_{\phi}/m_{\rho}$ as functions
of the lattice spacing $a$, in units of $1/m_{\rho}$.  The straight
lines are extraploations to zero lattice spacing, the error bars at zero
lattice spacing are uncertainties in the extrapolated ratios, and the
points at zero lattice spacing are observed values.}
\label{fig:aexkstars4}
\end{figure}

\begin{figure}
\epsfxsize=\textwidth \epsfbox{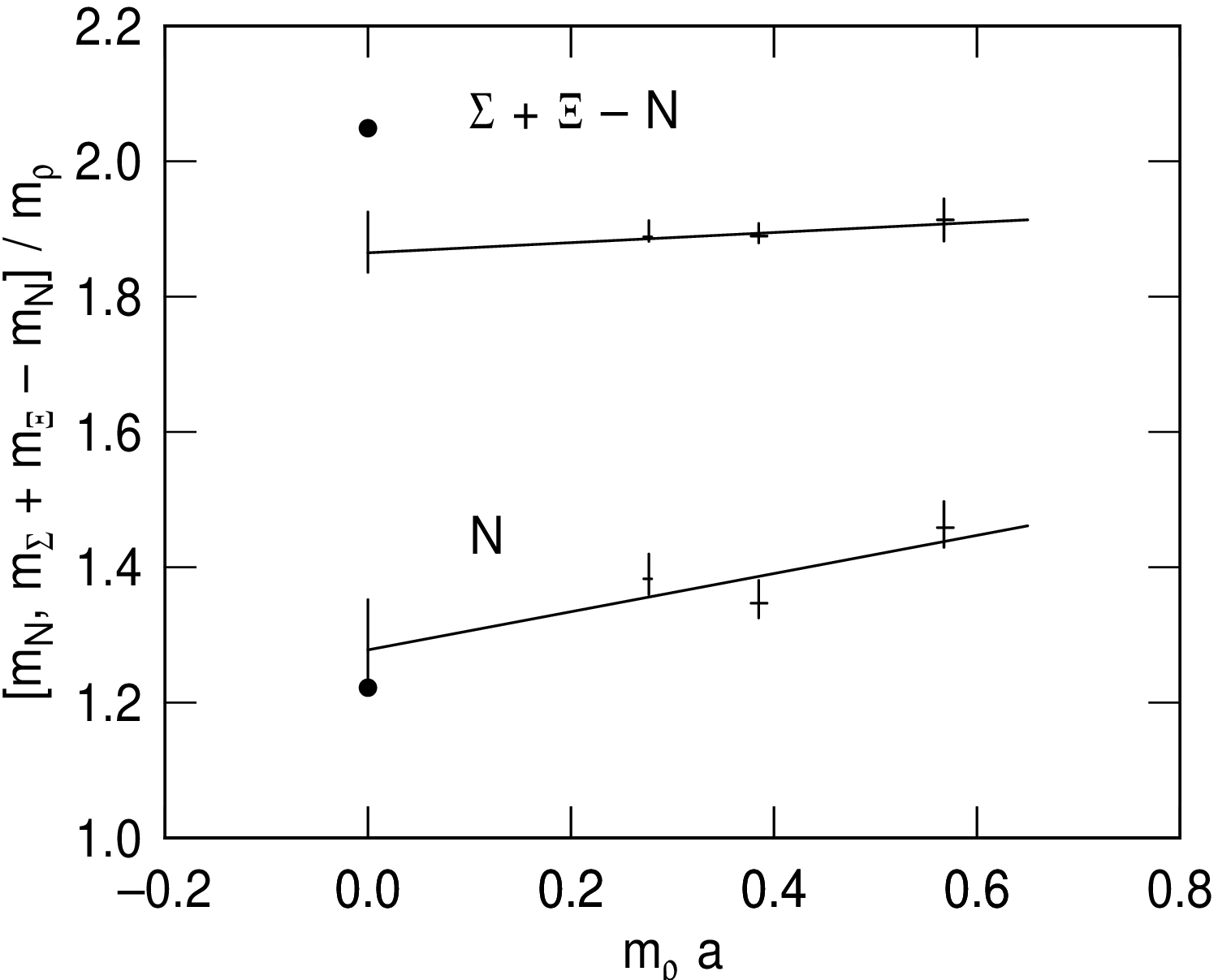}
\caption{ 
For sinks 0, 1 and 2 combined, $m_N/m_{\rho}$ and $(m_{\Xi} + m_{\Sigma}
- m_N) /m_{\rho}$ as functions of the lattice spacing $a$, in units of
$1/m_{\rho}$.  The straight lines are extraploations to zero lattice
spacing, the error bars at zero lattice spacing are uncertainties in the
extrapolated ratios, and the points at zero lattice spacing are observed
values.}
\label{fig:aexNs012}
\end{figure}

\begin{figure}
\epsfxsize=\textwidth \epsfbox{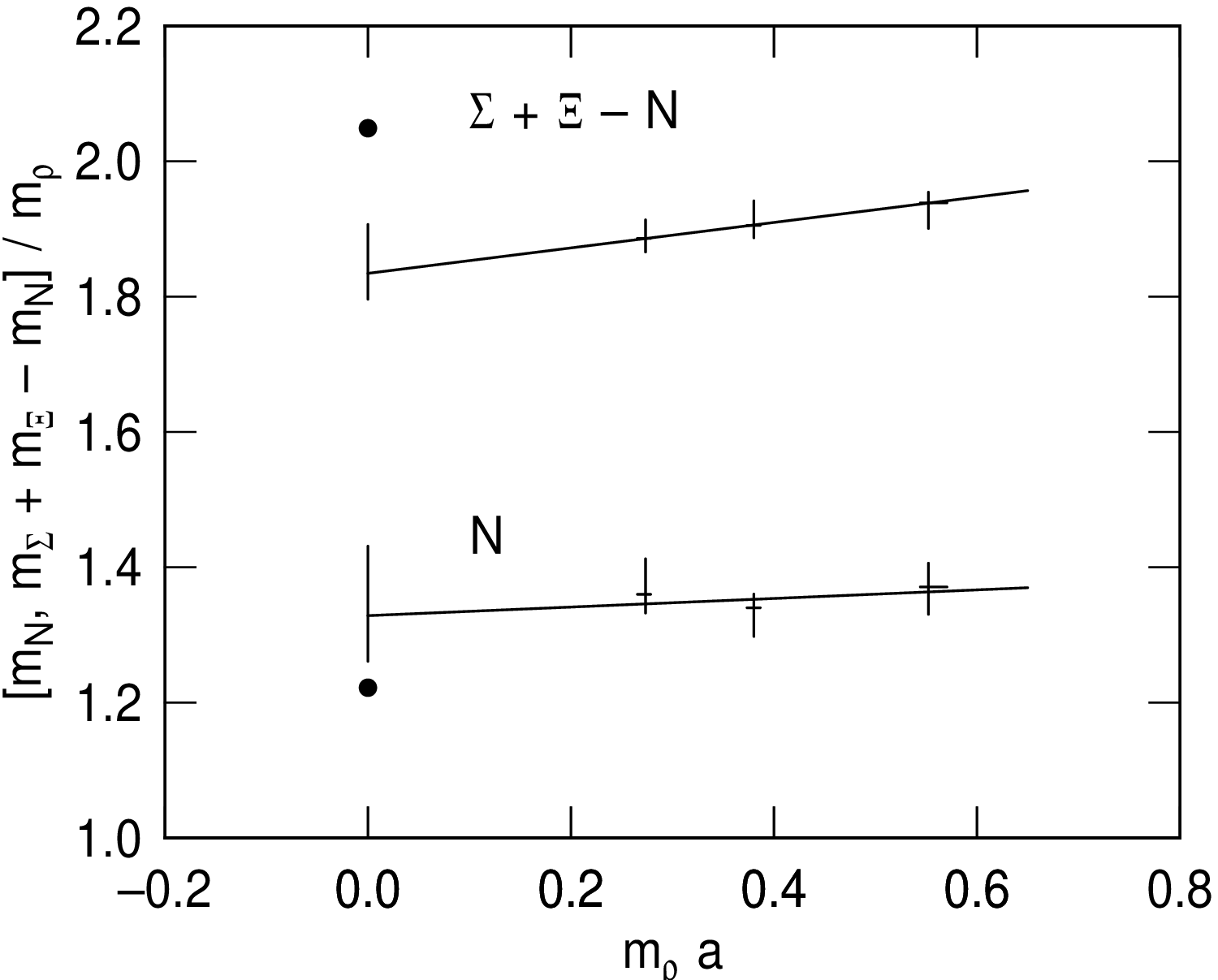}
\caption{ 
For sink 4, $m_N/m_{\rho}$ and $(m_{\Xi} + m_{\Sigma}
- m_N) /m_{\rho}$ as functions of the lattice spacing $a$, in units of
$1/m_{\rho}$.  The straight lines are
extraploations to zero lattice spacing, the error bars at zero lattice
spacing are uncertainties in the extrapolated ratios, and the points at
zero lattice spacing are observed values.}
\label{fig:aexNs4}
\end{figure}

\begin{figure}
\epsfxsize=\textwidth \epsfbox{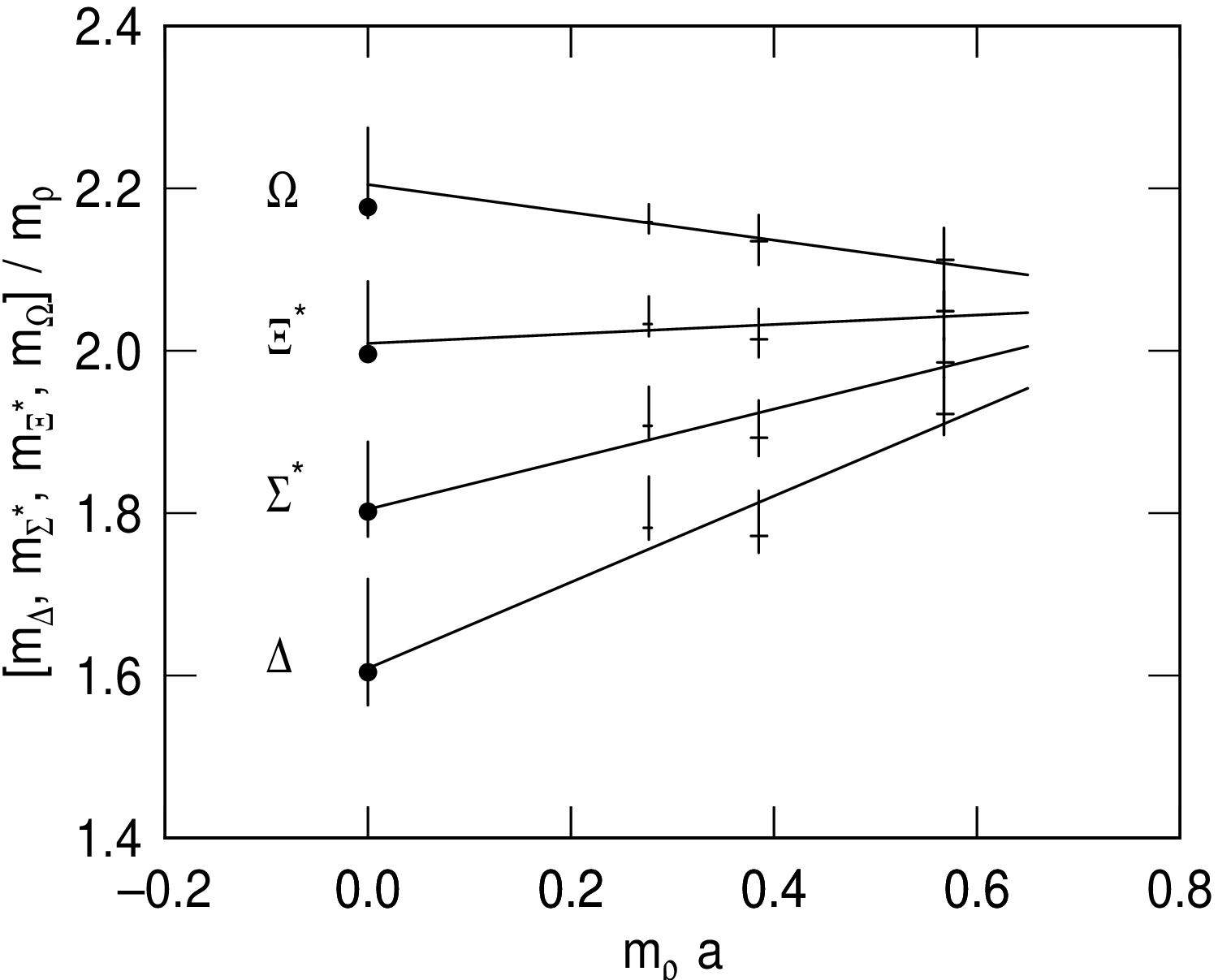}
\caption{ 
For sinks 0, 1 and 2 combined, $m_{\Delta}/m_{\rho}$,
$m_{\Sigma^{*}}/m_{\rho}$, $m_{\Xi^{*}}/m_{\rho}$ and
$m_{\Omega}/m_{\rho}$ as functions of the lattice spacing $a$, in units
of $1/m_{\rho}$.  The straight lines are extraploations to zero lattice
spacing, the error bars at zero lattice spacing are uncertainties in the
extrapolated ratios, and the points at zero lattice spacing are observed
values.}
\label{fig:aexDs012}
\end{figure}

\begin{figure}
\epsfxsize=\textwidth \epsfbox{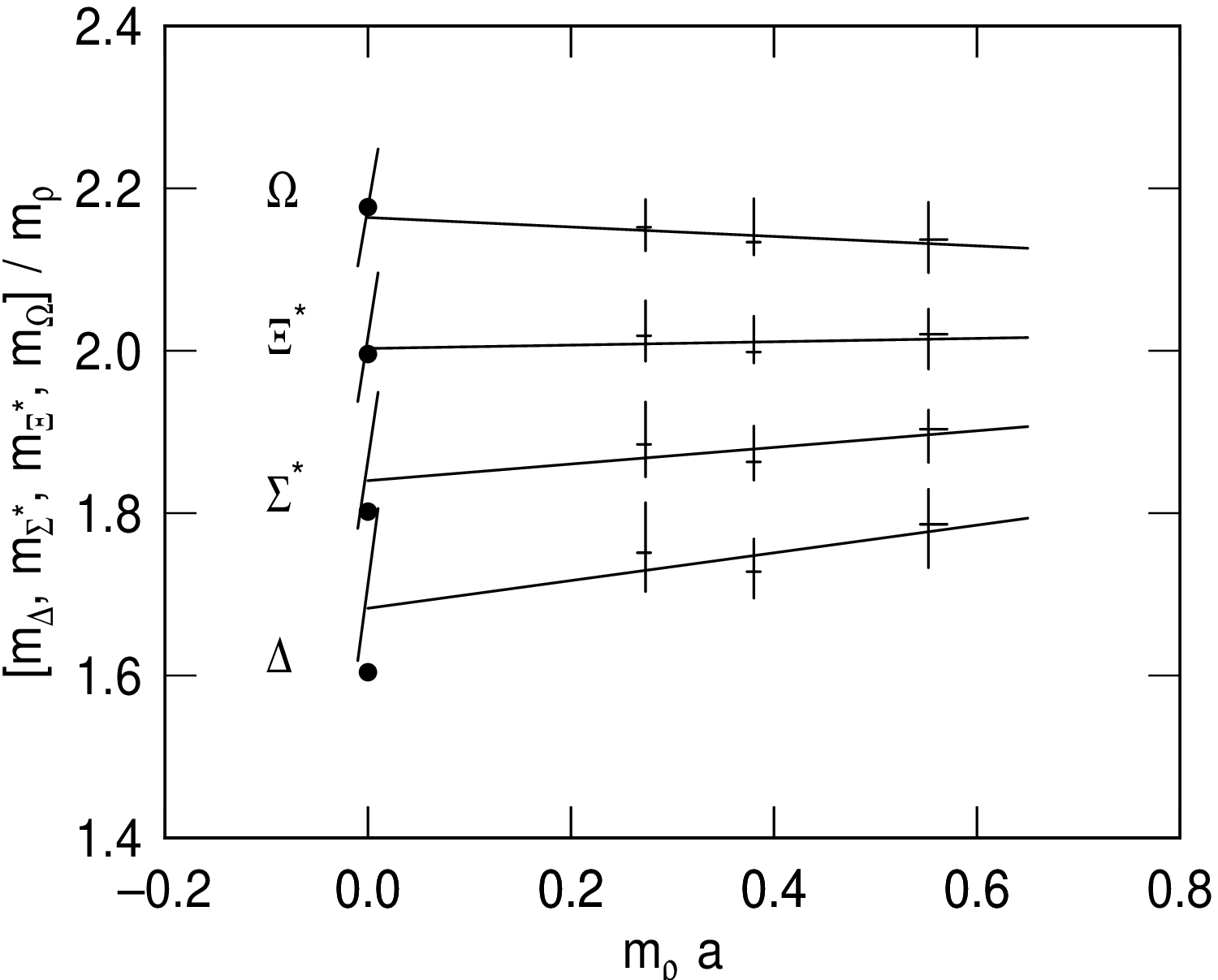}
\caption{ 
For sink 4, 
$m_{\Delta}/m_{\rho}$, $m_{\Sigma^{*}}/m_{\rho}$, $m_{\Xi^{*}}/m_{\rho}$
and $m_{\Omega}/m_{\rho}$  
as functions of the lattice spacing $a$, in units of
$1/m_{\rho}$.  The straight lines are extraploations to zero lattice
spacing, the error bars at zero lattice spacing are uncertainties in the
extrapolated ratios, and the points at zero lattice spacing are
observed values.}
\label{fig:aexDs4}
\end{figure}

For sufficiently small values of lattice spacing, the leading lattice
spacing dependence of hadron mass ratios is expected to be linear in
$a$. Figures~\ref{fig:aexkstars012} - \ref{fig:aexDs4} show mass ratios
for the lattices $16^3 \times 32$, $24^3 \times 36$ and $30 \times 32^2
\times 40$, listed in Tables~\ref{tab:mexx16},
\ref{tab:mexx24t36} and \ref{tab:mexx32}, as a function of lattice spacing
measured in units of $m_{\rho}( m_n)$. The lines are linear fits to
these points. The vertical bars at zero lattice spacing are the
statistical uncertainties in the linear extrapolations of hadron mass
ratios to zero lattice spacing. The dots at zero lattice spacing
represent the observed physical values of ratios. The extrapolated
ratios, corresponding observed values and $\chi^2$ per degree of freedom
of the linear fits are given in Table~\ref{tab:results}.

The finite volume continuum limits shown in Table~\ref{tab:results} for
simultaneous fits to sinks 0, 1 and 2 and for fits to sink 4 are
consistent and all lie within 1 standard deviation of each other. The
size of the statistical errors for the simultaneous fits to sinks 0, 1
and 2 are smaller than those for sink 4. In general, for an increasing
sequence of Monte Carlo statistical uncertainties, the uncertainties in
these uncertainties increase more rapidly. Thus we believe the
simultaneous fits to sinks 0, 1 and 2 and their error bars are more
reliable than the corresponding numbers found from sink 4.  With the
exception of the value of $(m_{\Xi} + m_{\Sigma} - m_N)/m_{\rho}$, the
predicted numbers lie within 1.7 standard deviations of the observed
results and are statistically consistent with the observed results. The
prediction for $(m_{\Xi} + m_{\Sigma} - m_N)/m_{\rho}$ differs from
experiment by 4.0 standard deviations.  We will show in
Section~\ref{sec:vollim}, however, that the agreement between the
prediction for $(m_{\Xi} + m_{\Sigma} - m_N)/m_{\rho}$ and observation
becomes much better when a correction to obtain infinite volume results is
applied.

\begin{figure}
\epsfxsize=\textwidth \epsfbox{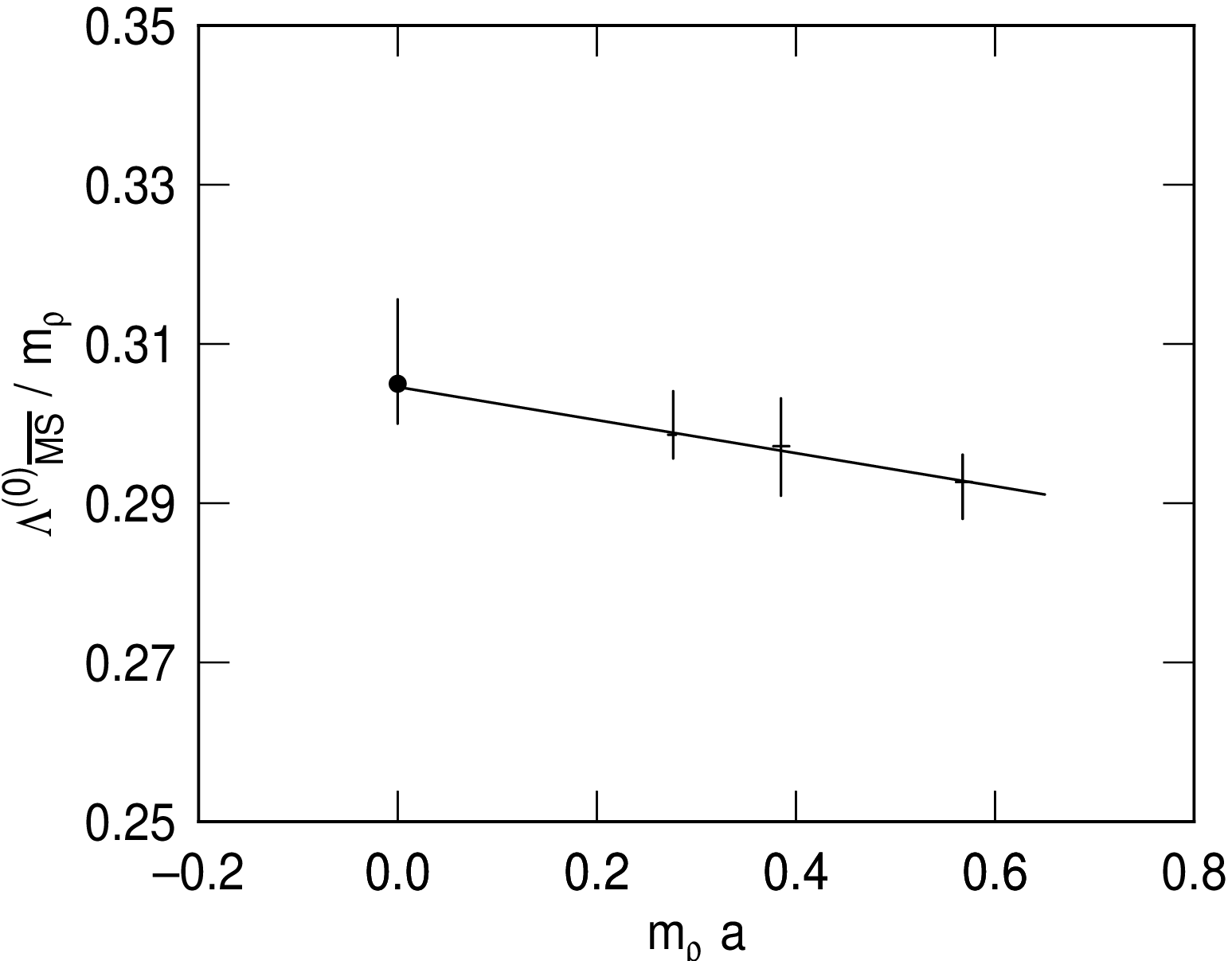}
\caption{ 
For sinks 0, 1 and 2 combined, 
$\Lambda^{(0)}_{\overline{MS}} / m_{\rho}$
as a function of the lattice spacing $a$, in units of
$1/m_{\rho}$.  The straight line is an extraploations to zero lattice
spacing, the error bars at zero lattice spacing are the uncertainty in the
extrapolated ratio, and the point at zero lattice spacing is another
groups result.}
\label{fig:aexlams012}
\end{figure}

\begin{figure}
\epsfxsize=\textwidth \epsfbox{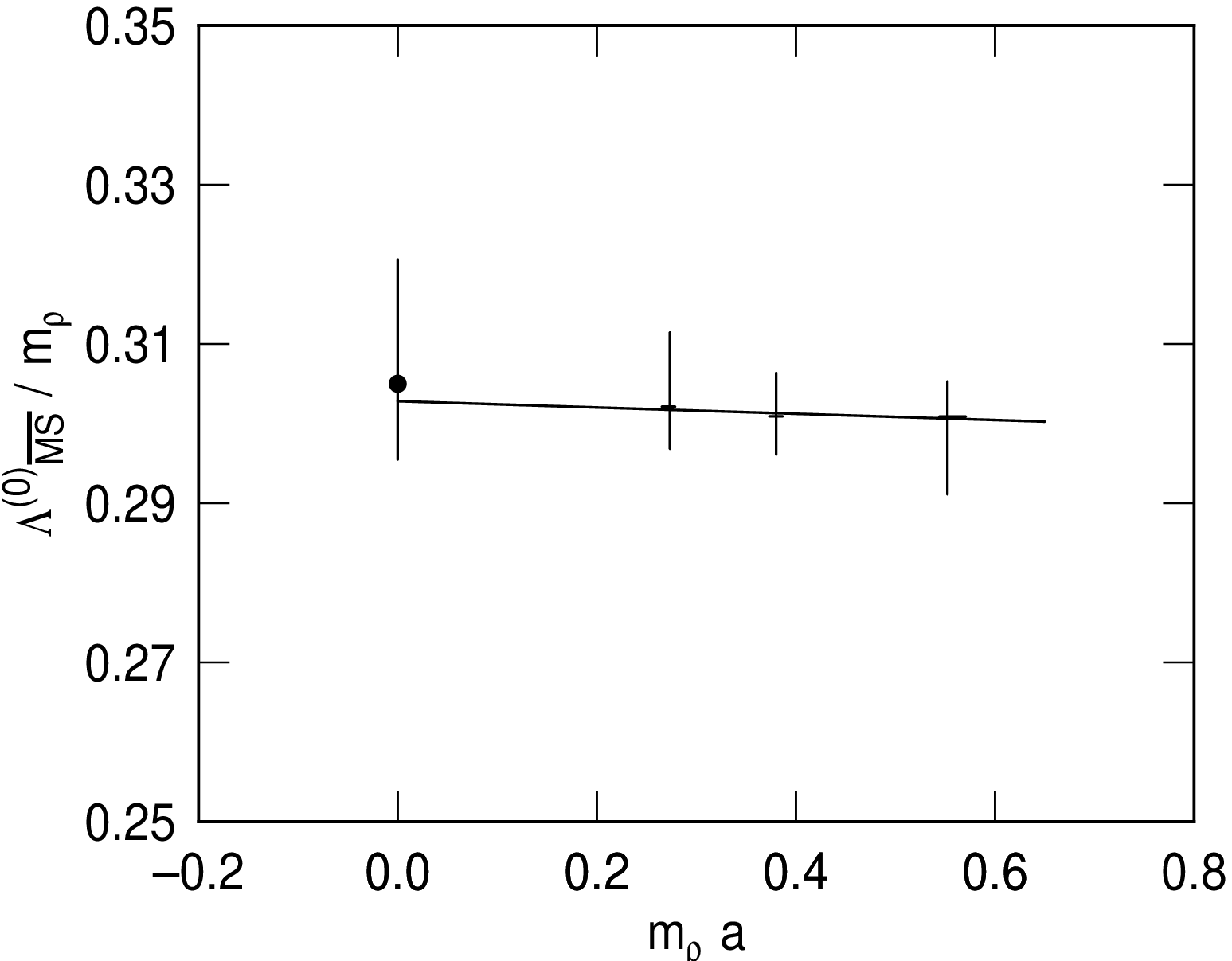}
\caption{ 
For sink 4, 
$\Lambda^{(0)}_{\overline{MS}} / m_{\rho}$
as a function of the lattice spacing $a$, in units of
$1/m_{\rho}$.  The straight line is an extraploations to zero lattice
spacing, the error bars at zero lattice spacing are the uncertainty in the
extrapolated ratio, and the point at zero lattice spacing is another
groups result.}
\label{fig:aexlams4}
\end{figure}

Figures~\ref{fig:aexlams012} and \ref{fig:aexlams4} show linear
extrapolations of  $\Lambda^{(0)}_{\overline{MS}} / m_{\rho}(m_n)$ to the
continuum limit, for $m_{\rho}( m_n)$ determined from sinks 0, 1, 2 and
from sink 4, respectively. The dot at zero lattice spacing in these
figures is the value determined from heavy quark spectroscopy in
Ref.~\cite{Fermilab}. Numerical values of the continuum limits shown in
these figures are listed in Table~\ref{tab:results} along with the
result from Ref.~\cite{Fermilab}.

\begin{figure}
\epsfxsize=\textwidth \epsfbox{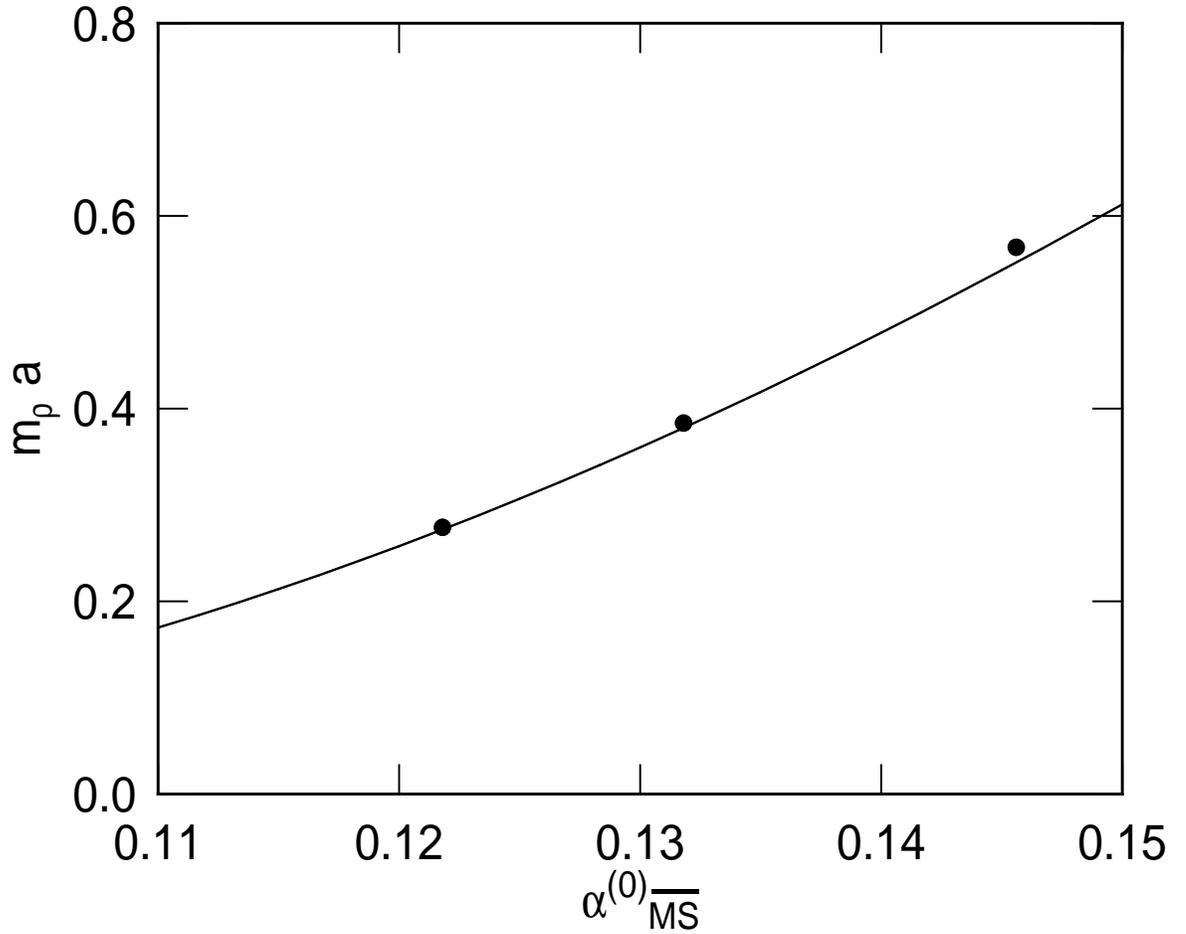}
\caption{ 
For sinks 0, 1 and 2, $m_{\rho} a$ as a function of
$\alpha^{(0)}_{\overline{MS}}$ in comparison to the prediction of the
Callan-Symanzik equation using the two-loop beta function and the
physical value of $\Lambda^{(0)}_{\overline{MS}}$ found from the
continuum limit of $\Lambda^{(0)}_{\overline{MS}} / m_{\rho} a$.}
\label{fig:asymscale}
\end{figure}

In both Figures~\ref{fig:aexlams012} and \ref{fig:aexlams4}, the slope
of the linear fit is statistically consistent with zero. Thus it follows
that the rho mass in lattice units $m_{\rho}(m_n) a$ depends on
$g^{(0)}_{\overline{MS}}$ as predicted by the Callan-Symanzik equation
using the two-loop beta function.  Figure~\ref{fig:asymscale} shows
$m_{\rho}( m_n) a$, for sinks 0, 1, 2, as a function of
$\alpha^{(0)}_{\overline{MS}}$ for the lattices $16^3 \times 32$, $24^3
\times 36$ and $30 \times 32^2 \times 40$. The line in this figure is 
the prediction of the Callan-Synamzik equation with 
$\Lambda^{(0)}_{\overline{MS}} / m_{\rho}(m_n)$ given by the continuum
limit value in Table~\ref{tab:results} found by linear extrapolation.
The data in Figure~\ref{fig:asymscale} appears to be quite close to the
asymptotic scaling curve. This evidence for asymptotic scaling tends to
support the reliability of the linear extrapolations we have used to
find continuum mass ratios.

The fits in Figures~\ref{fig:aexkstars012} - \ref{fig:aexDs4} were done
by minimizing the $\chi^2$ from the full correlation matrix among the
fitted data.  Both the x and y coordinates of each of the three fitted
points on each line have statistical uncertainties. We therefore
evaluated $\chi^2$ among all six pieces of data and chose as fitting
parameters the slope and intercept of the line along with the x
coordinate of each point.  The correlation matrices which we used were
found by the bootstrap method as were the statistical uncertainties of
the extrapolated predictions.  The correlation matrices used in fits for
each bootstrap ensemble were taken, for convenience, from the full
ensemble and not recalculated on each bootstrap ensemble independently.

\section{INFINITE VOLUME LIMIT} \label{sec:vollim}

The continuum limits found so far are for a finite volume lattice.  We
now consider the infinite volume limit of these finite volume continuum
predictions.

As a first step, we compare masses found on the lattices $16^3 \times
32$ and $24^3 \times 32$ at $\beta$ of 5.70. For the values of $k$ at
which we did direct mass calculations, Table~\ref{tab:massvoldep} shows
percent changes in masses from $16^3 \times 32$ to $24^3 \times 32$ at
$\beta$ of 5.70.  Data is shown for simultaneous fits to sinks 0, 1 and
2 and for fits to sink 4.  For masses from fits to sinks 0, 1 and 2
there is some indication, with marginal statistical significance, of
decreases of up to about 5\%. For masses from fits to sink 4, there is
still weaker evidence for decreases in mass ranging up to about 3\%.
Overall, it appears to us the data in Table~\ref{tab:massvoldep} is best
taken as evidence for an upper bound of 5\% on the change in masses from
$16^3 \times 32$ to $24^3 \times 32$ at $\beta$ of 5.70 for the values
of $k$ at which we calculated hadron masses directly. For hadron masses
extrapolated to physical quark mass, measured in units of $m_{\rho}(
m_n)$, Table~\ref{tab:ratiovoldep} provides a bound of about 5\% on
volume dependence both for mass ratios from simultaneous fits to sinks 0, 1 and 2
and for masses from fits to sink 4.

From the change between a hadron's mass evaluated on a lattice $16^3
\times 32$ and on a lattice $24^3 \times 32$, an estimate can be made of
the change from $24^3 \times 32$ to infinite volume if we have
sufficient information concerning the form of the dependence of hadron
masses on lattice volume.  A simple non-relativistic potential model
implies that the error in a particle's mass due to calculation in a
finite volume $L^3$ falls at large $L$ as $C e^{- L/R}$, where $R$ is
the radius of the probability density of the hadron's wave function and
$C$ is either constant or monotonically decreasing.  At $\beta$ of 5.70, $R$
for each of the hadrons we consider is then less than about 4 lattice
units. On the other hand, a model \cite{Fukugita} based, in part, on a
rigorous argument \cite{Lusch} gives a volume dependent error of the
form $D L^{-3}$ with a constant $D$ for the range of $L$ we consider.
At still larger $L$ this model yields an error falling exponentially. 

Assuming volume dependent errors of the form $C e^{- L/4}$, the changes
we have found between masses on a lattice $16^3 \times 32$ and on a
lattice $24^3 \times 32$ imply changes in mass from $24^3 \times 32$ to
infinite volume of less than 1\%.  Assuming volume dependent errors of
the form $D L^{-3}$, the changes we have found between $16^3 \times 32$
and $24^3 \times 32$ imply changes in mass from $24^3
\times 32$ to infinite volume of less than 2\%.

Estimates can now be made of the corrections needed to obtain infinite
volume continuum mass ratios from our finite volume values. For the
ratio of any hadron mass to the rho mass $m_h / m_{\rho}$, define the
finite volume correction term $\delta_h(a, L)$ to be
\begin{eqnarray}
\label{defdelta}
\delta_h( a, L) = \frac{m_h}{m_{\rho}}( a, \infty) - 
\frac{m_h}{m_{\rho}}( a, L).
\end{eqnarray}
The quantity which we would like to determine is $\delta_h( 0,
9/m_{\rho})$, since $9/m_{\rho}$ is $L$ for the lattices we used to find
continuum limit masses. In Section \ref{sec:contlim} $m_h / m_{\rho}$ for
all $h$ we considered shows a relative change of less than 20\% as $a$
goes from its value $a_{5.7}$ at $\beta$ of 5.7 down to 0.  Thus we
would expect a corresponding error of less than 20\% of $\delta_h( 0,
9/m_{\rho})$ for the approximation
\begin{eqnarray}
\label{approxina}
\delta_h( 0, \frac{9}{m_{\rho}}) \approx \delta_h( a_{5.7}, \frac{9}{m_{\rho}}).
\end{eqnarray}
On the other hand, from our estimate of the volume dependent error in
masses found on $24^3 \times 32$, for which $L$ is $13.5 /
m_{\rho}$, it follows that with an additional of less than
1\% or 2\% of $m_h / m_{\rho}$ we have
\begin{eqnarray}
\label{approxinL}
\delta_h( a_{5.7}, \frac{9}{m_{\rho}}) \approx 
\frac{m_h}{m_{\rho}}( a_{5.7}, \frac{13.5}{m_{\rho}}) - 
\frac{m_h}{m_{\rho}}( a_{5.7}, \frac{9}{m_{\rho}}), 
\end{eqnarray}
Finally, for each of the hadrons we consider, we have already found that the right
side of Eq.~(\ref{approxinL}) is less than 5\% of $m_h / m_{\rho}$.
Combining Eqs.~(\ref{defdelta}) - (\ref{approxinL}), we obtain the approximation
\begin{eqnarray}
\label{approxfinal}
\frac{m_h}{m_{\rho}}( 0, \infty) \approx
\frac{m_h}{m_{\rho}}( 0, \frac{9}{m_{\rho}}) + 
\frac{m_h}{m_{\rho}}( a_{5.7}, \frac{13.5}{m_{\rho}}) - 
\frac{m_h}{m_{\rho}}( a_{5.7}, \frac{9}{m_{\rho}}), 
\end{eqnarray}
with two contributions to the error. One contribution is less than 1\%
or 2\% of $m_h/m_{\rho}$, and the other less than 20\% of 5\% of
$m_h/m_{\rho}$, which is 1\% of $m_h/m_{\rho}$.  Thus overall the error
in Eq.~(\ref{approxfinal}) should be less than about 2\% of $m_h/m_{\rho}$.

Table~\ref{tab:results} shows infinite volume continuum limit mass
ratios obtained from the finite volume values using
Eq.~(\ref{approxfinal}). Results again are shown both for masses from
simultaneous fits to sinks 0, 1 and 2 and from fits to sink 4. The two
sets of numbers are statistically consistent but the errors for the fits
to sinks 0, 1 and 2 are smaller. As before, we consider both the
predicted values for sinks 0, 1 and 2 and the statistical errors on
these numbers to be more reliable than the corresponding numbers for
sink 4.  The main effect of the correction to infinite volume is an
increase in standard deviations by about a factor of 1.5. The shift in
central values in all cases is small and less than about 1.2 infinite
volume standard deviations and 1.9 finite volume standard deviations.
The only significant effect of the infinite volume correction on the
comparison between predicted numbers and experiment is for the value of
$(m_{\Xi} + m_{\Sigma} - m_N)/m_{\rho}$. The finite volume prediction
for sinks 0, 1 and 2 differs from experiment by 4.0 standard deviations
while the infinite volume number differs from experiment by only 1.6
standard deviations.

The infinite volume continuum limit predictions for sinks 0, 1 and 2 in
Table~\ref{tab:results} are statistically consistent with experiment.
The predicted values differ from experiment by amounts ranging up to 1.6
standard deviations.  As a fraction of the observed results, the errors
range up to 6\% with statistical uncertainties ranging up to 8\%. The
infinite volume continuum limit predictions for sink 4 are also
statistically consistent with experiment but do not agree with
experiment as well as do the predictions from sinks 0, 1 and 2. The
errors for the sink 4 predictions range up to 1.6 standard deviations,
with the exception of a single error of 2.5 standard deviations. As a
fraction of observed results, the errors go up to 10\% with
uncertainties up to 11\%.  In place of an observed value for
$\Lambda^{(0)}_{\overline{MS}}$ with which to compare our corresponding
prediction in Table~\ref{tab:results}, we use an infinite volume
continuum limit result obtained by a lattice QCD calculation combined
with the observed value of a charmonium mass splitting \cite{Fermilab}.

\section{PREDICTIONS WITHOUT EXTRAPOLATION TO SMALL QUARK MASS} 
\label{sec:nomextrap}

Calculations based on chiral perturbation theory \cite{Bernard, Sharpe}
show that at sufficiently small quark mass a peculiarity of the valence
approximation might lead to significant deviations from the linearity
between hadron masses and quark mass which we find for quark masses
above $0.3 m_s$.  In Ref.~\cite{Weingar94} evidence is discussed which
suggests that this potential difficulty occurs primarily at very small
quark mass and probably does not have a significant effect on our
extrapolations from $0.3 m_s$ down to $m_n$.  We now consider, however,
an alternate interpretation of our results which does not depend on the
extrapolation of hadron masses beyond the interval within which we have
direct evidence for linearity.

The linearity of real world hadron masses as a function of quark mass
found in Section~\ref{sec:test} implies that all masses in each hadron
multiplet are determined by the first two coefficients of a Taylor
series expansion around any quark mass between $m_n$ and $m_s$.  A
convenient expansion point, for example, is $m_{ns}$ defined to be
$(m_n + m_s)/2$.  We now reanalyze our data to obtain predictions
for five of the eight significant coefficients in Taylor expansions of 
$m_{\pi}( m_q)$,
$m_{\rho}( m_q)$, $m_N( m_q)$ and $m_{\Delta}( m_q)$ as
functions of $m_q - m_{ns}$. The three coefficients not predicted are,
in effect, taken from experiment and used to fix the three free parameters
of lattice QCD.

For each of the four lattices $16^3 \times 32$, $24^3 \times 32$, $24^3
\times 36$ and $30 \times 32^2 \times 40$, the fits in
Section~\ref{sec:mextrap} of $m_{\pi}(m_q)^2$, $m_{\rho}(m_q)$, $m_N(m_q)$ and
$m_{\Delta}(m_q)$ to linear functions of $m_q a$ we interpret, using
Eq.~\ref{defmq}, as fits to linear functions of $1/(2k)$.  We thereby
avoid the implicit dependence on extrapolation built into the original fits
as a consequence of the definition of $m_q a$ by Eq.~(\ref{defmq}). The critical
hopping constant $k_c$ entering Eq.~(\ref{defmq}) is found, in effect,
by extrapolation.  We then determine the hopping constant $k_{ns}$
corresponding to $m_{ns}$ by requiring $m_{\pi}(m_{ns}) /
m_{\rho}(m_{ns})$ to agree with the real world value of $m_K / m_{K^*}$
as expected according to Eqs.~(\ref{kamasseq}) and (\ref{kstarmasseq}).
For any $k$ we define the quark mass difference
\begin{eqnarray}
\label{defdeltmns}
m_q a - m_{ns} a = \frac{1}{2 k_{ns}} - \frac{1}{2 k}.
\end{eqnarray} 
The hopping constant $k_{2ns}$ corresponding to $2 m_{ns}$ we find by
requiring $m_{\pi}(2 m_{ns}) / m_{\rho}(m_{ns})$ to be equal to
$\sqrt{2} m_K / m_{K^*}$ as expected according to Eq.~(\ref{pilin}).  Both
$k_{ns}$ and $k_{2ns}$ for all four lattices lie within the interval for
which the fits in Section~\ref{sec:mextrap} give direct evidence, without
extrapolation, of linearity of $m_{\pi}( m_q)^2$, $m_{\rho}( m_q)$,
$m_N( m_q)$ and $m_{\Delta}( m_q)$ in $1/(2 k)$. The
determination of $k_{ns}$ and $k_{2ns}$ and the determination of hadron
masses at these points requires only interpolation of our data, not
extrapolation.  The value in lattice units $m_{ns} a$ is found from
Eq.~(\ref{defdeltmns}) to be
\begin{eqnarray}
\label{defmns}
m_{ns} a = \frac{1}{2 k_{ns}} - \frac{1}{2 k_{2ns}}.
\end{eqnarray} 
A definition of $m_q a$ which does not depend on extrapolation can be
found by combing Eqs.~(\ref{defdeltmns}) and (\ref{defmns}). For the
present discussion, however, we do not need a definition of $m_q a$
itself but only the difference $m_q a - m_{ns} a$.

Our fits $m_{\pi}( m_q)^2$, $m_{\rho}( m_q)$, $m_N( m_q)$ and
$m_{\Delta}( m_q)$ to linear functions of $1/(2k)$ we now reinterpret,
using Eq.~(\ref{defdeltmns}), as fits to linear functions of $m_q a -
m_{ns} a$.  Of the eight coefficients entering these fits, two have been
chosen in the course of determining $k_{ns}$ and $k_{2ns}$. A third is
needed to determine the lattice scale. We are left with five predictions
of Taylor coefficients at the point at which $m_q a - m_{ns} a$ is 0.
For the four lattices $16^3 \times 32$, $24^3 \times 32$, $24^3
\times 36$ and $30 \times 32^2 \times 40$, these five predictions are
shown in Tables~\ref{tab:mnxx16} - \ref{tab:mnxx32}. We show also
predictions for $\Lambda^{(0)}_{\overline{MS}} / m_{\rho}(m_{ns})$.
Numbers are again given both for fits to sinks 0, 1 and 2, and for fits
to sink 4. The two sets of numbers are statistically consistent, but the
error bars for the fits to sink 4 are significantly larger than those for the
fits to sinks 0, 1 and 2. The last column in each table gives the
$\chi^2$ per degree of freedom of the corresponding linear fit 
from which each mass or mass derivative was determined.

\begin{figure}
\epsfxsize=\textwidth \epsfbox{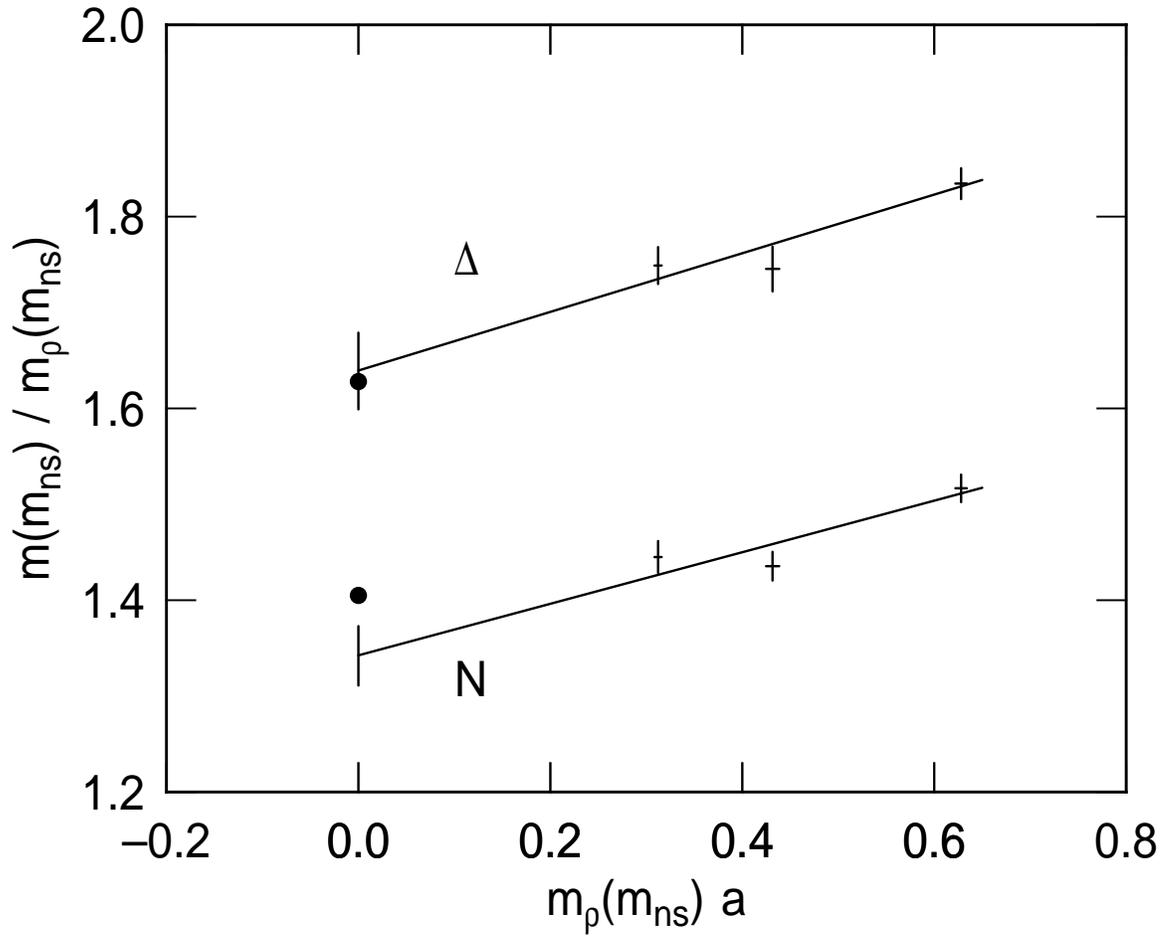}
\caption{ 
For sinks 0, 1 and 2, hadron masses evaluated at quark
mass $m_{ns}$ as a function of lattice spacing in units of
$1/m_{\rho}(m_{ns})$.  The straight lines are extraploations to zero lattice
spacing, the error bars at zero lattice spacing are uncertainties in the
extrapolated ratios, and the points at zero lattice spacing are
observed values.}
\label{fig:m_taylor}
\end{figure}

\begin{figure}
\epsfxsize=\textwidth \epsfbox{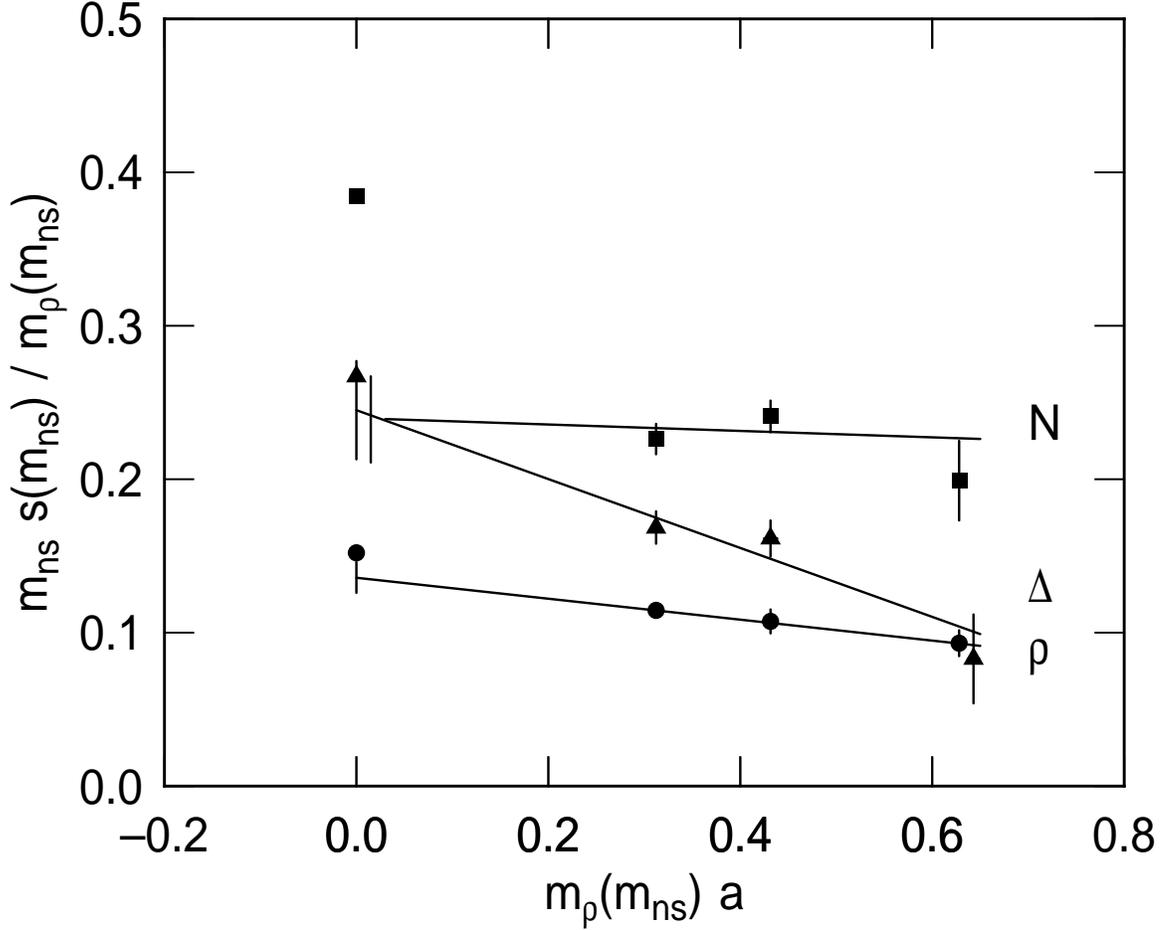}
\caption{ 
For sinks 0, 1 and 2, hadron masses slopes $s$ defined as 
$\partial m(m_q) / \partial m_q$ evaluated at quark
mass $m_{ns}$ as a function of lattice spacing in units of
$1/m_{\rho}(m_{ns})$.  The straight lines are extraploations to zero lattice
spacing, the error bars at zero lattice spacing are uncertainties in the
extrapolated ratios, and the points at zero lattice spacing are
observed values.}
\label{fig:s_taylor}
\end{figure}

Table~\ref{tab:nexresults} shows the continuum limits of these
predictions taken with physical volume held fixed, the continuum limits
corrected to infinite volume, and the corresponding observed values.
The continuum limit predictions are found as in
Section~\ref{sec:contlim} by linear extrapolation to zero lattice
spacing.  We now use $m_{\rho}(m_{ns}) a$ as the measure of lattice
spacing, rather than $m_{\rho}(m_n) a$.  The $\chi^2$ per degree of
freedom for each linear fit used to find a continuum limit with physical
volume held fixed is shown in the last column of
Table~\ref{tab:nexresults}. Figures~\ref{fig:m_taylor} and
\ref{fig:s_taylor} show the extrapolations to zero lattice spacing with
physical volume held fixed.  The infinite volume predictions shown in
Table~\ref{tab:nexresults} are found following Section~\ref{sec:vollim}.
Again the results for simultaneous fits to sinks 0, 1 and 2 are
statistically consistent with the results for sink 4, but the
statistical uncertainties for the fits to sink 4 are significantly
larger than those for fits to 0, 1 and 2.  We consider the 0, 1, 2
predictions to be more reliable.  

Observed numbers in Table~\ref{tab:nexresults} are determined as
discussed in Section~\ref{sec:test}.  The infinite volume continuum
limits of the predictions for sinks 0, 1 and 2 are statistically
consistent with the corresponding observed values. Four of the five
predicted values are within 1.6 standard deviations of experiment and
the fifth differs from experiment by less than 2.0 standard deviations.
The mass predictions are within about 6.5\% of experiment with
statistical uncertainties of up to 3.3\% of experiment.  The slope
predictions are within 22\% of experiment with statistical uncertainties
of up to 22\% of experiment.  The slope predictions are obtained, in
effect, from comparatively small differences between predicted masses
for hadrons composed of quarks with different masses and therefore have
larger relative statistical errors than the mass predictions.  As in
Table~\ref{tab:results}, in place of an observed value for
$\Lambda^{(0)}_{\overline{MS}}$ in Table~\ref{tab:nexresults}, we use an
infinite volume continuum limit result obtained by a lattice QCD
calculation combined with the observed value of a charmonium mass
splitting \cite{Fermilab}.

\section{ACKNOWLEDGEMENT}

We would like to thank Claude Bernard, Martin Golterman, Jim Labrenz and
Steve Sharpe for conversations.  We are grateful to Mike Cassera and
Dave George for their work in putting GF11 into operation, to Chi Chai
Huang, of Compunetics Inc., for his contributions to bringing GF11 up to
full power and to its continued maintenance, and to Molly Elliott and Ed
Nowicki for their work on GF11's disk software.

\newpage

\begin{table}
\begin{center}
\begin{tabular}{lllrrr}     \hline
\multicolumn{1}{c}{lattice} & \multicolumn{1}{c}{$\beta$} & 
\multicolumn{1}{c}{k}     & \multicolumn{1}{c}{$m_q$} & skip & count \\ \hline
 $8^3 \times 32$ & 5.70 & 0.1400 & $839 \pm 12$ & 1000 & 2439 \\ 
                  &     & 0.1450 & $672 \pm 9$  & 1000 & 2439 \\ 
                  &     & 0.1500 & $517 \pm 7$  & 1000 & 2439 \\ 
                  &     & 0.1550 & $371 \pm 5$  & 1000 & 2439 \\ 
                  &     & 0.1600 & $235 \pm 3$  & 1000 & 2439 \\ 
                  &     & 0.1650 & $107 \pm 1$  & 1000 & 2439 \\ \hline
 $16^3 \times 32$ & 5.70 & 0.1400& $839 \pm 12$ & 2000 & 47   \\ 
                  &     & 0.1450 & $672 \pm 9$  & 2000 & 47   \\ 
                  &     & 0.1500 & $517 \pm 7$  & 2000 & 47   \\ 
                  &     & 0.1550 & $371 \pm 5$  & 2000 & 47   \\ 
                  &     & 0.1600 & $235 \pm 3$  & 2000 & 219  \\
                  &     & 0.1650 & $107 \pm 1$  & 2000 & 219  \\ 
                  &     & 0.16625& $ 76 \pm 1$  & 2000 & 219  \\ 
                  &     & 0.1675 & $ 45 \pm 1$  & 2000 & 219  \\ \hline
 $24^3 \times 32$ & 5.70 & 0.1600 & $244 \pm 4$ & 4000 & 92    \\ 
                  &     & 0.1650  & $109 \pm 2$ & 4000 & 92   \\ 
                  &     & 0.1663  & $ 76 \pm 1$ & 4000 & 58    \\ 
                  &     & 0.1675  & $ 45 \pm 1$ & 4000 & 92    \\ \hline
 $24^3 \times 36$ & 5.93 & 0.1543 & $189 \pm 4$ & 4000 & 210   \\ 
                  &     & 0.1560  & $119 \pm 2$ & 4000 & 210  \\ 
                  &     & 0.1573  & $ 66 \pm 1$ & 4000 & 210   \\ 
                  &     & 0.1581  & $ 34 \pm 1$ & 4000 & 210   \\ \hline
 $32^2 \times 30 \times 40$ & 6.17& 0.1500 & $226 \pm 3$ & 6000 & 219    \\ 
                  &     & 0.1519  & $111 \pm 2$ & 6000 & 219   \\ 
                  &     & 0.1526  & $ 69 \pm 1$ & 6000 & 219    \\ 
                  &     & 0.1532  & $ 33 \pm 1$ & 6000 & 219    \\ \hline
\end{tabular}
\caption{Number of configurations analyzed for various lattices sizes
and parameters. Quark masses $m_q$ are given in $MeV/c^2$.
\label{tab:lattices}}
\end{center}
\end{table}

\newpage

\begin{table}
\begin{center}
\begin{tabular}{lllllr}     \hline
\multicolumn{1}{c}{lattice} & \multicolumn{1}{c}{$\beta$} & 
algorithm & \multicolumn{1}{c}{k}     & sweeps \\ \hline
 $16^3 \times 32$ & 5.70 & conj. grad. & 0.1600  &  500 \\
                  &     &     & 0.1650  & 1000 \\ 
                  &     &     & 0.16625 & 1500 \\ 
                  &     &     & 0.1675  & 4000 \\ \hline
 $24^3 \times 32$ & 5.70 &  precond. &  0.1600  & 150  \\ 
                  &      &  conj. grad. &  0.1650   & 340 \\ 
                  &      &     &  0.1663   & 511 \\ 
                  &      &     &  0.1675   & 998 \\ \hline
 $24^3 \times 36$ & 5.93 &  precond. & 0.1543  & 195 \\ 
                  &      &  conj. grad. & 0.1560  & 240 \\ 
                  &      &      & 0.1573  & 380 \\ 
                  &      &      & 0.1581  & 724 \\ \hline
 $32^2 \times 30 \times 40$ & 6.17 & precond. & 0.1500  &  184  \\
                  &         & min. resid. & 0.1519  &  350  \\
                  &         &            & 0.1526  &  709  \\
                  &         &            & 0.1532  & 1389 \\ \hline
\end{tabular}
\caption{Algorithms and average number of sweeps used to find quark propagators.
\label{tab:sweeps}}
\end{center}
\end{table}

\newpage

\begin{table}
\begin{center}
\begin{tabular}{cccccc}     \hline
particle  & k       & sink & $m a$     & t & $\chi^2$ \\ 
\hline
$\pi$     & 0.1400  & 0     &  $1.4106 \pm 0.0008$  & 12 - 15 & 0.1 \\
          & 0.1450  & 0     &  $1.2408 \pm 0.0008$  & 11 - 15 & 0.4 \\
          & 0.1500  & 0     &  $1.0666 \pm 0.0009$  & 9  - 12 & 0.5 \\
          & 0.1550  & 0     &  $0.8845 \pm 0.0012$  & 9  - 12 & 0.2 \\
          & 0.1600  & 0     &  $0.6883 \pm 0.0015$  & 9  - 12 & 0.2 \\
          & 0.1650  & 0     &  $0.4523 \pm 0.0038$  & 10 - 14 & 0.1 \\
\hline
$\rho$    & 0.1400  & 0     &  $1.4361 \pm 0.0010$  & 12 - 15 & 0.1 \\
          & 0.1450  & 0     &  $1.2766 \pm 0.0009$  & 12 - 15 & 0.0 \\
          & 0.1500  & 0     &  $1.1177 \pm 0.0011$  & 12 - 15 & 0.1 \\
          & 0.1550  & 0     &  $0.9602 \pm 0.0019$  & 12 - 15 & 0.3 \\
          & 0.1600  & 0     &  $0.8023 \pm 0.0024$  & 8  - 12 & 1.9 \\
          & 0.1650  & 0     &  $0.6558 \pm 0.0071$  & 8  - 11 & 0.5 \\
\hline
$N$       & 0.1400  & 0     &  $2.2990 \pm 0.0035$  & 11 - 14 & 0.1 \\
          & 0.1450  & 0     &  $2.0553 \pm 0.0039$  & 11 - 14 & 0.5 \\
          & 0.1500  & 0     &  $1.8110 \pm 0.0050$  & 11 - 14 & 1.2 \\
          & 0.1550  & 0     &  $1.5697 \pm 0.0070$  & 12 - 15 & 1.1 \\
          & 0.1600  & 0     &  $1.3307 \pm 0.0105$  & 9  - 12 & 0.5 \\
          & 0.1650  & 0     &  $1.0874 \pm 0.0230$  & 7  - 10 & 0.2 \\
\hline
$\Delta$  & 0.1400  & 0     &  $2.3140 \pm 0.0037$  & 11 - 14 & 0.1 \\  
          & 0.1450  & 0     &  $2.0764 \pm 0.0046$  & 11 - 14 & 0.2 \\  
          & 0.1500  & 0     &  $1.8416 \pm 0.0053$  & 11 - 14 & 0.6 \\  
          & 0.1550  & 0     &  $1.6146 \pm 0.0076$  & 12 - 15 & 0.5 \\  
          & 0.1600  & 0     &  $1.3959 \pm 0.0159$  & 10 - 13 & 0.3 \\  
          & 0.1650  & 0     &  $1.2476 \pm 0.0337$  & 7  - 10 & 2.4 \\  
\hline
\end{tabular}
\caption{Masses from an $8^3 \times 32$ lattice at $\beta$ of 5.70 using 2439 configurations.
\label{tab:allmx8}}
\end{center}
\end{table}

\begin{table}
\begin{center}
\begin{tabular}{cccccc}     \hline
particle  & k       & sink & $m a$     & t & $\chi^2$ \\ 
\hline
$\pi$     & 0.1400  & 0     &  $1.4109 \pm 0.0057$ & 10-13 & 0.0 \\
          & 0.1450  & 0     &  $1.2423 \pm 0.0052$ & 10-13 & 0.0 \\
          & 0.1500  & 0     &  $1.0691 \pm 0.0047$ & 10-13 & 0.0 \\
          & 0.1550  & 0     &  $0.8891 \pm 0.0040$ & 9 -12 & 0.1 \\
\hline
$\rho$    & 0.1400  & 0     &  $1.4351 \pm 0.0072$ & 10-13 & 0.0 \\ 
          & 0.1450  & 0     &  $1.2761 \pm 0.0068$ & 10-13 & 0.0 \\ 
          & 0.1500  & 0     &  $1.1183 \pm 0.0064$ & 9 -13 & 0.1 \\ 
          & 0.1550  & 0     &  $0.9562 \pm 0.0066$ & 12-15 & 0.2 \\ 
\hline
$N$       & 0.1400  & 0     &  $2.2763 \pm 0.0235$ & 8 -11 & 1.0 \\ 
          & 0.1450  & 0     &  $2.0307 \pm 0.0207$ & 8 -11 & 0.7 \\ 
          & 0.1500  & 0     &  $1.7844 \pm 0.0205$ & 8 -11 & 0.4 \\ 
          & 0.1550  & 0     &  $1.5358 \pm 0.0227$ & 8 -11 & 0.1 \\ 
\hline
$\Delta$  & 0.1400  & 0     &  $2.2765 \pm 0.0252$ & 11-14 & 1.1 \\ 
          & 0.1450  & 0     &  $2.0473 \pm 0.0244$ & 8 -11 & 0.8 \\ 
          & 0.1500  & 0     &  $1.8088 \pm 0.0227$ & 8 -11 & 0.6 \\ 
          & 0.1550  & 0     &  $1.5736 \pm 0.0256$ & 8 -11 & 0.4 \\ 
\hline
\end{tabular}
\caption{
Masses from a $16^3 \times 32$ lattice at $\beta$ of 5.70 using 47 configurations.
\label{tab:himx16}}
\end{center}
\end{table}

\begin{table}
\begin{center}
\begin{tabular}{ccccc}     \hline
k       & sink & $m_{\pi} a$     & t & $\chi^2$ \\ 
\hline
0.1600  & 0     &  $0.6889 \pm 0.0013$  & 11 - 14 & 0.2 \\  
        & 1     &  $0.6891 \pm 0.0015$  & 10 - 14 & 0.2 \\  
        & 2     &  $0.6904 \pm 0.0021$  & 10 - 13 & 0.4 \\  
        & 3     &  $0.6917 \pm 0.0028$  & 10 - 13 & 0.3 \\  
        & 4     &  $0.6929 \pm 0.0032$  & 10 - 13 & 0.2 \\  
\hline 
0.1650  & 0     &  $0.4584 \pm 0.0018$  & 11 - 14 & 1.0 \\  
        & 1     &  $0.4590 \pm 0.0019$  & 10 - 13 & 0.4 \\  
        & 2     &  $0.4601 \pm 0.0021$  & 10 - 13 & 0.3 \\  
        & 3     &  $0.4582 \pm 0.0030$  & 11 - 14 & 0.5 \\  
        & 4     &  $0.4579 \pm 0.0034$  & 11 - 14 & 0.2 \\  
\hline 
0.16625  & 0     &  $0.3866 \pm 0.0022$  & 11 - 14 & 0.5 \\  
         & 1     &  $0.3869 \pm 0.0021$  & 10 - 13 & 0.4 \\  
         & 2     &  $0.3880 \pm 0.0025$  & 10 - 13 & 0.2 \\  
         & 3     &  $0.3865 \pm 0.0029$  & 11 - 14 & 0.3 \\  
         & 4     &  $0.3865 \pm 0.0033$  & 11 - 14 & 0.1 \\  
\hline 
0.1675  & 0     &  $0.2977 \pm 0.0043$  & 11 - 14 & 0.1 \\  
        & 1     &  $0.2979 \pm 0.0040$  & 11 - 14 & 0.1 \\  
        & 2     &  $0.2979 \pm 0.0040$  & 10 - 14 & 0.1 \\  
        & 3     &  $0.2981 \pm 0.0046$  & 11 - 14 & 0.2 \\  
        & 4     &  $0.2991 \pm 0.0051$  & 11 - 14 & 0.1 \\  
\hline 
\end{tabular}
\caption{Values of $m_{\pi} a$ from a $16^3 \times 32$ lattice at $\beta$
of 5.70 using 219 configurations.
\label{tab:pimx16}}
\end{center}
\end{table}

\begin{table}
\begin{center}
\begin{tabular}{ccccc}     \hline
k       & sink & $m_{\rho} a$     & 
t & $\chi^2$ \\ \hline

0.1600  &  0     &  $0.8044 \pm 0.0029$  & 9 - 13 & 0.5 \\  
        &  1     &  $0.8061 \pm 0.0030$  & 10 - 13 & 0.0 \\  
        &  2     &  $0.8090 \pm 0.0035$  & 10 - 13 & 0.0 \\  
        &  3     &  $0.8113 \pm 0.0058$  & 11 - 14 & 0.0 \\  
        &  4     &  $0.8133 \pm 0.0070$  & 11 - 14 & 0.1 \\  \hline 
0.1650  &  012   &  $0.6642 \pm 0.0052$  &  7 - 10 & 0.9 \\  
        &  0     &  $0.6621 \pm 0.0062$  &  7 - 11 & 0.8 \\  
        &  1     &  $0.6660 \pm 0.0050$  &  6 - 9  & 0.5 \\  
        &  2     &  $0.6627 \pm 0.0054$  &  6 - 9  & 0.0 \\  
        &  3     &  $0.6598 \pm 0.0074$  & 6 - 9  & 0.2 \\  
        &  4     &  $0.6705 \pm 0.0070$  & 2 - 6  & 0.7 \\  \hline 
0.16625  & 012   &  $0.6321 \pm 0.0062$  & 7 - 10 & 0.4 \\  
         & 0     &  $0.6306 \pm 0.0066$  & 7 - 10 & 0.3 \\  
         & 1     &  $0.6302 \pm 0.0064$  & 7 - 10 & 0.2 \\  
         & 2     &  $0.6311 \pm 0.0063$  & 6 - 9  & 0.0 \\  
         & 3     &  $0.6281 \pm 0.0080$  & 6 - 9  & 0.1 \\  
         & 4     &  $0.6262 \pm 0.0087$  & 4 - 8  & 0.4 \\  \hline 
0.1675  & 012   &  $0.6068 \pm 0.0051$  & 5 - 8  & 0.4 \\  
        & 0     &  $0.6050 \pm 0.0062$  & 5 - 8  & 0.1 \\  
        & 1     &  $0.6053 \pm 0.0061$  & 4 - 7  & 0.2 \\  
        & 2     &  $0.6017 \pm 0.0075$  & 5 - 8  & 0.3 \\  
        & 3     &  $0.5940 \pm 0.0085$  & 4 - 7  & 0.8 \\  
        & 4     &  $0.6006 \pm 0.0077$  & 2 - 6  & 0.7 \\  \hline 
\end{tabular}
\caption{Values of $m_{\rho} a$ from a $16^3 \times 32$ lattice at $\beta$
of 5.70 using 219 configurations.
\label{tab:rhmx16}}
\end{center}
\end{table}

\begin{table}
\begin{center}
\begin{tabular}{ccccc}     \hline
k       & sink & $m_N a$     & 
t & $\chi^2$ \\ \hline

0.1600  & 0     &  $1.3043 \pm 0.0080$  & 8  - 12 & 0.6 \\  
        & 1     &  $1.3015 \pm 0.0077$  & 7  - 10 & 0.2 \\  
        & 2     &  $1.3004 \pm 0.0090$  & 6  - 10 & 0.1 \\  
        & 3     &  $1.2905 \pm 0.0162$  & 6  - 9  & 0.4 \\  
        & 4     &  $1.2985 \pm 0.0141$  & 2  - 5  & 0.8 \\  \hline 
0.1650  & 012   &  $1.0312 \pm 0.0125$  & 8  - 11 & 0.8 \\  
        & 0     &  $1.0276 \pm 0.0143$  & 8  - 11 & 0.9 \\  
        & 1     &  $1.0293 \pm 0.0182$  & 7  - 10 & 0.7 \\  
        & 2     &  $1.0293 \pm 0.0176$  & 6  - 10 & 0.7 \\  
        & 3     &  $1.0419 \pm 0.0198$  & 3  - 8  & 0.7 \\  
        & 4     &  $1.0322 \pm 0.0154$  & 1  - 4  & 0.1 \\  \hline 
0.16625  & 012   &  $0.9838 \pm 0.0103$  & 5  - 8  & 1.2 \\  
         & 0     &  $0.9654 \pm 0.0138$  & 7  - 10 & 1.4 \\  
         & 1     &  $0.9867 \pm 0.0188$  & 4  - 7  & 0.3 \\  
         & 2     &  $0.9857 \pm 0.0148$  & 3  - 6  & 0.4 \\  
         & 3     &  $0.9708 \pm 0.0124$  & 3  - 8  & 0.4 \\  
         & 4     &  $0.9576 \pm 0.0178$  & 1  - 4  & 0.1 \\  \hline 
0.1675  & 012   &  $0.9052 \pm 0.0123$  & 5  - 8  & 0.5 \\  
        & 0     &  $0.9100 \pm 0.0115$  & 5  - 8  & 0.1 \\  
        & 1     &  $0.9120 \pm 0.0110$  & 5  - 8  & 0.0 \\  
        & 2     &  $0.9065 \pm 0.0124$  & 3  - 7  & 0.3 \\  
        & 3     &  $0.8906 \pm 0.0128$  & 2  - 5  & 0.3 \\  
        & 4     &  $0.8610 \pm 0.0154$  & 2  - 5  & 0.1 \\  \hline

\end{tabular}
\caption{Values of $m_N a$ from a $16^3 \times 32$ lattice at $\beta$
of 5.70 using 219 configurations.
\label{tab:numx16}}
\end{center}
\end{table}

\begin{table}
\begin{center}
\begin{tabular}{ccccc}     \hline
k       & sink & $m_{\Delta} a$     & 
t & $\chi^2$ \\ \hline

0.1600  & 0     &  $1.3804 \pm 0.0088$  & 7  - 12 & 1.4 \\  
        & 1     &  $1.3713 \pm 0.0088$  & 7  - 10 & 0.4 \\  
        & 2     &  $1.3687 \pm 0.0173$  & 6  - 10 & 0.3 \\  
        & 3     &  $1.3980 \pm 0.0130$  & 2  - 6  & 1.0 \\  
        & 4     &  $1.3703 \pm 0.0149$  & 2  - 5  & 0.5 \\  \hline 
0.1650  & 012   &  $1.1704 \pm 0.0169$  & 7  - 11 & 1.6 \\  
        & 0     &  $1.1597 \pm 0.0132$  & 7  - 10 & 1.1 \\  
        & 1     &  $1.1519 \pm 0.0268$  & 7  - 10 & 0.5 \\  
        & 2     &  $1.1845 \pm 0.0261$  & 4  - 7  & 1.0 \\  
        & 3     &  $1.1876 \pm 0.0149$  & 2  - 6  & 0.6 \\  
        & 4     &  $1.1511 \pm 0.0211$  & 1  - 5  & 0.3 \\  \hline 
0.16625  & 012   &  $1.1214 \pm 0.0180$  & 7  - 10 & 1.4 \\  
         & 0     &  $1.1119 \pm 0.0144$  & 7  - 10 & 0.4 \\  
         & 1     &  $1.1054 \pm 0.0273$  & 7  - 10 & 0.2 \\  
         & 2     &  $1.1520 \pm 0.0171$  & 3  - 6  & 0.5 \\  
         & 3     &  $1.1481 \pm 0.0183$  & 1  - 4  & 0.3 \\  
         & 4     &  $1.1012 \pm 0.0215$  & 1  - 5  & 0.3 \\  \hline 
0.1675  &  012   &  $1.1349 \pm 0.0159$  & 5  - 8  & 2.1 \\  
        &  0     &  $1.1208 \pm 0.0141$  & 5  - 8  & 1.0 \\  
        &  1     &  $1.1164 \pm 0.0134$  & 4  - 7  & 0.2 \\  
        &  2     &  $1.1067 \pm 0.0141$  & 4  - 7  & 0.1 \\  
        &  3     &  $1.0874 \pm 0.0170$  & 2  - 5  & 0.1 \\  
        &  4     &  $1.0526 \pm 0.0167$  & 1  - 5  & 0.3 \\  \hline 

\end{tabular}
\caption{Values of $m_{\Delta} a$ from a $16^3 \times 32$ lattice at $\beta$
of 5.70 using 219 configurations.
\label{tab:demx16}}
\end{center}
\end{table}

\clearpage

\begin{table}
\begin{center}
\begin{tabular}{ccccc}     \hline
k       & sink & $m_{\pi} a$     & t & $\chi^2$ \\ \hline
0.1600  & 0     &  $0.6887 \pm 0.0011$  & 11 - 14 & 0.0 \\  
        & 1     &  $0.6884 \pm 0.0012$  & 11 - 14 & 0.1 \\  
        & 2     &  $0.6894 \pm 0.0016$  & 9  - 12 & 0.2 \\  
        & 3     &  $0.6900 \pm 0.0023$  & 9  - 12 & 0.2 \\  
        & 4     &  $0.6905 \pm 0.0031$  & 9  - 12 & 0.1 \\  \hline 
0.1650  & 0     &  $0.4576 \pm 0.0013$  & 10 - 14 & 0.2 \\  
        & 1     &  $0.4574 \pm 0.0012$  & 10 - 13 & 0.0 \\  
        & 2     &  $0.4577 \pm 0.0014$  & 9  - 12 & 0.0 \\  
        & 3     &  $0.4582 \pm 0.0019$  & 9  - 12 & 0.1 \\  
        & 4     &  $0.4589 \pm 0.0022$  & 9  - 12 & 0.1 \\  \hline 
0.1663  & 0     &  $0.3799 \pm 0.0016$  & 8  - 11 & 0.0 \\  
        & 1     &  $0.3800 \pm 0.0016$  & 8  - 11 & 0.1 \\  
        & 2     &  $0.3822 \pm 0.0019$  & 11 - 14 & 0.1 \\  
        & 3     &  $0.3825 \pm 0.0022$  & 11 - 14 & 0.2 \\  
        & 4     &  $0.3829 \pm 0.0026$  & 9  - 14 & 0.3 \\  \hline 
0.1675  & 0     &  $0.2943 \pm 0.0021$  & 9  - 12 & 0.1 \\  
        & 1     &  $0.2946 \pm 0.0023$  & 9  - 12 & 0.0 \\  
        & 2     &  $0.2949 \pm 0.0023$  & 9  - 12 & 0.1 \\  
        & 3     &  $0.2953 \pm 0.0019$  & 10 - 13 & 0.1 \\  
        & 4     &  $0.2955 \pm 0.0024$  & 10 - 13 & 0.1 \\  \hline 
\end{tabular}
\caption{Values of $m_{\pi} a$ from a $24^3 \times 32$ lattice at $\beta$
of 5.70 using 92 configurations.
\label{tab:pimx24}}
\end{center}
\end{table}

\begin{table}
\begin{center}
\begin{tabular}{ccccc}     \hline
k       & sink & $m_{\rho} a$     & t & $\chi^2$ \\ \hline
0.1600  & 0     &  $0.8062 \pm 0.0018$  & 9  - 14 & 0.3 \\  
        & 1     &  $0.8046 \pm 0.0023$  & 11 - 14 & 0.2 \\  
        & 2     &  $0.8058 \pm 0.0028$  & 10 - 13 & 0.7 \\  
        & 3     &  $0.8076 \pm 0.0047$  & 9  - 12 & 0.8 \\  
        & 4     &  $0.8022 \pm 0.0056$  & 6  - 10 & 0.7 \\  \hline 
0.1650  & 012   &  $0.6601 \pm 0.0031$  & 6  - 11 & 1.3 \\  
        & 0     &  $0.6585 \pm 0.0036$  & 9  - 12 & 0.3 \\  
        & 1     &  $0.6596 \pm 0.0045$  & 9  - 12 & 0.5 \\  
        & 2     &  $0.6514 \pm 0.0050$  & 7  - 10 & 1.5 \\  
        & 3     &  $0.6511 \pm 0.0071$  & 6  - 9  & 0.8 \\  
        & 4     &  $0.6491 \pm 0.0073$  & 6  - 9  & 0.2 \\  \hline 
0.1663  & 012   &  $0.6225 \pm 0.0042$  & 6  - 9  & 0.6 \\  
        & 0     &  $0.6246 \pm 0.0051$  & 6  - 9  & 0.8 \\  
        & 1     &  $0.6213 \pm 0.0064$  & 7  - 10 & 0.2 \\  
        & 2     &  $0.6191 \pm 0.0084$  & 7  - 10 & 0.0 \\  
        & 3     &  $0.6191 \pm 0.0086$  & 7  - 10 & 0.0 \\  
        & 4     &  $0.6206 \pm 0.0103$  & 7  - 10 & 0.0 \\  \hline 
0.1675  & 012   &  $0.5867 \pm 0.0062$  & 6  - 9  & 0.3 \\  
        & 0     &  $0.5875 \pm 0.0068$  & 6  - 9  & 0.7 \\  
        & 1     &  $0.5863 \pm 0.0089$  & 6  - 9  & 0.4 \\  
        & 2     &  $0.5842 \pm 0.0117$  & 6  - 9  & 0.2 \\  
        & 3     &  $0.5822 \pm 0.0110$  & 6  - 9  & 0.3 \\  
        & 4     &  $0.5912 \pm 0.0125$  & 4  - 8  & 0.2 \\  \hline 
\end{tabular}
\caption{Values of $m_{\rho} a$ from a $24^3 \times 32$ lattice at $\beta$
of 5.70 using 92 configurations.
\label{tab:rhmx24}}
\end{center}
\end{table}

\begin{table}
\begin{center}
\begin{tabular}{ccccc}     \hline
k       & sink & $m_N a$     & t & $\chi^2$ \\ \hline
0.1600  & 0     &  $1.3068 \pm 0.0051$  & 8  - 11 & 0.5 \\  
        & 1     &  $1.3077 \pm 0.0063$  & 9  - 12 & 0.6 \\  
        & 2     &  $1.3062 \pm 0.0065$  & 6  - 9  & 0.2 \\  
        & 3     &  $1.3088 \pm 0.0091$  & 7  - 10 & 0.3 \\  
        & 4     &  $1.3124 \pm 0.0135$  & 7  - 10 & 0.4 \\  \hline 
0.1650  & 012   &  $1.0337 \pm 0.0067$  & 7  - 11 & 0.5 \\  
        & 0     &  $1.0358 \pm 0.0083$  & 6  - 9  & 0.1 \\  
        & 1     &  $1.0313 \pm 0.0098$  & 8  - 11 & 0.2 \\  
        & 2     &  $1.0295 \pm 0.0102$  & 8  - 11 & 0.1 \\  
        & 3     &  $1.0323 \pm 0.0098$  & 4  - 7  & 0.1 \\  
        & 4     &  $1.0301 \pm 0.0104$  & 4  - 7  & 0.2 \\  \hline 
0.1663  & 012   &  $0.9402 \pm 0.0117$  & 7  - 10 & 0.6 \\  
        & 0     &  $0.9488 \pm 0.0109$  & 6  - 9  & 0.4 \\  
        & 1     &  $0.9440 \pm 0.0114$  & 6  - 9  & 0.0 \\  
        & 2     &  $0.9459 \pm 0.0118$  & 5  - 8  & 0.1 \\  
        & 3     &  $0.9472 \pm 0.0135$  & 4  - 8  & 0.1 \\  
        & 4     &  $0.9421 \pm 0.0131$  & 3  - 8  & 0.1 \\  \hline 
0.1675  & 012   &  $0.8634 \pm 0.0135$  & 6  - 9  & 1.3 \\  
        & 0     &  $0.8571 \pm 0.0118$  & 6  - 9  & 0.0 \\  
        & 1     &  $0.8527 \pm 0.0137$  & 6  - 9  & 0.2 \\  
        & 2     &  $0.8579 \pm 0.0133$  & 5  - 8  & 0.0 \\  
        & 3     &  $0.8587 \pm 0.0158$  & 5  - 8  & 0.1 \\  
        & 4     &  $0.8668 \pm 0.0177$  & 4  - 8  & 0.1 \\  \hline 
\end{tabular}
\caption{Values of $m_N a$ from a $24^3 \times 32$ lattice at $\beta$
of 5.70 using 92 configurations.
\label{tab:numx24}}
\end{center}
\end{table}

\begin{table}
\begin{center}
\begin{tabular}{ccccc}     \hline
k       & sink & $m_{\Delta} a$     & t & $\chi^2$ \\ \hline
0.1600  & 0     &  $1.3799 \pm 0.0062$  & 8  - 11 & 0.2 \\  
        & 1     &  $1.3815 \pm 0.0061$  & 9  - 12 & 0.2 \\  
        & 2     &  $1.3826 \pm 0.0066$  & 5  - 8  & 0.0 \\  
        & 3     &  $1.3827 \pm 0.0104$  & 5  - 8  & 0.1 \\  
        & 4     &  $1.3823 \pm 0.0163$  & 7  - 10 & 0.3 \\  \hline 
0.1650  & 012   &  $1.1718 \pm 0.0055$  & 6  - 10 & 0.6 \\  
        & 0     &  $1.1614 \pm 0.0101$  & 7  - 10 & 0.6 \\  
        & 1     &  $1.1693 \pm 0.0128$  & 5  - 8  & 0.6 \\  
        & 2     &  $1.1673 \pm 0.0067$  & 4  - 7  & 0.1 \\  
        & 3     &  $1.1633 \pm 0.0107$  & 5  - 8  & 0.0 \\  
        & 4     &  $1.1524 \pm 0.0150$  & 4  - 7  & 0.1 \\  \hline 
0.1663  & 012   &  $1.1170 \pm 0.0138$  & 6  - 10 & 0.7 \\  
        & 0     &  $1.1120 \pm 0.0163$  & 7  - 10 & 0.4 \\  
        & 1     &  $1.1226 \pm 0.0119$  & 5  - 8  & 0.7 \\  
        & 2     &  $1.1170 \pm 0.0115$  & 4  - 8  & 0.8 \\  
        & 3     &  $1.1032 \pm 0.0164$  & 3  - 6  & 0.1 \\  
        & 4     &  $1.0842 \pm 0.0177$  & 3  - 6  & 0.0 \\  \hline 
0.1675  & 012   &  $1.0817 \pm 0.0213$  & 6  - 9  & 1.0 \\  
        & 0     &  $1.0917 \pm 0.0132$  & 5  - 9  & 1.1 \\  
        & 1     &  $1.0783 \pm 0.0178$  & 5  - 9  & 0.9 \\  
        & 2     &  $1.0770 \pm 0.0128$  & 4  - 7  & 0.5 \\  
        & 3     &  $1.0695 \pm 0.0198$  & 3  - 7  & 0.1 \\  
        & 4     &  $1.0523 \pm 0.0214$  & 4  - 7  & 0.0 \\  \hline 
\end{tabular}
\caption{Values of $m_{\Delta} a$ from a $24^3 \times 32$ lattice at $\beta$
of 5.70 using 92 configurations.
\label{tab:demx24}}
\end{center}
\end{table}

\begin{table}
\begin{center}
\begin{tabular}{ccccc}     \hline
k       & sink & $m_{\pi} a$     & t & $\chi^2$ \\ \hline
0.1543  & 0     &  $0.4570 \pm 0.0016$  & 12 - 16 & 0.8 \\  
        & 1     &  $0.4556 \pm 0.0020$  & 14 - 17 & 1.0 \\  
        & 2     &  $0.4580 \pm 0.0021$  & 11 - 14 & 0.4 \\  
        & 3     &  $0.4575 \pm 0.0022$  & 11 - 14 & 0.7 \\  
        & 4     &  $0.4572 \pm 0.0026$  & 11 - 14 & 1.0 \\  \hline 
0.1560  & 0     &  $0.3570 \pm 0.0015$  & 12 - 15 & 0.4 \\  
        & 1     &  $0.3583 \pm 0.0017$  & 11 - 14 & 0.9 \\  
        & 2     &  $0.3580 \pm 0.0016$  & 11 - 14 & 0.4 \\  
        & 3     &  $0.3576 \pm 0.0017$  & 11 - 14 & 0.3 \\  
        & 4     &  $0.3573 \pm 0.0019$  & 11 - 14 & 0.4 \\  \hline 
0.1573  & 0     &  $0.2653 \pm 0.0017$  & 12 - 15 & 0.0 \\  
        & 1     &  $0.2658 \pm 0.0019$  & 10 - 14 & 0.2 \\  
        & 2     &  $0.2653 \pm 0.0018$  & 10 - 13 & 0.1 \\  
        & 3     &  $0.2648 \pm 0.0020$  & 10 - 13 & 0.0 \\  
        & 4     &  $0.2641 \pm 0.0025$  & 10 - 13 & 0.0 \\  \hline 
0.1581  & 0     &  $0.1906 \pm 0.0023$  & 12 - 15 & 0.1 \\  
        & 1     &  $0.1902 \pm 0.0023$  & 11 - 14 & 0.0 \\  
        & 2     &  $0.1898 \pm 0.0024$  & 10 - 13 & 0.0 \\  
        & 3     &  $0.1892 \pm 0.0026$  & 10 - 13 & 0.0 \\  
        & 4     &  $0.1885 \pm 0.0031$  & 10 - 13 & 0.0 \\  \hline 
\end{tabular}
\caption{Values of $m_{\pi} a$ from a $24^3 \times 36$ lattice at $\beta$
of 5.93 using 210 configurations.
\label{tab:pimx24t36}}
\end{center}
\end{table}

\begin{table}
\begin{center}
\begin{tabular}{ccccc}     \hline
k       & sink & $m_{\rho} a$     & t & $\chi^2$ \\ \hline
0.1543  & 0     &  $0.5505 \pm 0.0029$  & 12 - 16 & 0.8 \\  
        & 1     &  $0.5528 \pm 0.0024$  & 11 - 14 & 0.5 \\  
        & 2     &  $0.5519 \pm 0.0027$  & 11 - 14 & 0.4 \\  
        & 3     &  $0.5565 \pm 0.0041$  & 8  - 11 & 0.4 \\  
        & 4     &  $0.5527 \pm 0.0040$  & 7  - 10 & 0.0 \\  \hline 
0.1560  & 012   &  $0.4843 \pm 0.0033$  & 12 - 15 & 1.0 \\  
        & 0     &  $0.4829 \pm 0.0042$  & 12 - 15 & 1.3 \\  
        & 1     &  $0.4889 \pm 0.0033$  & 9  - 12 & 0.9 \\  
        & 2     &  $0.4846 \pm 0.0036$  & 11 - 14 & 0.3 \\  
        & 3     &  $0.4891 \pm 0.0045$  & 8  - 11 & 0.1 \\  
        & 4     &  $0.4864 \pm 0.0042$  & 7  - 10 & 0.0 \\  \hline 
0.1573  & 012   &  $0.4399 \pm 0.0047$  & 9  - 12 & 0.3 \\  
        & 0     &  $0.4397 \pm 0.0044$  & 9  - 12 & 0.6 \\  
        & 1     &  $0.4393 \pm 0.0047$  & 9  - 12 & 0.5 \\  
        & 2     &  $0.4404 \pm 0.0048$  & 8  - 11 & 0.3 \\  
        & 3     &  $0.4398 \pm 0.0045$  & 7  - 10 & 0.0 \\  
        & 4     &  $0.4369 \pm 0.0048$  & 6  - 9  & 0.0 \\  \hline 
0.1581  & 012   &  $0.4091 \pm 0.0063$  & 9  - 12 & 0.5 \\  
        & 0     &  $0.4110 \pm 0.0058$  & 9  - 12 & 0.7 \\  
        & 1     &  $0.4154 \pm 0.0057$  & 8  - 11 & 0.6 \\  
        & 2     &  $0.4146 \pm 0.0045$  & 7  - 10 & 0.0 \\  
        & 3     &  $0.4121 \pm 0.0042$  & 7  - 10 & 0.2 \\  
        & 4     &  $0.4071 \pm 0.0057$  & 6  - 9  & 0.1 \\  \hline 
\end{tabular}
\caption{Values of $m_{\rho} a$ from a $24^3 \times 36$ lattice at $\beta$
of 5.93 using 210 configurations.
\label{tab:rhmx24t36}}
\end{center}
\end{table}

\clearpage

\begin{table}
\begin{center}
\begin{tabular}{ccccc}     \hline
k       & sink & $m_N a$     & t & $\chi^2$ \\ \hline
0.1543  & 0     &  $0.8628 \pm 0.0056$  & 13 - 16 & 0.6 \\  
        & 1     &  $0.8662 \pm 0.0054$  & 11 - 14 & 0.0 \\  
        & 2     &  $0.8631 \pm 0.0048$  & 11 - 14 & 0.9 \\  
        & 3     &  $0.8792 \pm 0.0100$  & 8  - 11 & 0.6 \\  
        & 4     &  $0.8674 \pm 0.0102$  & 6  - 9  & 0.1 \\  \hline 
0.1560  & 012   &  $0.7438 \pm 0.0053$  & 12 - 15 & 1.0 \\  
        & 0     &  $0.7434 \pm 0.0064$  & 11 - 14 & 0.4 \\  
        & 1     &  $0.7427 \pm 0.0059$  & 11 - 14 & 0.0 \\  
        & 2     &  $0.7404 \pm 0.0062$  & 11 - 14 & 0.9 \\  
        & 3     &  $0.7514 \pm 0.0064$  & 7  - 10 & 0.1 \\  
        & 4     &  $0.7448 \pm 0.0099$  & 7  - 10 & 0.1 \\  \hline 
0.1573  & 012   &  $0.6398 \pm 0.0068$  & 11 - 14 & 0.9 \\  
        & 0     &  $0.6391 \pm 0.0087$  & 11 - 14 & 0.3 \\  
        & 1     &  $0.6389 \pm 0.0082$  & 11 - 14 & 0.2 \\  
        & 2     &  $0.6489 \pm 0.0079$  & 8  - 11 & 0.1 \\  
        & 3     &  $0.6469 \pm 0.0063$  & 7  - 10 & 0.2 \\  
        & 4     &  $0.6423 \pm 0.0080$  & 6  - 10 & 0.1 \\  \hline 
0.1581  & 012   &  $0.5759 \pm 0.0099$  & 9  - 12 & 1.8 \\  
        & 0     &  $0.5718 \pm 0.0074$  & 9  - 12 & 0.3 \\  
        & 1     &  $0.5707 \pm 0.0077$  & 9  - 12 & 0.1 \\  
        & 2     &  $0.5714 \pm 0.0068$  & 8  - 11 & 0.1 \\  
        & 3     &  $0.5701 \pm 0.0086$  & 7  - 10 & 0.5 \\  
        & 4     &  $0.5652 \pm 0.0092$  & 6  - 10 & 0.2 \\  \hline 
\end{tabular}
\caption{Values of $m_N a$ from a $24^3 \times 36$ lattice at $\beta$
of 5.93 using 210 configurations.
\label{tab:numx24t36}}
\end{center}
\end{table}

\begin{table}
\begin{center}
\begin{tabular}{ccccc}     \hline
k       & sink & $m_{\Delta} a$     & t & $\chi^2$ \\ \hline
0.1543  & 0     &  $0.9216 \pm 0.0071$  & 13 - 16 & 0.3 \\  
        & 1     &  $0.9280 \pm 0.0080$  & 11 - 14 & 0.4 \\  
        & 2     &  $0.9240 \pm 0.0076$  & 11 - 14 & 0.7 \\  
        & 3     &  $0.9406 \pm 0.0088$  & 8  - 11 & 0.4 \\  
        & 4     &  $0.9276 \pm 0.0117$  & 6  - 9  & 0.1 \\  \hline 
0.1560  & 012   &  $0.8314 \pm 0.0062$  & 12 - 15 & 1.1 \\  
        & 0     &  $0.8304 \pm 0.0085$  & 11 - 15 & 1.1 \\  
        & 1     &  $0.8264 \pm 0.0082$  & 11 - 15 & 0.5 \\  
        & 2     &  $0.8225 \pm 0.0082$  & 11 - 14 & 0.3 \\  
        & 3     &  $0.8349 \pm 0.0081$  & 7  - 10 & 0.0 \\  
        & 4     &  $0.8272 \pm 0.0115$  & 7  - 10 & 0.1 \\  \hline 
0.1573  & 012   &  $0.7564 \pm 0.0103$  & 11 - 14 & 1.5 \\  
        & 0     &  $0.7594 \pm 0.0098$  & 10 - 14 & 1.0 \\  
        & 1     &  $0.7566 \pm 0.0106$  & 10 - 13 & 0.5 \\  
        & 2     &  $0.7620 \pm 0.0100$  & 8  - 11 & 0.2 \\  
        & 3     &  $0.7581 \pm 0.0097$  & 7  - 10 & 0.0 \\  
        & 4     &  $0.7500 \pm 0.0108$  & 7  - 10 & 0.0 \\  \hline 
0.1581  & 012   &  $0.7234 \pm 0.0114$  & 9  - 12 & 1.2 \\  
        & 0     &  $0.7161 \pm 0.0094$  & 9  - 12 & 0.3 \\  
        & 1     &  $0.7126 \pm 0.0113$  & 9  - 12 & 0.4 \\  
        & 2     &  $0.7105 \pm 0.0113$  & 8  - 11 & 0.2 \\  
        & 3     &  $0.7083 \pm 0.0114$  & 7  - 10 & 0.2 \\  
        & 4     &  $0.6978 \pm 0.0141$  & 7  - 10 & 0.1 \\  \hline 
\end{tabular}
\caption{Values of $m_{\Delta} a$ from a $24^3 \times 36$ lattice at $\beta$
of 5.93 using 210 configurations.
\label{tab:demx24t36}}
\end{center}
\end{table}

\begin{table}
\begin{center}
\begin{tabular}{ccccc}     \hline
k       & sink & $m_{\pi} a$     & t & $\chi^2$ \\ \hline
0.1500  & 0     &  $0.3870 \pm 0.0009$  & 14 - 17 & 0.6 \\  
        & 1     &  $0.3870 \pm 0.0010$  & 14 - 17 & 1.0 \\  
        & 2     &  $0.3863 \pm 0.0009$  & 14 - 19 & 1.2 \\  
        & 3     &  $0.3866 \pm 0.0010$  & 13 - 18 & 0.9 \\  
        & 4     &  $0.3866 \pm 0.0012$  & 13 - 18 & 0.7 \\  \hline 
0.1519  & 0     &  $0.2633 \pm 0.0012$  & 14 - 17 & 0.3 \\  
        & 1     &  $0.2633 \pm 0.0012$  & 14 - 17 & 0.7 \\  
        & 2     &  $0.2630 \pm 0.0012$  & 14 - 18 & 0.6 \\  
        & 3     &  $0.2631 \pm 0.0011$  & 14 - 18 & 0.4 \\  
        & 4     &  $0.2631 \pm 0.0012$  & 14 - 18 & 0.3 \\  \hline 
0.1526  & 0     &  $0.2065 \pm 0.0017$  & 14 - 17 & 0.4 \\  
        & 1     &  $0.2066 \pm 0.0016$  & 14 - 17 & 0.3 \\  
        & 2     &  $0.2066 \pm 0.0014$  & 14 - 18 & 0.2 \\  
        & 3     &  $0.2067 \pm 0.0014$  & 14 - 18 & 0.1 \\  
        & 4     &  $0.2064 \pm 0.0015$  & 15 - 18 & 0.1 \\  \hline 
0.1532  & 0     &  $0.1445 \pm 0.0020$  & 12 - 15 & 0.2 \\  
        & 1     &  $0.1453 \pm 0.0020$  & 15 - 18 & 0.4 \\  
        & 2     &  $0.1451 \pm 0.0019$  & 13 - 17 & 0.1 \\  
        & 3     &  $0.1453 \pm 0.0020$  & 13 - 16 & 0.0 \\  
        & 4     &  $0.1455 \pm 0.0020$  & 13 - 16 & 0.0 \\  \hline 
\end{tabular}
\caption{Values of $m_{\pi} a$ from a $32 \times 30 \times 32 \times 40$ 
lattice at $\beta$ of 6.17 using 219 configurations.
\label{tab:pimx32}}
\end{center}
\end{table}

\begin{table}
\begin{center}
\begin{tabular}{ccccc}     \hline
k       & sink & $m_{\rho} a$     & t & $\chi^2$ \\ \hline
0.1500  & 0     &  $0.4462 \pm 0.0012$  & 14 - 17 & 0.1 \\  
        & 1     &  $0.4461 \pm 0.0011$  & 14 - 17 & 0.2 \\  
        & 2     &  $0.4458 \pm 0.0012$  & 12 - 16 & 0.6 \\  
        & 3     &  $0.4458 \pm 0.0014$  & 13 - 16 & 0.5 \\  
        & 4     &  $0.4458 \pm 0.0018$  & 13 - 16 & 0.4 \\  \hline 
0.1519  & 012   &  $0.3581 \pm 0.0018$  & 13 - 16 & 0.3 \\  
        & 0     &  $0.3577 \pm 0.0017$  & 14 - 17 & 0.1 \\  
        & 1     &  $0.3577 \pm 0.0018$  & 13 - 16 & 0.2 \\  
        & 2     &  $0.3573 \pm 0.0018$  & 12 - 16 & 0.2 \\  
        & 3     &  $0.3573 \pm 0.0022$  & 13 - 16 & 0.3 \\  
        & 4     &  $0.3572 \pm 0.0026$  & 13 - 16 & 0.3 \\  \hline 
0.1526  & 012   &  $0.3265 \pm 0.0022$  & 13 - 16 & 0.4 \\  
        & 0     &  $0.3254 \pm 0.0027$  & 13 - 16 & 0.1 \\  
        & 1     &  $0.3252 \pm 0.0026$  & 13 - 16 & 0.0 \\  
        & 2     &  $0.3247 \pm 0.0027$  & 12 - 16 & 0.1 \\  
        & 3     &  $0.3246 \pm 0.0032$  & 13 - 16 & 0.1 \\  
        & 4     &  $0.3245 \pm 0.0039$  & 13 - 16 & 0.1 \\  \hline 
0.1532  & 012   &  $0.3024 \pm 0.0048$  & 13 - 16 & 0.4 \\  
        & 0     &  $0.2991 \pm 0.0059$  & 13 - 16 & 0.2 \\  
        & 1     &  $0.2987 \pm 0.0059$  & 12 - 15 & 0.1 \\  
        & 2     &  $0.2969 \pm 0.0066$  & 13 - 16 & 0.1 \\  
        & 3     &  $0.2965 \pm 0.0076$  & 13 - 16 & 0.0 \\  
        & 4     &  $0.2965 \pm 0.0088$  & 13 - 16 & 0.0 \\  \hline 
\end{tabular}
\caption{Values of $m_{\rho} a$ from a $32 \times 30 \times 32 \times 40$ 
lattice at $\beta$ of 6.17 using 219 configurations.
\label{tab:rhmx32}}
\end{center}
\end{table}

\begin{table}
\begin{center}
\begin{tabular}{ccccc}     \hline
k       & sink & $m_N a$     & t & $\chi^2$ \\ \hline

0.1500  & 0     &  $0.7004 \pm 0.0028$  & 14 - 17 & 0.3 \\  
        & 1     &  $0.6995 \pm 0.0025$  & 14 - 17 & 0.7 \\  
        & 2     &  $0.6975 \pm 0.0029$  & 14 - 18 & 0.9 \\  
        & 3     &  $0.6970 \pm 0.0030$  & 14 - 18 & 0.7 \\  
        & 4     &  $0.6966 \pm 0.0040$  & 14 - 18 & 0.6 \\  \hline 
0.1519  & 012   &  $0.5427 \pm 0.0038$  & 15 - 18 & 0.9 \\  
        & 0     &  $0.5444 \pm 0.0037$  & 14 - 17 & 0.2 \\  
        & 1     &  $0.5437 \pm 0.0035$  & 14 - 17 & 0.8 \\  
        & 2     &  $0.5424 \pm 0.0036$  & 14 - 18 & 1.0 \\  
        & 3     &  $0.5428 \pm 0.0038$  & 14 - 18 & 0.6 \\  
        & 4     &  $0.5460 \pm 0.0052$  & 13 - 17 & 0.4 \\  \hline 
0.1526  & 012   &  $0.4845 \pm 0.0053$  & 13 - 16 & 1.5 \\  
        & 0     &  $0.4819 \pm 0.0050$  & 14 - 17 & 0.9 \\  
        & 1     &  $0.4826 \pm 0.0048$  & 12 - 16 & 0.6 \\  
        & 2     &  $0.4824 \pm 0.0050$  & 13 - 16 & 0.3 \\  
        & 3     &  $0.4831 \pm 0.0042$  & 13 - 16 & 0.2 \\  
        & 4     &  $0.4848 \pm 0.0068$  & 13 - 17 & 0.1 \\  \hline 
0.1532  & 012   &  $0.4234 \pm 0.0065$  & 13 - 16 & 1.1 \\  
        & 0     &  $0.4187 \pm 0.0059$  & 12 - 15 & 0.5 \\  
        & 1     &  $0.4188 \pm 0.0087$  & 12 - 17 & 0.7 \\  
        & 2     &  $0.4199 \pm 0.0088$  & 13 - 16 & 0.5 \\  
        & 3     &  $0.4215 \pm 0.0080$  & 13 - 16 & 0.2 \\  
        & 4     &  $0.4097 \pm 0.0078$  & 10 - 13 & 0.9 \\  \hline 
\end{tabular}
\caption{Values of $m_N a$ from a $32 \times 30 \times 32 \times 40$ 
lattice at $\beta$ of 6.17 using 219 configurations.
\label{tab:numx32}}
\end{center}
\end{table}

\begin{table}
\begin{center}
\begin{tabular}{ccccc}     \hline
k       & sink & $m_{\Delta} a$ & t & $\chi^2$ \\ \hline
0.1500  &  0     &  $0.7406 \pm 0.0034$  & 14 - 17 & 0.5 \\  
        &  1     &  $0.7398 \pm 0.0030$  & 14 - 17 & 0.7 \\  
        &  2     &  $0.7379 \pm 0.0036$  & 14 - 18 & 0.6 \\  
        &  3     &  $0.7368 \pm 0.0038$  & 14 - 18 & 0.4 \\  
        &  4     &  $0.7358 \pm 0.0045$  & 14 - 18 & 0.3 \\  \hline 
0.1519  &  012   &  $0.6126 \pm 0.0041$  & 13 - 17 & 0.6 \\  
        &  0     &  $0.6107 \pm 0.0040$  & 14 - 17 & 0.2 \\  
        &  1     &  $0.6101 \pm 0.0039$  & 14 - 17 & 0.7 \\  
        &  2     &  $0.6087 \pm 0.0047$  & 12 - 16 & 0.5 \\  
        &  3     &  $0.6085 \pm 0.0046$  & 13 - 16 & 0.5 \\  
        &  4     &  $0.6083 \pm 0.0058$  & 13 - 16 & 0.4 \\  \hline 
0.1526  &  012   &  $0.5656 \pm 0.0046$  & 13 - 17 & 0.4 \\  
        &  0     &  $0.5637 \pm 0.0049$  & 14 - 17 & 0.0 \\  
        &  1     &  $0.5626 \pm 0.0053$  & 14 - 17 & 0.2 \\  
        &  2     &  $0.5610 \pm 0.0056$  & 12 - 16 & 0.5 \\  
        &  3     &  $0.5608 \pm 0.0063$  & 13 - 16 & 0.4 \\  
        &  4     &  $0.5598 \pm 0.0073$  & 13 - 16 & 0.4 \\  \hline 
0.1532  &  012   &  $0.5267 \pm 0.0081$  & 13 - 17 & 0.9 \\  
        &  0     &  $0.5233 \pm 0.0088$  & 14 - 17 & 0.1 \\  
        &  1     &  $0.5188 \pm 0.0111$  & 14 - 17 & 0.0 \\  
        &  2     &  $0.5133 \pm 0.0104$  & 14 - 17 & 0.5 \\  
        &  3     &  $0.5197 \pm 0.0101$  & 11 - 16 & 0.9 \\  
        &  4     &  $0.5120 \pm 0.0101$  & 11 - 15 & 0.7 \\  \hline 
\end{tabular}
\caption{Values of $m_{\Delta} a$ from a $32 \times 30 \times 32 \times 40$ 
lattice at $\beta$ of 6.17 using 219 configurations.
\label{tab:demx32}}
\end{center}
\end{table}

\clearpage

\begin{table}
\begin{center}
\begin{tabular}{r@{}lllccc}     \hline
 \multicolumn{2}{c}{lattice}          & $\beta$ & sink & $m_n a$
& $m_s a$ & $m_{\rho} a$ \\
 \hline
 $16^3$           & $\:\times \: 32$ & 5.70 & 012 &  $0.00390 \pm 0.00012$ & $0.09662 \pm 0.00291$ 
& $0.5676 \pm 0.0079$ \\ 
                  &                  &      & 4   &  $0.00372 \pm 0.00018$ & $0.09223 \pm 0.00434$ 
& $0.5522 \pm 0.0132$ \\
 $24^3$           & $\:\times \: 32$ & 5.70 & 012 &  $0.00348 \pm 0.00012$ & $0.08620 \pm 0.00303$ 
& $0.5409 \pm 0.0089$ \\ 
                  &                  &      & 4   &  $0.00366 \pm 0.00029$ & $0.09070 \pm 0.00710$ 
& $0.5560 \pm 0.0212$ \\
 $24^3$           & $\:\times \: 36$ & 5.93 & 012 &  $0.00223 \pm 0.00009$ & $0.05536 \pm 0.00226$ 
& $0.3851 \pm 0.0079$ \\
                  &                  &      & 4   &  $0.00216 \pm 0.00008$ & $0.05364 \pm 0.00191$ 
& $0.3803 \pm 0.0064$ \\
 $30 \times 32^2$ & $\:\times \: 40$ & 6.17 & 012 &  $0.00141 \pm 0.00004$ & $0.03503 \pm 0.00096$ 
& $0.2768 \pm 0.0039$ \\
                  &                  &      & 4   &  $0.00139 \pm 0.00007$ & $0.03451 \pm 0.00164$ 
& $0.2736 \pm 0.0065$ \\
 \hline
\end{tabular}
\caption{The normal quark mass $m_n a$ and strange quark mass $m_s a$, and the rho mass extrapolated to 
quark mass $m_n a$.
\label{tab:mq_mrho}}
\end{center}
\end{table}

\begin{table}
\begin{center}
\begin{tabular}{clrr}     \hline
    & sink &  & $\chi^2$ \\ \hline

 $m_{K^*} / m_{\rho}$      &  012 & $1.106 \pm 0.009$ & 0.6 \\	&    4 & $1.123 \pm 0.015$ & 0.7 \\       
 $m_{\Phi} / m_{\rho}$     &  012 & $1.213 \pm 0.017$ & 0.6 \\	&    4 & $1.246 \pm 0.030$ & 0.7 \\       
 $m_{N} / m_{\rho}$        &  012 & $1.459 \pm 0.034$ & 1.3 \\	&    4 & $1.371 \pm 0.038$ & 1.9 \\       
 $\Delta m / m_{\rho}$     &  012 & $1.914 \pm 0.031$ & 1.3 \\	&    4 & $1.939 \pm 0.027$ & 1.9 \\       
 $m_{\Delta} / m_{\rho}$   &  012 & $1.922 \pm 0.036$ & 3.4 \\	&    4 & $1.786 \pm 0.048$ & 0.0 \\       
 $m_{\Sigma^*} / m_{\rho}$ &  012 & $1.986 \pm 0.024$ & 3.4 \\	&    4 & $1.903 \pm 0.032$ & 0.0 \\       
 $m_{\Xi^*} / m_{\rho}$    &  012 & $2.049 \pm 0.030$ & 3.4 \\	&    4 & $2.020 \pm 0.037$ & 0.0 \\       
 $m_{\Omega} / m_{\rho}$   &  012 & $2.112 \pm 0.050$ & 3.4 \\	&    4 & $2.137 \pm 0.043$ & 0.0 \\       
 $\Lambda^{(0)}_{\overline{MS}} / m_{\rho}$ 
			   &  012 & $0.301 \pm 0.007$ & --- \\	&    4 & $0.301 \pm 0.007$ & --- \\        
 \hline
\end{tabular}
\caption{Hadron masses extrapolated to physical quark mass measured in
units of $m_{\rho}$ extrapolated to physical quark mass from a $16^3 \times 32$
lattice at $\beta$ of 5.70.
The mass difference $\Delta m$ is $m_{\Xi} + m_{\Sigma} - m_N$.
\label{tab:mexx16}}
\end{center}
\end{table}

\begin{table}
\begin{center}
\begin{tabular}{clrr}     \hline
   & sink &  & $\chi^2$ \\ \hline

 $m_{K^*} / m_{\rho}$      &  012 & $1.124 \pm 0.007$ & 0.0 \\	&    4 & $1.099 \pm 0.023$ & 0.0 \\       
 $m_{\Phi} / m_{\rho}$     &  012 & $1.249 \pm 0.014$ & 0.0 \\	&    4 & $1.198 \pm 0.046$ & 0.0 \\       
 $m_{N} / m_{\rho}$        &  012 & $1.397 \pm 0.036$ & 0.2 \\	&    4 & $1.373 \pm 0.064$ & 0.0 \\       
 $\Delta m / m_{\rho}$     &  012 & $1.975 \pm 0.015$ & 0.2 \\	&    4 & $1.940 \pm 0.035$ & 0.0 \\       
 $m_{\Delta} / m_{\rho}$   &  012 & $1.879 \pm 0.052$ & 0.4 \\	&    4 & $1.775 \pm 0.090$ & 0.9 \\       
 $m_{\Sigma^*} / m_{\rho}$ &  012 & $1.987 \pm 0.033$ & 0.4 \\	&    4 & $1.890 \pm 0.070$ & 0.9 \\       
 $m_{\Xi^*} / m_{\rho}$    &  012 & $2.095 \pm 0.022$ & 0.4 \\	&    4 & $2.005 \pm 0.056$ & 0.9 \\       
 $m_{\Omega} / m_{\rho}$   &  012 & $2.203 \pm 0.028$ & 0.4 \\	&    4 & $2.119 \pm 0.055$ & 0.9 \\       
 $\Lambda^{(0)}_{\overline{MS}} / m_{\rho}$ 
			   &  012 & $0.299 \pm 0.011$ & --- \\	&    4 & $0.299 \pm 0.011$ & --- \\        
 \hline
\end{tabular}
\caption{Hadron masses extrapolated to physical quark mass measured in
units of $m_{\rho}$ extrapolated to physical quark mass from a $24^3 \times 32$
lattice at $\beta$ of 5.70.
The mass difference $\Delta m$ is $m_{\Xi} + m_{\Sigma} - m_N$.
\label{tab:mexx24}}
\end{center}
\end{table}

\begin{table}
\begin{center}
\begin{tabular}{clrr}     \hline
   & sink &  & $\chi^2$ \\ \hline

 $m_{K^*} / m_{\rho}$      &  012 & $1.121 \pm 0.008$ & 0.4 \\	&    4 & $1.126 \pm 0.007$ & 0.0 \\       
 $m_{\Phi} / m_{\rho}$     &  012 & $1.242 \pm 0.016$ & 0.4 \\	&    4 & $1.252 \pm 0.013$ & 0.0 \\       
 $m_{N} / m_{\rho}$        &  012 & $1.347 \pm 0.028$ & 0.0 \\	&    4 & $1.340 \pm 0.031$ & 1.9 \\       
 $\Delta m / m_{\rho}$     &  012 & $1.890 \pm 0.014$ & 0.0 \\	&    4 & $1.905 \pm 0.027$ & 1.9 \\       
 $m_{\Delta} / m_{\rho}$   &  012 & $1.772 \pm 0.038$ & 0.6 \\	&    4 & $1.728 \pm 0.036$ & 0.2 \\       
 $m_{\Sigma^*} / m_{\rho}$ &  012 & $1.893 \pm 0.034$ & 0.6 \\	&    4 & $1.863 \pm 0.033$ & 0.2 \\       
 $m_{\Xi^*} / m_{\rho}$    &  012 & $2.014 \pm 0.030$ & 0.6 \\	&    4 & $1.998 \pm 0.029$ & 0.2 \\       
 $m_{\Omega} / m_{\rho}$   &  012 & $2.135 \pm 0.031$ & 0.6 \\	&    4 & $2.134 \pm 0.035$ & 0.2 \\       
 $\Lambda^{(0)}_{\overline{MS}} / m_{\rho}$ 
			   &  012 & $0.301 \pm 0.005$ & --- \\	&    4 & $0.301 \pm 0.005$ & --- \\        
 \hline
\end{tabular}
\caption{Hadron masses extrapolated to physical quark mass measured in
units of $m_{\rho}$ extrapolated to physical quark mass from a $24^3 \times 36$
lattice at $\beta$ of 5.93.
The mass difference $\Delta m$ is $m_{\Xi} + m_{\Sigma} - m_N$.
\label{tab:mexx24t36}}
\end{center}
\end{table}

\begin{table}
\begin{center}
\begin{tabular}{clrr}     \hline
    & sink & & $\chi^2$ \\ \hline

 $m_{K^*} / m_{\rho}$      &  012 & $1.128 \pm 0.004$ & 0.6 \\	&    4 & $1.131 \pm 0.004$ & 0.0 \\       
 $m_{\Phi} / m_{\rho}$     &  012 & $1.256 \pm 0.007$ & 0.6 \\	&    4 & $1.262 \pm 0.009$ & 0.0 \\       
 $m_{N} / m_{\rho}$        &  012 & $1.383 \pm 0.030$ & 2.2 \\	&    4 & $1.360 \pm 0.040$ & 4.6 \\       
 $\Delta m / m_{\rho}$     &  012 & $1.888 \pm 0.015$ & 2.2 \\	&    4 & $1.886 \pm 0.024$ & 4.6 \\       
 $m_{\Delta} / m_{\rho}$   &  012 & $1.782 \pm 0.039$ & 0.0 \\	&    4 & $1.751 \pm 0.055$ & 0.3 \\       
 $m_{\Sigma^*} / m_{\rho}$ &  012 & $1.907 \pm 0.031$ & 0.0 \\	&    4 & $1.885 \pm 0.046$ & 0.3 \\       
 $m_{\Xi^*} / m_{\rho}$    &  012 & $2.033 \pm 0.025$ & 0.0 \\	&    4 & $2.018 \pm 0.037$ & 0.3 \\       
 $m_{\Omega} / m_{\rho}$   &  012 & $2.158 \pm 0.018$ & 0.0 \\	&    4 & $2.152 \pm 0.032$ & 0.3 \\       
 $\Lambda^{(0)}_{\overline{MS}} / m_{\rho}$ 
			   &  012 & $0.302 \pm 0.007$ & --- \\	&    4 & $0.302 \pm 0.007$ & --- \\        
 \hline
\end{tabular}
\caption{Hadron masses extrapolated to physical quark mass measured in
units of $m_{\rho}$ extrapolated to physical quark mass from a $30
\times 32^2 \times 40$
lattice at $\beta$ of 6.17.
The mass difference $\Delta m$ is $m_{\Xi} + m_{\Sigma} - m_N$.
\label{tab:mexx32}}
\end{center}
\end{table}

\begin{table}
\begin{center}
\begin{tabular}{cllr@{}l@{$\pm$}lr@{}l@{$\pm$}lr@{}l}     
 \hline
 & sink & $\chi^2$ & \multicolumn{3}{c}{fin. vol.} & 
 \multicolumn{3}{c}{inf. vol.} 
  & \multicolumn{2}{c}{obs.}\\ \hline
 
 $m_{K^*} / m_{\rho}$           & 012 & 0.01 & 1     &.149	& 0.009	& 1     &.166	& 0.016	& 1 & .164\\ 
                                & 4   & 0.03 & 1     &.140      & 0.016 & 1     &.116   & 0.032 & \\
 $m_{\Phi} / m_{\rho}$          & 012 & 0.01 & 1     &.297	& 0.019	& 1     &.333	& 0.032	& 1 &  .327\\ 
                                & 4   & 0.03 & 1     &.280      & 0.032 & 1     &.232   & 0.064 & \\
 $m_{N} / m_{\rho}$             & 012 & 2.61 & 1     &.278	& 0.068	& 1     &.216	& 0.104	& 1 &  .222\\ 
                                & 4   & 0.27 & 1     &.328      & 0.085 & 1     &.330   & 0.131 & \\
 $\Delta m / m_{\rho}$          & 012 & 0.12 & 1     &.865	& 0.045	& 1     &.927	& 0.074	& 2 &  .047\\
\hline
\end{tabular}
\caption{
Calculated values of hadron mass ratios at physical quark masses,
extrapolated to zero lattice spacing in finite volume, then corrected to
infinite volume, compared with observed values.
The mass difference $\Delta m$ is $m_{\Xi} + m_{\Sigma} - m_N$.
\label{tab:results}}
\end{center}
\end{table}

\begin{table}
\begin{center}
\begin{tabular}{crrr}     \hline
   & k & sink 012 & sink 4 \\ \hline

$m_{\pi}$       & 0.1650  & $ -0.2 	\pm 0.5 \% $ 	& $ 0.2 	\pm 0.8 \% $ \\
                & 0.1663  & $ -1.8 	\pm 0.7 \% $ 	& $ -0.9 	\pm 1.1 \% $ \\
                & 0.1675  & $ -1.1 	\pm 1.5 \% $ 	& $ -1.2 	\pm 1.9 \% $ \\\hline
$m_{\rho}$      & 0.1650  & $ -0.6 	\pm 0.9 \% $ 	& $ -3.3 	\pm 1.5 \% $ \\
                & 0.1663  & $ -1.5 	\pm 1.4 \% $ 	& $ -0.9 	\pm 2.1 \% $ \\
                & 0.1675  & $ -3.4 	\pm 1.4 \% $ 	& $ -1.6 	\pm 2.6 \% $ \\\hline
$m_{N}$         & 0.1650  & $ 0.2 	\pm 1.4 \% $ 	& $ -0.2 	\pm 1.8 \% $ \\
                & 0.1663  & $ -4.6 	\pm 1.8 \% $ 	& $ -1.6 	\pm 2.0 \% $ \\
                & 0.1675  & $ -4.8 	\pm 2.0 \% $ 	& $ 0.7 	\pm 2.4 \% $ \\\hline
$m_{\Delta}$    & 0.1650  & $ 0.1 	\pm 1.6 \% $ 	& $ 0.1 	\pm 2.1 \% $ \\
                & 0.1663  & $ -0.4 	\pm 2.0 \% $ 	& $ -1.6 	\pm 2.6 \% $ \\
                & 0.1675  & $ -4.9 	\pm 2.2 \% $ 	& $ -0.0 	\pm 2.3 \% $ \\\hline
\end{tabular}
\caption{
Changes in masses from 
$16^3 \times 32$ to $24^3 \times 32$ 
at $\beta = 5.70$. 
\label{tab:massvoldep}}
\end{center}
\end{table}

\begin{table}
\begin{center}
\begin{tabular}{crr}     \hline
   & sink 012 & sink 4 \\ \hline

 $m_{K^*} / m_{\rho}$         & $ 1.6 	\pm 0.9 \% $ 	& $ -2.2 	\pm 2.2 \% $ \\
 $m_{\Phi} / m_{\rho}$        & $ 2.9 	\pm 1.6 \% $ 	& $ -4.0 	\pm 4.0 \% $ \\
 $m_{N} / m_{\rho}$           & $ -4.5 	\pm 3.7 \% $ 	& $ 0.2 	\pm 5.8 \% $ \\
 $\Delta m / m_{\rho}$        & $ 3.1 	\pm 2.1 \% $ 	& $ 0.0 	\pm 2.5 \% $ \\
 $m_{\Delta} / m_{\rho}$      & $ -2.3 	\pm 3.3 \% $ 	& $ -0.6 	\pm 5.8 \% $ \\
 $m_{\Sigma^*} / m_{\rho}$    & $ 0.1 	\pm 2.1 \% $ 	& $ -0.7 	\pm 4.4 \% $ \\
 $m_{\Xi^*} / m_{\rho}$       & $ 2.2 	\pm 1.7 \% $ 	& $ -0.8 	\pm 3.3 \% $ \\
 $m_{\Omega} / m_{\rho}$      & $ 4.1 	\pm 2.4 \% $ 	& $ -0.8 	\pm 3.2 \% $ \\
 $\Lambda^{(0)}_{\overline{MS}} / m_{\rho}$   & $ 4.7 	\pm 2.1 \% $ & $ -0.7 	\pm 4.3 \% $ \\  \hline
\end{tabular}
\caption{
Changes in mass ratios from $16^3 \times 32$ to $24^3 \times 32$ at
$\beta = 5.70$ The mass difference $\Delta m$ is $m_{\Xi} + m_{\Sigma} -
m_N$. \label{tab:ratiovoldep}}
\end{center}
\end{table}

\begin{table}
\begin{center}
\begin{tabular}{clcc}     \hline
    & sink &  & $\chi^2$ \\ \hline

 $\frac{m_{ns}}{m_{\rho}(m_{ns})}
  \frac{\partial m_{\rho}}{\partial m_q}(m_{ns})$ &  012 & $0.093 \pm 0.008$ & 0.6 \\
                                                  &  4   & $0.109 \pm 0.015$ & 0.7 \\       
 $\frac{m_{N}(m_{ns})}{m_{\rho}(m_{ns})}$               &  012 & $1.517 \pm 0.014$ & 1.3 \\	
                                                  &  4   & $1.465 \pm 0.022$ & 1.9 \\       
 $\frac{m_{ns}}{m_{\rho}(m_{ns})}                
  \frac{\partial m_N}{\partial m_q}(m_{ns})$      &  012 & $0.199 \pm 0.026$ & 1.3 \\	
                                                  &  4   & $0.253 \pm 0.017$ & 1.9 \\       
 $\frac{m_{\Delta}(m_{ns})}{ m_{\rho}(m_{ns})}$          &  012 & $1.834 \pm 0.016$ & 3.4 \\	
                                                  &  4   & $1.750 \pm 0.024$ & 0.0 \\       
 $\frac{m_{ns}}{m_{\rho}(m_{ns})}                
  \frac{\partial m_{\Delta}}{\partial m_q}(m_{ns})$ &  012 & $0.083 \pm 0.029$ & 3.4 \\	
                                                  &  4   & $0.156 \pm 0.021$ & 0.0 \\  
 $\frac{\Lambda^{(0)}_{\overline{MS}}}{m_{\rho}(m_{ns})}$
                                                  &  012 & $0.2677 \pm 0.0027$ & $-$ \\
                                                  &  4   & $0.2745 \pm 0.0039$ & $-$ \\
 \hline
\end{tabular}
\caption{Hadron mass parameters determined without extrapolation in
quark mass for the lattice $16^3 \times 32$
lattice at $\beta$ of 5.70.
\label{tab:mnxx16}}
\end{center}
\end{table}

\begin{table}
\begin{center}
\begin{tabular}{clcc}     \hline
    & sink &  & $\chi^2$ \\ \hline
 
 $\frac{m_{ns}}{m_{\rho}(m_{ns})}                
  \frac{\partial m_{\rho}}{\partial m_q}(m_{ns})$ &  012 & $0.111 \pm 0.007$ & 0.0 \\
                                                  &  4   & $0.086 \pm 0.022$ & 0.0 \\       
 $\frac{m_{N}(m_{ns})}{m_{\rho}(m_{ns})}$               &  012 & $1.492 \pm 0.017$ & 0.2 \\	
                                                  &  4   & $1.490 \pm 0.026$ & 0.0 \\       
 $\frac{m_{ns}}{m_{\rho}(m_{ns})}                
  \frac{\partial m_N}{\partial m_q}(m_{ns})$      &  012 & $0.258 \pm 0.017$ & 0.2 \\	
                                                  &  4   & $0.246 \pm 0.027$ & 0.0 \\       
 $\frac{m_{\Delta}(m_{ns})}{ m_{\rho}(m_{ns})}$          &  012 & $1.820 \pm 0.023$ & 0.4 \\	
                                                  &  4   & $1.772 \pm 0.039$ & 0.9 \\       
 $\frac{m_{ns}}{m_{\rho}(m_{ns})}                
  \frac{\partial m_{\Delta}}{\partial m_q}(m_{ns})$ &  012 & $0.144 \pm 0.027$ & 0.4 \\	
                                                  &  4   & $0.149 \pm 0.033$ & 0.9 \\       
 $\frac{\Lambda^{(0)}_{\overline{MS}}}{m_{\rho}(m_{ns})}$
                                   &  012 & $0.2759 \pm 0.0029$ & $-$ \\
                                   &    4 & $0.2752 \pm 0.0047$ & $-$ \\
\hline
\end{tabular}
\caption{Hadron mass parameters determined without extrapolation in
quark mass for the lattice $24^3 \times 32$
lattice at $\beta$ of 5.70.
\label{tab:mnxx24}}
\end{center}
\end{table}

\begin{table}
\begin{center}
\begin{tabular}{clcc}     \hline
    & sink &  & $\chi^2$ \\ \hline

 $\frac{m_{ns}}{m_{\rho}(m_{ns})}                 
  \frac{\partial m_{\rho}}{\partial m_q}(m_{ns})$ &  012 & $0.107 \pm 0.008$ & 0.4 \\
                                                  &    4 & $0.113 \pm 0.007$ & 0.0 \\       
 $\frac{m_{N}(m_{ns})}{m_{\rho}(m_{ns})}$               &  012 & $1.435 \pm 0.015$ & 0.0 \\	
                                                  &    4 & $1.433 \pm 0.017$ & 1.9 \\       
 $\frac{m_{ns}}{m_{\rho}(m_{ns})}                
  \frac{\partial m_N}{\partial m_q}(m_{ns})$      &  012 & $0.241 \pm 0.010$ & 0.0 \\	
                                                  &    4 & $0.252 \pm 0.018$ & 1.9 \\       
 $\frac{m_{\Delta}(m_{ns})}{ m_{\rho}(m_{ns})}$          &  012 & $1.745 \pm 0.023$ & 0.6 \\	
                                                  &    4 & $1.715 \pm 0.024$ & 0.2 \\       
 $\frac{m_{ns}}{m_{\rho}(m_{ns})}                
  \frac{\partial m_{\Delta}}{\partial m_q}(m_{ns})$ &  012 & $0.161 \pm 0.012$ & 0.6 \\	
                                                  &    4 & $0.181 \pm 0.012$ & 0.2 \\       
 $\frac{\Lambda^{(0)}_{\overline{MS}}}{m_{\rho}(m_{ns})}$
                                   &  012 & $0.2679 \pm 0.0041$ & $-$ \\
                                   &    4 & $0.2698 \pm 0.0034$ & $-$ \\
 \hline
\end{tabular}
\caption{Hadron mass parameters determined without extrapolation in
quark mass for the lattice $24^3 \times 36$
lattice at $\beta$ of 5.93.
\label{tab:mnxx24t36}}
\end{center}
\end{table}

\begin{table}
\begin{center}
\begin{tabular}{clrr}     \hline
    & sink &  & $\chi^2$ \\ \hline
 
 $\frac{m_{ns}}{m_{\rho}(m_{ns})}                
  \frac{\partial m_{\rho}}{\partial m_q}(m_{ns})$ &  012 & $0.115 \pm 0.004$ & 0.6 \\
                                                  &    4 & $0.118 \pm 0.005$ & 0.0 \\       
 $\frac{m_{N}(m_{ns})}{m_{\rho}(m_{ns})}$               &  012 & $1.445 \pm 0.017$ & 2.2 \\	
                                                  &    4 & $1.430 \pm 0.024$ & 4.6 \\       
 $\frac{m_{ns}}{m_{\rho}(m_{ns})}                
  \frac{\partial m_N}{\partial m_q}(m_{ns})$      &  012 & $0.226 \pm 0.010$ & 2.2 \\	
                                                  &    4 & $0.236 \pm 0.012$ & 4.6 \\       
 $\frac{m_{\Delta}(m_{ns})}{ m_{\rho}(m_{ns})}$          &  012 & $1.749 \pm 0.019$ & 0.0 \\	
                                                  &    4 & $1.727 \pm 0.033$ & 0.3 \\       
 $\frac{m_{ns}}{m_{\rho}(m_{ns})}                
  \frac{\partial m_{\Delta}}{\partial m_q}(m_{ns})$ &  012 & $0.169 \pm 0.011$ & 0.0 \\	
                                                  &    4 & $0.180 \pm 0.012$ & 0.3 \\       
 $\frac{\Lambda^{(0)}_{\overline{MS}}}{m_{\rho}(m_{ns})}$
                                   &  012 & $0.2671 \pm 0.0032$ & $-$ \\
                                   &    4 & $0.2695 \pm 0.0057$ & $-$ \\
 \hline
\end{tabular}
\caption{Hadron mass parameters determined without extrapolation in
quark mass for the lattice $30 \times 32^2 \times 40$
lattice at $\beta$ of 6.17.
\label{tab:mnxx32}}
\end{center}
\end{table}

\begin{table}
\begin{center}
\begin{tabular}{clcccc}     \hline
    & sink &  $\chi^2$ & \multicolumn{1}{c}{fin. vol.} & 
    \multicolumn{1}{c}{inf. vol.} & \multicolumn{1}{c}{obs.} \\ \hline

 $\frac{m_{ns}}{m_{\rho}(m_{ns})}                
  \frac{\partial m_{\rho}}{\partial m_q}(m_{ns})$ &  012 & 0.0 & $0.136 \pm 0.010$ & $0.154 \pm 0.016$& 0.152\\
                                                  &    4 & 0.0 & $0.127 \pm 0.017$ & $0.103 \pm 0.032$&   \\ 
 $\frac{m_{N}(m_{ns})}{m_{\rho}(m_{ns})}$         &  012 & 3.2 & $1.342 \pm 0.031$ & $1.317 \pm 0.045$& 1.405\\	
                                                  &    4 & 0.2 & $1.384 \pm 0.047$ & $1.409 \pm 0.071$&   \\
 $\frac{m_{ns}}{m_{\rho}(m_{ns})}                
  \frac{\partial m_N}{\partial m_q}(m_{ns})$      &  012 & 2.5 & $0.239 \pm 0.028$ & $0.298 \pm 0.055$& 0.384 \\
                                                  &    4 & 0.2 & $0.221 \pm 0.028$ & $0.215 \pm 0.062$&   \\ 
 $\frac{m_{\Delta}(m_{ns})}{ m_{\rho}(m_{ns})}$   &  012 & 1.6 & $1.639 \pm 0.040$ & $1.625 \pm 0.053$& 1.628 \\
                                                  &    4 & 0.4 & $1.684 \pm 0.053$ & $1.707 \pm 0.082$&   \\ 
 $\frac{m_{ns}}{m_{\rho}(m_{ns})}                
  \frac{\partial m_{\Delta}}{\partial m_q}(m_{ns})$ & 012 & 1.5 & $0.245 \pm 0.032$ & $0.306 \pm 0.059$& 0.267\\
                                                  &    4 & 0.2 & $0.205 \pm 0.041$ & $0.198 \pm 0.075$&   \\ 
 $\frac{\Lambda^{(0)}_{\overline{MS}}}{m_{\rho}(m_{ns})}$ 
                                   &  012 & $-$ & $0.2669 \pm 0.0057$ &
                                      $0.2750 \pm 0.0077$& $0.262 \pm 0.015$ \\
                                   &    4 & $-$ & $0.2698 \pm 0.0034$ & $0.2728 \pm 0.0120$& \\
 \hline
\end{tabular}
\caption{
Calculated values of hadron mass parameters determined without
extrapolation in quark mass,  
extrapolated to zero lattice spacing in finite volume, then corrected to
infinite volume, compared with observed values. \label{tab:nexresults}}
\end{center}
\end{table}


\clearpage

\clearpage

\clearpage

\clearpage


\clearpage

\clearpage

\clearpage

\clearpage


\clearpage

\clearpage

\clearpage

\clearpage

\clearpage

\clearpage


\clearpage

\clearpage

\clearpage

\clearpage

\clearpage

\clearpage


\clearpage

\clearpage

\clearpage

\clearpage

\clearpage

\clearpage


\clearpage

\clearpage

\clearpage

\clearpage

\clearpage

\clearpage


\clearpage

\clearpage

\clearpage

\clearpage

\clearpage

\clearpage

\clearpage

\clearpage

\clearpage

\clearpage

\clearpage

\clearpage

\clearpage

\clearpage


\clearpage


\clearpage

\end{document}